\documentclass[prd,aps,nofootinbib,floatfix,superscriptaddress]{revtex4}
\usepackage[T1]{fontenc}
\usepackage{graphicx}
\usepackage{epsfig}
\usepackage{rotating}
\usepackage{amssymb}
\usepackage{subfigure}
\usepackage{dsfont}
\usepackage{psfrag}
\usepackage{amsmath,euscript,array,mathrsfs}
\usepackage{bbold}
\usepackage{epsf}
\usepackage{color}

\newcommand{\comm}[1]{} % Comment blocks

\newcommand{\beq}{\begin{equation}}
\newcommand{\eeq}{\end{equation}}
\newcommand{\beqs}{\begin{eqnarray}}
\newcommand{\eeqs}{\end{eqnarray}}

\newcommand{\gsim}{\mathrel{\raisebox{-
.6ex}{$\stackrel{\textstyle>}{\sim}$}}}

\renewcommand{\L}{{\cal L}}

\def\hbar{\hspace{0pt}\raisebox{1pt}{$-$} \hspace{-7pt} h}

\def\di{\mbox{d}}
\def\r{\rho}

\newcommand{\be}{\begin{equation}}
\newcommand{\ee}{\end{equation}}
\newcommand{\bea}{\begin{eqnarray}}
\newcommand{\eea}{\end{eqnarray}}
\newcommand{\nn}{\nonumber}

\def\lbldef#1#2{\expandafter\gdef\csname #1\endcsname {#2}}

\def\href#1#2{#2}
\def\half{{1 \over 2}}

\newcommand{\ber}{\begin{eqnarray}}
\newcommand{\eer}{\end{eqnarray}}

\newcommand{\beqar}{\begin{eqnarray}}

\newcommand{\cF}{{\cal F}}

\newcommand{\eeqar}{\end{eqnarray}}

\newcommand{\dsl}{\kern.06em\hbox{\raise.15ex\hbox{$/$}\kern-.56em\hbox{$\partial$}}}

\newcommand{\eeqarr}{\end{eqnarray}}
\newcommand{\ZZ}{{\rm \kern 0.275em Z \kern -0.92em Z}\;}

\def\CC{{\mathchoice
{\rm C\mkern-8mu\vrule height1.45ex depth-.05ex
width.05em\mkern9mu\kern-.05em}
{\rm C\mkern-8mu\vrule height1.45ex depth-.05ex
width.05em\mkern9mu\kern-.05em}
{\rm C\mkern-8mu\vrule height1ex depth-.07ex
width.035em\mkern9mu\kern-.035em}
{\rm C\mkern-8mu\vrule height.65ex depth-.1ex
width.025em\mkern8mu\kern-.025em}}}

\def\RR{{\rm I\kern-1.6pt {\rm R}}}

\def\ZZ{{\rm Z}\kern-3.8pt {\rm Z} \kern2pt}
\def\IB{\relax{\rm I\kern-.18em B}}
\def\ID{\relax{\rm I\kern-.18em D}}
\def\II{\relax{\rm I\kern-.18em I}}
\def\IP{\relax{\rm I\kern-.18em P}}

\newcommand{\bear}{\begin{eqnarray}}
\newcommand{\eear}{\end{eqnarray}}

\def\to{\rightarrow}

\def\to{\rightarrow}

\def\c{\gamma}

\def\f{\phi}

\def\l{\lambda}

\def\r{\rho}
\def\s{\sigma}

\def\6{\partial}

\def\bea{\begin{eqnarray}}
\def\eea{\end{eqnarray}}

\def\beqx{\begin{displaymath}}
\def\eeqx{\end{displaymath}}

\newcommand{\bmat}{\left(\begin{array}}
\newcommand{\emat}{\end{array}\right)}

\def\half{\frac{1}{2}}

\def\c{\chi}

\def\f{\phi}

\def\l{\lambda}

\def\p{\pi}

\def\r{\rho}
\def\s{\sigma}

\def\L{\Lambda}

\def\cf{{\cal F}}

\def\bo{{\raise-.3ex\hbox{\large$\Box$}}}               % D'Alembertian
\def\face{{\raise.2ex\hbox{$\displaystyle \bigodot$}\mskip-2.2mu \llap {$\ddot
        \smile$}}}                                   % happy face
\def\>{\rangle}                                      %right angle
\def\<{\langle}                                      %left angle

\def\leftrightarrowfill{$\mathsurround=0pt \mathord\leftarrow \mkern-6mu
        \cleaders\hbox{$\mkern-2mu \mathord- \mkern-2mu$}\hfill
        \mkern-6mu \mathord\rightarrow$}        % <--> double differential
\def\dvec#1{\vbox{\ialign{##\crcr
        \leftrightarrowfill\crcr\noalign{\kern-1pt\nointerlineskip}
        $\hfil\displaystyle{#1}\hfil$\crcr}}}           % <--> accent
\def\-{\hphantom{-}}
\newcommand{\dd}{\mbox{d}}

\begin{document}

\title{Dilatonic states near holographic phase transitions}

\author{Daniel Elander}
\affiliation{Laboratoire Charles Coulomb (L2C), University of Montpellier, CNRS, Montpellier, France}

\author{Maurizio Piai}
\affiliation{Department of Physics, College of Science, Swansea University,
Singleton Park, SA2 8PP, Swansea, Wales, UK}

\author{John Roughley}
\affiliation{Department of Physics, College of Science, Swansea University,
Singleton Park, SA2 8PP, Swansea, Wales, UK}

\date{\today}

\begin{abstract}
The spectrum of bound states of special strongly coupled confining field theories might include a parametrically light dilaton, associated with the formation of enhanced condensates that break (approximate) scale invariance spontaneously. It has been suggested in the literature that such a state may arise in connection with the theory being close to the unitarity bound in holographic models. We extend these ideas to cases where the background geometry is non-AdS, and the gravity description of the dual confining field theory has a top-down origin in supergravity.

We exemplify this programme by studying the circle compactification of Romans six-dimensional half-maximal supergravity. We uncover a rich space of solutions, many of which were previously unknown in the literature. We compute the bosonic spectrum of excitations, and identify a tachyonic instability in a region of parameter space for a class of regular background solutions. A tachyon only exists along an energetically disfavoured (unphysical) branch of solutions of the gravity theory; we find evidence of a first-order phase transition that separates this region of parameter space from the physical one. Along the physical branch of regular solutions, one of the lightest scalar particles is approximately a dilaton, and it is associated with a condensate in the underlying theory. Yet, because of the location of the phase transition, its mass is not parametrically small, and it is, coincidentally, the next-to-lightest scalar bound state, rather than the lightest one.
\end{abstract}

\maketitle

\tableofcontents

\section{Introduction}
\label{Sec:Introduction}

The Standard Model (SM) of particle physics is likely to be replaced by
a more complete theory above some unknown new physics scale $\Lambda$.
Yet, the discovery by the LHC collaborations
of the Higgs particle~\cite{Aad:2012tfa,Chatrchyan:2012xdj}, with
mass $m_h\simeq 126$ GeV, has not been
accompanied by convincing signals of new phenomena in (direct and indirect) searches,
further pushing the hypothetical scale $\Lambda$ into the multi-TeV range.
This observation hints at a difficulty in the application
of effective field theory (EFT) ideas to high energy particle physics.
If the fundamental theory  completing the Standard Model
above $\Lambda$ plays a role (even indirectly) in electroweak symmetry
breaking (EWSB) and Higgs physics, it is technically difficult to implement the
hierarchy $m_h \ll \Lambda$,  and justify it
inside the general low-energy EFT paradigm.
The resulting low-energy description requires  fine-tuning;
in the literature, this tension is referred to as  the
{\it little hierarchy problem}.

The Higgs particle
might emerge  at the dynamical scale $\Lambda$ from strongly-coupled new physics.
If one could dial
the effects of explicit breaking of  scale invariance to be smaller than those associated
with its spontaneous breaking, the Higgs boson could be identified with
the pseudo-Nambu-Goldstone boson (pNGB) associated with the spontaneous breaking of
dilatation invariance: the {\it dilaton}.
If furthermore its mass $m_h$ could be made small enough to yield the hierarchy  $m_h \ll \Lambda$,
and without  fine-tuning,
then the little hierarchy problem would be solved.

The properties of the EFT description of the dilaton are the subject of a vast body of  literature (see for instance Refs.~\cite{Matsuzaki:2013eva, Golterman:2016lsd,Kasai:2016ifi,Golterman:2016hlz,Hansen:2016fri,Golterman:2016cdd,Appelquist:2017wcg,Appelquist:2017vyy,Golterman:2018mfm,Cata:2019edh,Appelquist:2019lgk,Cata:2018wzl,Brown:2019ipr}), which includes well known studies dating from a long time ago~\cite{Migdal:1982jp,Coleman:1985rnk}. The details of how this idea is implemented in phenomenologically relevant models of EWSB are the subject of many studies (see for example Refs.~\cite{Goldberger:2008zz,Hong:2004td,Dietrich:2005jn,Hashimoto:2010nw,Appelquist:2010gy,Vecchi:2010gj, Chacko:2012sy, Bellazzini:2012vz, Abe:2012eu,Eichten:2012qb,Bellazzini:2013fga,Hernandez-Leon:2017kea}),
and some date back to earlier days of dynamical EWSB symmetry breaking and walking technicolor~\cite{Leung:1985sn,Bardeen:1985sm,Yamawaki:1985zg}.

The main limitation to the study from first principles of dilaton dynamics with strongly-coupled origin
comes from calculability. For example, lattice studies  have
started to uncover evidence that an anomalously light scalar particle appears
in confining gauge theories that are believed to be close to the edge of the conformal window,
namely in  $SU(3)$ gauge theories with either $N_f=8$ fundamental Dirac
fermions~\cite{Aoki:2014oha,Appelquist:2016viq,Aoki:2016wnc,Gasbarro:2017fmi,Appelquist:2018yqe},
or with $N_f=2$ Dirac fermions transforming in the two-index symmetric
representation~\cite{Fodor:2012ty,Fodor:2015vwa,Fodor:2016pls,Fodor:2017nlp,Fodor:2019vmw,Fodor:2020niv}.
It may be premature to conclude that these works have uncovered firm evidence that
the scalar particle is a dilaton, but EFT-based studies yield
encouraging indications in this direction (see for example Refs.~\cite{Appelquist:2017vyy,Fodor:2019vmw,Golterman:2020tdq}).

A complementary approach exploits holography and
gauge-gravity correspondences~\cite{Maldacena:1997re,Gubser:1998bc,
	Witten:1998qj}. The properties of some special strongly-coupled field theories can be derived from weakly-coupled gravity theories in higher dimensions
(see the introductory review in Ref.~\cite{Aharony:1999ti}).
A systematic prescription exists for calculating correlation functions and condensates,
via holographic renormalisation~\cite{Bianchi:2001kw} (see  the
lecture notes in Refs.~\cite{Skenderis:2002wp,Papadimitriou:2004ap}).
The study of simplified toy models, implementing the Goldberger-Wise stabilisation
mechanism~\cite{Goldberger:1999uk,DeWolfe:1999cp,Goldberger:1999un,Csaki:2000zn,
	ArkaniHamed:2000ds,Rattazzi:2000hs,Kofman:2004tk}, shows the
presence of a light dilaton in the spectrum.
Attempts at constructing phenomenologically more realistic models,
while disregarding the fundamental origin of the higher-dimensional theory
(the {\it bottom-up} approach to holography)
yield similarly encouraging results~\cite{Elander:2011aa,Elander:2012fk,Lawrance:2012cg,
	Kutasov:2012uq,Goykhman:2012az,Evans:2013vca,
	Megias:2014iwa,Elander:2015asa,Pomarol:2019aae}.

Evidence that strong dynamics can lead to the formation of a light
dilaton has been confirmed also
in the context of less realistic, but
more rigorous holographic models 
built starting from
supergravity
({\it top-down} approach)~\cite{Elander:2017cle,Elander:2017hyr}
(see also Refs.~\cite{Nunez:2008wi,Elander:2009pk,
	Elander:2012yh,Elander:2014ola}).
So far, this has been shown to be true only inside the special framework of
a particular five-dimensional sigma-model coupled to gravity, the solutions of which
lift to backgrounds with geometry
related to the conifold~\cite{Candelas:1989js,Chamseddine:1997nm,Klebanov:1998hh,
	Klebanov:2000hb,Maldacena:2000yy,Butti:2004pk}.
These backgrounds are related to
confining gauge theories. The known existence of a moduli space
(along the baryonic branch of the Klebanov-Strassler system),
and of a tunable parameter appearing in some of the condensates of the gauge theory,
provide a non-trivial dynamical explanation for the existence of a light state,
which is tempting to identify with the dilaton.

Along a parallel line of investigation, we are intrigued by the ideas exposed in Ref.~\cite{Pomarol:2019aae}
(and in Refs.~\cite{Gorbenko:2018ncu,Gorbenko:2018dtm}), which are closely related to
the discussions in Ref.~\cite{Kaplan:2009kr}.
The present paper is a first step towards transferring these ideas from the bottom-up context to that of rigorous
holographic models built within the top-down approach, and hence testing them within  known supergravity theories. Within the bottom-up
approach to holography, the authors of Refs.~\cite{Kaplan:2009kr,Pomarol:2019aae} 
start by identifying the Breitenlohner-Freedman
(BF) unitarity bound~\cite{Breitenlohner:1982jf} 
as a marker of the transition between conformal and
non-conformal behaviour of the dual gauge theory.
The BF bound is related to the dimension
$\Delta$ of an operator ${\cal O}$ in the dual conformal field theory (CFT).
In the case of five dimensional gravity theories, the BF bound selects $\Delta=2$,
which agrees with the arguments discussed in the context of the
Schwinger-Dyson equations and their approximation~\cite{Cohen:1988sq}, according to which,
in gauge theories with fermion matter field content,
the  ${\cal O}=\bar{\psi}\psi$ operator acquires the non-perturbative dimension
$\Delta=2$ precisely at the edge of the conformal window.
In recent dilaton EFT studies,
$\Delta$ is measured by fitting the aforementioned $SU(3)$ lattice data
to yield $\Delta\simeq 2$~\cite{Appelquist:2017wcg,Appelquist:2017vyy,Appelquist:2019lgk}.

Ref.~\cite{Pomarol:2019aae}  discusses the dynamics  in
proximity of the BF bound, particularly in relation to the dilaton mass.
It adopts a bottom-up simplified
model  to describe this scenario and to test it.
The spectrum of bound states of the putative dual theory is then calculated.
It is found that, when dialling the bulk mass to approach the BF  bound~\cite{Pomarol:2019aae}:
\begin{center}
	\it{ `...the dilaton is always the lightest resonance, although \\ not parametrically lighter than the others.'}
\end{center}
In this paper, we consider a holographic model that realises a physical system
sharing {   several core features }  with those advocated in Ref.~\cite{Pomarol:2019aae}, but
within the context  of top-down holography. 
We see this paper as a precursor to a broad research programme of
exploration of supergravity backgrounds.
We now describe how we can develop this programme, and
anticipate our main results for the one
model we focus upon in the body of the paper.

The techniques that we use are applicable to systems for which
the gravity geometry is asymptotically anti-de Sitter
at large values of the holographic direction $\r$ (corresponding to the UV of the dual field theory).
This is best suited for the application of  holographic renormalisation,
as we want not only to compute the mass spectrum, but also
the free energy of the system, which plays a crucial role in the body of the paper.

In the bottom-up approach to holography, the mass gap of the dual theory can be introduced by adding 
by hand an end of space to the geometry in what corresponds to the IR in the dual field theory.
The presence of boundaries to the space provides additional freedom,
 and allows for the mass gap to emerge in a way that can often be arranged to
preserve (approximate) scale invariance arbitrarily close to the confinement scale.
Hence, the notion of scaling dimension and the associated BF bound may be well defined even in
close proximity of the end of space in the geometry. (Notice that the value of the mass to which one applies the BF bound must be calculated in reference to the AdS geometry, or critical point of the sigma model, that is closest to the end of space of the geometry along the dual renormalisation group (RG) flow.)

This is not the case within the top-down approach.
The backgrounds in the
higher-dimensional geometry depart from AdS$_D$, and in the case when the dual field theory confines, the geometry closes smoothly.
The linear behaviour
for the quark-antiquark static potential  is recovered
by considering open strings in the uplifted 10-dimensional geometry~\cite{Rey:1998ik,
	Maldacena:1998im}, and minimising the classical action~\cite{Kinar:1998vq,Brandhuber:1999jr,
	Avramis:2006nv,Nunez:2009da,Faedo:2013ota}.
In all known classical backgrounds that yield linear confinement in the dual theory in $D=4$ dimensions
(see for instance Refs.~\cite{Chamseddine:1997nm,Witten:1998zw,Klebanov:2000hb,Maldacena:2000yy,Wen:2004qh,Kuperstein:2004yf}
and the generalisations of these models), the geometry is manifestly quite different from AdS$_D$ in the proximity of its end of space.

By combining the fact that in supergravity
 the potential and mass of the bulk scalar fields are fixed and known,
 and cannot be arbitrarily dialled, together with the aforementioned
departure from AdS$_D$ of the geometry in the crucial region near the end of space, 
we conclude that the whole notion of 
proximity to the BF bound (central to Refs.~\cite{Kaplan:2009kr,Pomarol:2019aae})
needs to be generalised.
The BF bound in AdS$_D$ spaces is a
marker of classical instabilities, taking the form of non-unitary behaviour.
Within supergravity, it
is possible to consider classical backgrounds that
evolve near the end of space towards regions of instability.
The instability of the RG trajectory in the dual field theory 
eventually gives rise to tachyonic behaviour
for some of the lower-dimensional classical fluctuations of the background solutions.
We hence replace the notion of proximity to the BF bound (useful only in approximately AdS$_D$ models, but not applicable to the dual of confining theories),
with the proximity to such tachyonic backgrounds.
We will not dial the parameters in the action (related to the coefficients in the dual renormalisation group equations), but rather  the only allowed freedom: the UV boundary conditions satisfied by the gravity and scalar fields. 

We  want to obtain physically meaningful results, hence ideally we should focus on background
solutions that are regular. Nevertheless, we find that in order to explain 
some crucial features of the gravity dynamics we are compelled to
include in part of our analysis also singular solutions that
exhibit what Gubser in Ref.~\cite{Gubser:2000nd} called
a {\it good singularity},  that is characterised by the fact that the scalar potential
of the supergravity theory, evaluated along the classical solutions considered,
is bounded from above. As we shall see, some of these solutions exhibit a mild singular behaviour in $D = 10$ dimensions, that can be detected only in higher-order curvature invariants, not in the Ricci scalar.
We are pushed even further: we have to include also badly singular backgrounds 
in the study of the energetics.
We will clarify these notions eventually, in the body of the paper,
but we anticipate here that the reason why we introduce the singular solutions
in the study of the energetics is not that we are making use of their field theory interpretation (which does not exist),
but rather that for our treatment of the gravity theory to be self-contained and consistent, 
we must treat all the classical solutions on the same grounds, in order to select what 
are the features of the dynamics. If it turns out that a singular solution has free energy lower than the (known) regular solutions, it is not legitimate to discard it, as its contribution to the path integral in the gravity theory is actually dominant. This signals the incompleteness of the gravity description.

So far, we have introduced a quite general programme of research, which can be carried out systematically on the many known supergravity theories and their consistent truncations (see for instance Refs.~\cite{Samtleben:2008pe,Freedman:2012zz}). In this paper, we exemplify this study with one specific class of theories. We choose this class mostly on the grounds of simplicity---the model is a simple example of a gravity theory which provides the dual of a confining field theory, within the top-down approach to holography. To this end, we broaden the classes of backgrounds studied in earlier publications about this same special system~\cite{Wen:2004qh,Kuperstein:2004yf,Elander:2013jqa,Elander:2018aub}.

The gravity theory we consider is the half-maximal ${\cal N}=(2,2)$ supergravity in $D=6$ dimensions first
described by Romans~\cite{Romans:1985tw}. It has been studied in great detail
and for many different purposes~\cite{Romans:1985tz,Brandhuber:1999np,Cvetic:1999un,
	Hong:2018amk,Jeong:2013jfc,DAuria:2000afl,Andrianopoli:2001rs,Nishimura:2000wj,
	Ferrara:1998gv,Gursoy:2002tx,Nunez:2001pt,Karndumri:2012vh,Lozano:2012au,
	Wen:2004qh,Kuperstein:2004yf,Karndumri:2014lba,Chang:2017mxc,Gutperle:2018axv,Suh:2018tul,Suh:2018szn,Kim:2019fsg,Chen:2019qib}.
Its beauty lies in its simplicity: the model in $D=6$ dimensions
contains only one scalar field $\phi$ coupled to gravity, with a  known classical action
describing also four vectors and one 2-form.
We compactify one of the dimensions on a circle.
The backgrounds approach the  critical point $\phi=0$ at large values of the holographic coordinate $\rho$
(corresponding to the UV of the dual field theory).
The physics of confinement is captured by the fact that there are solutions of the
background equations in which the circle shrinks smoothly to zero size at some finite
point in the radial coordinate~\cite{Witten:1998zw}.
We extend the study with respect
to Refs.~\cite{Wen:2004qh,Kuperstein:2004yf} and~\cite{Elander:2013jqa,Elander:2018aub} and
look at additional  branches of solutions. In particular, we consider solutions in which
the scalar field assumes positive values,
$\phi>0$, for which the potential in six dimensions is unbounded from below,
and an instability arises in the system.
In parts of the study we also consider gravity solutions that do not have an interpretation in terms of
four-dimensional confining theories, either because the dual theory is genuinely five-dimensional
at all scales, or because a singularity emerges in the gravity description.

We expect that, as long as $\phi$ experiences just a small excursion away from $\phi=0$, the spectrum 
of fluctuations should not differ
substantially from the one computed elsewhere~\cite{Wen:2004qh,Kuperstein:2004yf,Elander:2013jqa,Elander:2018aub,Hoyos:2020fjx},
and resemble qualitatively that of a generic confining theory.
In particular, all the
fluctuations have positive mass squared, and there is no parametric separation of scales
visible in the spectrum. At the other extreme, if the scalar $\phi$ explores deep into the instability region with $\phi>0$, in which the potential of the six-dimensional gravity theory is unbounded,
we expect at least one of the states of the dual field theory to become tachyonic.
Under the assumption of continuity, somewhere in between one expects that the mass
squared of one of the
states in the spectrum will cross zero, for some special
choice of background solution.
We compute the spectra 
by making use of the
gauge-invariant formalism developed in Refs.~\cite{Bianchi:2003ug,Berg:2005pd,
	Berg:2006xy,Elander:2009bm,Elander:2010wd} and~\cite{Elander:2018aub},
and verify that all of these expectations are realised.
In close proximity of the aforementioned
tuned choice of background the lightest state is a scalar, and it can be made parametrically
light in comparison to all other states.
Furthermore, we show that this state can be characterised as an (approximate) dilaton,
by performing a non-trivial test on the spectrum~\cite{Elander:2020csd}, and by identifying the presence of an enhanced condensate in the vacuum. We use the term {\it approximate dilaton} to refer to a state that has significant mixing with the dilaton but is not necessarily light.

\begin{figure}[t]
\begin{center}
\includegraphics[width=9cm]{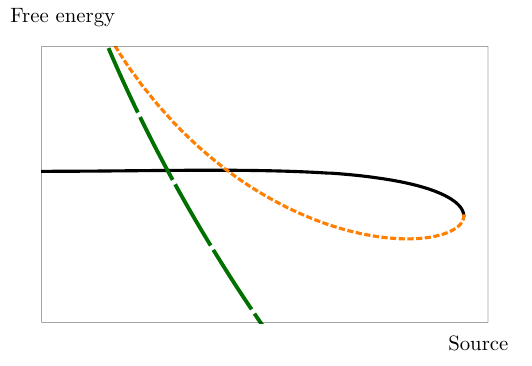}
\caption{Sketch of the free energy as a function of the source, for two of the different branches of classical solutions, showing the first order phase transition at the crossing between the black (solid) and dark-green (long-dashed) lines, as well as the region with a tachyonic instability depicted by the orange (short-dashed) line. The Reader should consult Fig.~\ref{Fig:YetAnotherPlot}, its caption and the text around it, for more details on the phase transition in the realistic model described in the main body of the paper.}
\label{Fig:Cartoon}
\end{center}
\end{figure}

But the attentive Reader certainly took notice of the phrase `... assumption of continuity...' in the
previous paragraph, and realised that this assumption may fail,
for example in the presence of a phase transition.
In fact, this assumption {\it must} fail: the very presence of classical instabilities, as indicated by the existence of a tachyon, in what is an otherwise
perfectly well defined physical system (a well known and established
classical supergravity),
demands the existence of a different branch of solutions,
which must be energetically favoured.
The light (approximate) dilaton and the tachyon appear along a branch of classical solutions that eventually ceases to be physically realised.
We uncover evidence of a first-order phase transition in the gravity theory,
and show that the physically realised solutions do not
come immediately close to the tachyonic ones, undermining the chain of implications from the
previous paragraph. This is illustrated in Fig.~\ref{Fig:Cartoon} which sketches the free energy as a function of the source of a relevant deformation in the field theory for two of the different branches of the classical solutions. As the source is increased, one encounters a first-order phase transition between a branch of regular solutions (in solid black) and a branch of singular solutions (in long-dashed dark green), which happens before the tachyonic region (in short-dashed orange) is reached. We will be more specific and precise in describing these phenomena in the body of the paper.

The conclusion of this exercise
is  almost identical to the one we explicitly  quoted earlier on in italics,
taken from Ref.~\cite{Pomarol:2019aae}. A distinctive element is that, in the physical part of parameter space, the state that is approximately a dilaton is the next-to-lightest state.
Furthermore, the connection with the study of phase transitions bridges between the physics
arguments in Ref.~\cite{Pomarol:2019aae} and those exposed in Refs.~\cite{Gorbenko:2018dtm,Gorbenko:2018ncu},
that are inferred from lower-dimensional statistical mechanics. The first-order phase
transition (to unphysical gravity configurations) signals the metastability of the 
branch of regular solutions along which the dilaton becomes light, and hence
precludes  us from
creating an arbitrarily large hierarchy between the mass of said 
dilaton and that of the other states along the stable branch.

We think this paper  exemplifies and clarifies a few important general points, and opens a new avenue for exploration. First of all, we confirm in the  context of top-down gauge-gravity
dualities that one of the lightest states arising in this way indeed overlaps significantly with the dilaton, and is not some accidentally light scalar particle with generic properties.
We do so by repeating the calculation of the spectrum by treating the scalars in the probe approximation,
which ignores the fluctuations of the five-dimensional metric,
and by comparing the results with the full gauge-invariant results.
We expand on this technical procedure in a different publication~\cite{Elander:2020csd}.

Secondly,
the techniques we adopted can be equally applied to many other backgrounds
that are asymptotically anti-de Sitter in the far-UV,
and evolve towards a classical instability.
In principle, there are countless such examples one can build within the
known catalogue of supergravity theories.
We expect the results we found here to hold generically: 
there will be  choices of parameters/solutions that make one of the scalars
arbitrarily light.
And there will be a first-order phase transition that prevents such solutions from being
physically realised.
Yet, there is no reason to expect that all phase transitions be equally strong;
the phase transition might take place in close proximity  of the tachyonic instability.
In this case, we would be able---by exploiting the formalism we are testing in this paper---to compute
whether a non-trivial hierarchy  appears in the mass spectrum,
as the dilaton state behaves differently from the rest of the spectrum.
Or, conversely,  it might turn out that our findings are  truly
universal, so that  no hierarchy of scale can be produced with this mechanism.
Even such a negative result would be an interesting finding.

The paper is organised as follows.
In Section~\ref{Sec:Model} we define the properties of the
model we study. We show the branches of solutions of interest
to our investigation in Section~\ref{Sec:Solutions}, and produce a
classification of non-trivial classical backgrounds based upon their asymptotic behaviour in 
proximity of the end of the space in the interior of the geometry.
All of the backgrounds share the same 
properties at large values of the radial direction $\rho$, corresponding to the UV of the dual theory.
Many such solutions had not been identified before in the literature.
In Section~\ref{Sec:Spectra} we present the spectra of fluctuations, restricting ourselves to the regular gravity backgrounds, and discuss their interpretation as bound states in the dual confining theory.
Section~\ref{Sec:Probe} contains one of the core parts of the analysis: by
comparing the results of the  probe approximation to those of the full calculation of the spectrum, we 
identify states that have an overlap with the dilaton, and we discuss how this relates to the magnitude of the
condensates in the dual theory.
We discuss the energetics, computing the 
free energy of the background configurations in Section~\ref{Sec:Free}.
Section~\ref{Sec:FreeAnalysis} shows evidence for the arising of a phase transition.
Given the length of the paper, and the fact that the model we consider is non-trivial,
we find it useful to summarise all our results in
Section~\ref{Sec:Summary}, and we conclude the paper with an 
outline of further avenues for future exploration
in Section~\ref{Sec:Conclusions}.
We relegate to the appendices some useful technical details.

\section{The model}
\label{Sec:Model}

In this section, we summarise the action of the six-dimensional supergravity written 
by Romans, adopt an ansatz in which one of the dimensions is a circle, 
perform the dimensional reduction of the theory on this circle, and write the resulting 
dimensionally-reduced action in $D=5$ space-time dimensions. 
Most of the material reported here can be found in the literature,
that we have already cited and will further refer to throughout the section. 
Yet, we find it convenient to collect all the useful background
 information in this section, both to make the exposition self-contained, 
as well as to fix the notation unambiguously.

\subsection{Action and formalism of the $D=6$ model}
\label{Sec:Action}

The six-dimensional (gauged) supergravity constructed by Romans~\cite{Romans:1985tw} describes 32 bosonic degrees of freedom (d.o.f.) (we ignore the fermions). We denote by indices $\hat{M}\in\{0,1,2,3,5,6\}$ the coordinates in $D=6$ dimensions. The field content (number of d.o.f.) is the following: one scalar $\phi$ ($1\times1$), one vector $A_{\hat{M}}$ ($1\times 4$) transforming
 as a singlet under $U(1)$, three vectors $A^{i}_{\hat{M}}$ ($3\times 4$)
  in the \textbf{3} representation of $SU(2)$, one
   2-form $B_{\hat{M}\hat{N}}$ ($1\times 6$), and the six-dimensional metric
    tensor $\hat{g}_{\hat{M}\hat{N}}$ ($1\times 9$). The action is given by     
{\small
	\begin{align}
	\mathcal S_{6}=\int_{}^{}\text{d}^{6}x\sqrt{-\hat g_{6}}
	\bigg(\frac{\mathcal{R}_{6}}{4}&- \hat g^{\hat{M}\hat{N}}\partial_{\hat{M}}\phi\partial_{\hat{N}}\phi -\mathcal{V}_{6}(\phi)  
	-\frac{1}{4}e^{-2\phi}\hat g^{\hat{M}\hat{R}}
	\hat g^{\hat{N}\hat{S}}\sum_{i}\hat F_{\hat{M}\hat{N}}^{i}\hat F_{\hat{R}\hat{S}}^{i}\, \notag \\
	&-\frac{1}{4}e^{-2\phi}\hat g^{\hat{M}\hat{R}}
	\hat g^{\hat{N}\hat{S}}\hat{\mathcal{H}}_{\hat{M}\hat{N}}\hat{\mathcal{H}}_{\hat{R}\hat{S}}
	-\frac{1}{12}e^{4\phi}\hat g^{\hat{M}\hat{R}}
	\hat g^{\hat{N}\hat{S}}\hat g^{\hat{T}\hat{U}}\hat G_{\hat{M}\hat{N}\hat{T}}\hat G_{\hat{R}\hat{S}\hat{U}} \bigg)\,,
	\label{ActionS6}
	\end{align} }
where summation over repeated indices is implied. The tensors are defined as follows: 
\begin{align}
\hat F_{\hat{M}\hat{N}}^{i} &\equiv \partial_{\hat{M}}A_{\hat{N}}^{i}-
\partial_{\hat{N}}A_{\hat{M}}^{i}+g\epsilon^{ijk}A_{\hat{M}}^{j}A_{\hat{N}}^{k}\,,\\
\hat F_{\hat{M}\hat{N}} &\equiv \partial_{\hat{M}}A_{\hat{N}}-
\partial_{\hat{N}}A_{\hat{M}}\,,\\
\hat{\mathcal{H}}_{\hat{M}\hat{N}} &\equiv \hat F_{\hat{M}\hat{N}} + mB_{\hat{M}\hat{N}}\,,\\
\hat G_{\hat{M}\hat{N}\hat{T}}&\equiv 3\partial_{[\hat{M}}B_{\hat{N}\hat{T}]}=
\partial_{\hat{M}}B_{\hat{N}\hat{T}}+
\partial_{\hat{N}}B_{\hat{T}\hat{M}}+
\partial_{\hat{T}}B_{\hat{M}\hat{N}}\,.
\end{align}
Here $\hat{g}_{6}$ is the determinant of the metric tensor and $\mathcal{R}_{6}\equiv g^{\hat{M}\hat{N}}R_{\hat{M}\hat{N}}$
 is the corresponding Ricci scalar. We return to the scalar potential 
 $\mathcal{V}_{6}(\phi)$ and its critical points
 in Section~\ref{Sec:6potential}.
 
 \subsection{The scalar potential in $D=6$ dimensions}
\label{Sec:6potential}

\begin{figure}[t]
\begin{center}
\includegraphics[width=10cm]{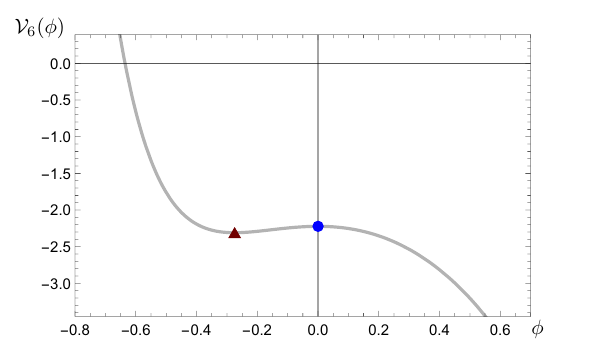}
\caption{The potential ${\cal V}_6(\phi)$ of Romans supergravity as a function of the one scalar $\phi$ in the sigma-model coupled to gravity in $D=6$ dimensions. We highlight the two critical points: the case $\phi=\phi_{UV}=0$ (blue disk) and the case $\phi=\phi_{IR}=-\frac{1}{4}\log(3)$ (dark-red triangle), both of which allow for AdS$_6$ background solutions.}
\label{Fig:potential}
\end{center}
\end{figure}

The potential for the real scalar field $\phi$ in the six-dimensional model is given by\footnote{In the language of Ref.~\cite{Jeong:2013jfc}, 
we adopted the choice of $g=\sqrt{8}$ and $m=\frac{2\sqrt{2}}{3}$, without loss of generality.}~\cite{Jeong:2013jfc}
\be
\label{Eq:V6}
\mathcal{V}_{6}(\phi)=\frac{1}{9}(e^{-6\phi}-9e^{2\phi}-12e^{-2\phi}) \,,
\ee
and is shown in Fig.~\ref{Fig:potential}. It admits two critical points:
\begin{equation}
\phi_{UV}=0 \hspace{1.90cm} \bigg(\mathcal{V}_{6}(\phi_{UV})=-\frac{20}{9}\bigg)\,,
\end{equation} and
\begin{equation}
\phi_{IR}=-\frac{\log(3)}{4} \-\ \hspace{0.45cm}\bigg(\mathcal{V}_{6}(\phi_{IR})=-\frac{4}{\sqrt{3}}\bigg)\,,
\end{equation}
with the former (latter) a global maximum (minimum) which preserves (breaks) supersymmetry. As we shall see, there exist numerical solutions to the equations of motion that interpolate between these two critical points, corresponding to a renormalisation group flow between a UV and IR fixed point in the dual field theory. In previous work \cite{Elander:2018aub}, we restricted $\f$ to the closed interval $\f\in[\f_{IR},\f_{UV}]$. For the purposes of this paper, we extend this domain by 
 allowing positive values of $\f$.
 
\subsection{Reduction from $D=6$ to $D=5$ dimensions}
\label{Sec:Reduction}

We compactify one of the external dimensions (described by the coordinate $\eta\in[0,2\p)$) of the Romans theory on a circle $S^{1}$, and parametrise the six-dimensional metric as follows:
 \begin{align}\label{Eq:6Dmetric}
 \text{d}s_{6}^{2}&=e^{-2\c}\text{d}s_{5}^{2} + e^{6\c}\big(\text{d}\eta+V_{M}\text{d}x^{M}\big)^{2}\notag\\
&=e^{-2\c}\big(e^{2A(r)}\text{d}x_{1,3}^{2} +\text{d}r^{2}\big)+ e^{6\c}\big(\text{d}\eta+V_{M}\text{d}x^{M}\big)^{2}\notag\\
&=e^{-2\c}\big(e^{2A(\r)}\text{d}x_{1,3}^{2} +e^{2\c}\text{d}\r^{2}\big)+ e^{6\c}\big(\text{d}\eta+V_{M}\text{d}x^{M}\big)^{2}\, ,
 \end{align}
where $\text{d}s_{5}^{2}$ is the five-dimensional line element so that $\text{det}(g_{MN})\equiv g_{5}=-e^{8A(r)}$, with warp factor $A(r)$, 
and we have adopted the ``mostly plus'' four-dimensional Minkowski metric signature, $\eta_{\mu\nu}=\text{diag}(-,+,+,+)$; 
indices run over $\mu,\nu\in\{0,1,2,3\}$ and $M,N\in\{0,1,2,3,5\}$. The third equality introduces the convenient redefinition of the radial coordinate $\text{d}r\equiv e^{\c}\text{d}\r$. In the background solutions that we will consider, each field of the supergravity model depends \emph{only} on the radial coordinate $\r$, and additionally \emph{only}  $\phi(\rho)$, $\chi(\rho)$, and $A(\rho)$ acquire non-zero radial profiles, thus ensuring Poincar\'{e} invariance along the Minkowski $x^{\mu}$ directions. We constrain the holographic coordinate to take values in the closed interval $\r\in[\r_{1},\r_{2}]$, for reasons to be discussed later, but it is understood that the physical results that apply to the dual field theory are recovered only after removing these restrictions.

After decomposing the fields, and some algebra (see Ref.~\cite{Elander:2018aub} for details), the action of the reduced five-dimensional model is given by 
{\small\begin{align}
	\mathcal S_{5}=\int_{}^{}\text{d}^{5}x\sqrt{-g_{5}}
	\bigg(\frac{\mathcal{R}_{5}}{4}-\half G_{ab} g^{MN}\partial_{M}\Phi^{a}\partial_{N}\Phi^{b} &-\mathcal{V}(\phi,\chi)  
	-\frac{1}{4}H_{AB}g^{MR}g^{NS}F_{MN}^{A}F_{RS}^{B} \,\notag\\
	-\frac{1}{4}e^{2\chi-2\phi}g^{MR}g^{NS}\mathcal{H}_{MN}\mathcal{H}_{RS}
	&-\frac{1}{12}e^{4\chi+4\phi}g^{MR}g^{NS}g^{TU}G_{MNT}G_{RSU} \,\notag\\
	-\frac{1}{2}e^{-6\chi-2\phi}g^{NS}\mathcal{H}_{6N}\mathcal{H}_{6S}
	&-\frac{1}{4}e^{-4\chi+4\phi}g^{NS}g^{TU}G_{6NT}G_{6SU}
	\bigg)\,,\label{Eq:S5ast}
\end{align}}%
where the 32 physical degrees of freedom are now carried by the following five-dimensional field content: six scalars $\{\phi,\chi,\pi^{i},A_{6}\}$ ($6\times1$), six vectors $\{A_{M},A^{i}_{M},B_{6N},V_{M}\}$ ($6\times3$), one 2-form $B_{MN}$ ($1\times3$), and the metric tensor $g_{MN}$ ($1\times5$) in $D=5$ dimensions. The dynamical scalar field $\chi$ parameterises the size of the compact $S^{1}$ (see Eq.~(\ref{Eq:6Dmetric})). The sigma-model scalars are $\Phi^a = \{ \phi, \chi, \pi^i \}$ with the metric $G_{ab} = {\rm diag} \left( 2, 6, e^{-6\chi-2\phi} \right)$, while the field strengths $\{ F^V, F^i \}$ have the metric $H_{AB} = {\rm diag} \left( \frac{1}{4} e^{8\chi}, e^{2\chi-2\phi} \right)$. The five-dimensional scalar potential appearing in the circle-reduced model is given by $\mathcal{V}(\phi,\chi)=e^{-2\chi}\mathcal{V}_{6}(\phi)$.

\subsection{Equations of motion}
\label{Sec:EOM}
The classical equations of motion can be obtained from $\mathcal S_{5}$, the action for the five-dimensional model, provided in Eq.~(\ref{Eq:S5ast}). We remind the Reader that all of the classical supergravity background fields are assumed to depend solely on the holographic coordinate $\r$. Hence, the equations of motion for the background functions are given by
\begin{align}
\partial^{2}_{\rho}\phi+(4\partial_{\rho}A-\partial_{\rho}\chi)\partial_{\rho}\phi&=\frac{1}{2}\frac{\partial\mathcal{V}_{6}}{\partial\phi}\, ,\label{Eq:eom1}\\
\partial^{2}_{\rho}\chi+(4\partial_{\rho}A-\partial_{\rho}\chi)\partial_{\rho}\chi&=-\frac{\mathcal{V}_{6}}{3}\, ,\label{Eq:eom2}\\
3(\partial_{\rho}A)^{2}-(\partial_{\rho}\phi)^{2}-3(\partial_{\rho}\chi)^{2}&=-\mathcal{V}_{6}\, ,\label{Eq:eom3}\\
3\partial^{2}_{\rho}A +6(\partial_{\rho}A)^{2}+2(\partial_{\rho}\phi)^{2}+6(\partial_{\rho}\chi)^{2}-3\partial_{\rho}A\partial_{\rho}\chi
&=-2\mathcal{V}_{6}\, .\label{Eq:eom4}
\end{align}
Only the first three are independent. These equations of motion can be reformulated using the following convenient redefinitions:
\begin{equation}
c\equiv 4A-\c\quad , \quad d\equiv A-4\c\, ,
\end{equation}
or equivalently
\begin{equation}
\c\equiv\frac{1}{15}(c-4d)\quad , \quad A\equiv\frac{1}{15}(4c-d)\, ,
\end{equation}
so that we can recast them in the following form:
\begin{align}
\partial^{2}_{\r}\f+\partial_{\r}c\partial_{\r}\f&=\frac{1}{2}\frac{\partial\mathcal{V}_{6}}{\partial\f}\, ,\label{Eq:EOMphic}\\
\partial^{2}_{\r}c+(\partial_{\r}c)^{2}&=-5\mathcal{V}_{6}\, ,\label{Eq:EOMc}\\
(\partial_{\r}c)^{2}-(\partial_{\r}d)^{2}-5(\partial_{\r}\f)^{2}\label{Eq:EOM3}
&=-5\mathcal{V}_{6}\,,\\
\partial^{2}_{\r}d+\partial_{\r}c\partial_{\r}d&=0\label{Eq:EOMd}\,.
\end{align} 
The last of these four equations can be derived from the previous three, yet we show it explicitly for a  reason we will explain shortly.
Having solved the coupled equations  (\ref{Eq:EOMphic}) and (\ref{Eq:EOMc}) to yield $\f$ and $c$, one can proceed then to solve
Eq.~(\ref{Eq:EOM3}) to determine $d$. Notice that for any given solution $d$, one finds that $-d$ is also admissible. We observe that Eq.~(\ref{Eq:EOMd}) can be rewritten as a vanishing total derivative, and hence we obtain the following useful relation:
\begin{equation}
\label{Eq:C}
e^{4A-\c}(4\partial_{\r}\c-\partial_{\r}A)=C\, ,
\end{equation}
for some background-dependent integration constant $C$. We will make use of this relation later in the paper. Finally, we also note that by combining Eq.~\eqref{Eq:EOMc} and Eq.~\eqref{Eq:EOM3}, one can derive the inequality
\beq
\label{eq:RGinequality}
	\partial_\rho^2 c \leq 0 \,,
\eeq
that constrains the RG flows of the dual field theory admitting a description based on the classical backgrounds.

\subsection{Superpotential formalism}    
\label{Sec:Super}

The conventions we are using in writing the action in Eq.~(\ref{ActionS6}) are such that \emph{if}
the potential of the model in $D$ dimensions ${\cal V}_D$ can be written in terms of a 
superpotential ${\cal W}$ that satisfies the following equation~\cite{Berg:2005pd}
\beqs
{\cal V}_D&=&\frac{1}{2}G^{\phi\phi}(\partial_{\phi}{\cal W})^2\,-\,\frac{D-1}{D-2}{\cal W}^2\,,
\label{Eq:super}
\eeqs
for the metric ansatz
\beqs
\di s^2_D &=&e^{2{\cal A}}\di x_{1,D-2}^2+\di \rho^2\,,
\eeqs
\emph{then} one finds that the solutions to a special set of first-order equations 
are also solutions to the second-order classical equations. The aforementioned first-order equations are the following:
\beqs
\partial_{\rho}{\cal A}&=&-\frac{2}{D-2}{\cal W}\,, \label{Eq:FirstOrderA}\\
\partial_{\rho}\phi&=&G^{\phi\phi}\partial_{\phi}{\cal W}\,.
\eeqs

As we are working with $D=6$, and given the potential ${\cal V}_6$ of Eq.~(\ref{Eq:V6}),
one finds~\cite{Gursoy:2002tx} the superpotential $\mathcal{W}=\mathcal{W}_{1}$, which together with the corresponding first-order equations is
\beqs
{\cal W}_{1}&=&-e^{\phi}-\frac{1}{3}e^{-3\phi}\,,\label{Eq:FirstOrder1}\\
\partial_{\rho}{\cal A}&=&-\frac{1}{2}{\cal W}_{1}\,=\,\frac{1}{2}\left(e^{\phi}+\frac{1}{3}e^{-3\phi}\right)\,,\label{Eq:FirstOrder2}\\
\partial_{\rho}\phi&=&\frac{1}{2}\partial_{\phi}{\cal W}_{1}\,=\,\frac{1}{2}\left(-e^{\phi}+\frac{}{}e^{-3\phi}\right)\,.\label{Eq:FirstOrder3}
\eeqs 
It is straightforward to verify that solutions to the previous two equations also solve the full equations of motion of the system, which after imposing the constraint $A=4\chi$ (and hence ${\cal A}=3\chi$) can be rewritten as
\begin{align}
4\partial^{2}_{\rho}\phi+15\partial_{\rho}A\partial_{\rho}\phi&=2\frac{\partial\mathcal{V}_{6}}{\partial\phi}\, ,\label{Eq:EOMc1}\\
3\partial^{2}_{\rho}A + 4(\partial_{\rho}\phi)^{2}
&=0\, ,\label{Eq:EOMc2}\\
45(\partial_{\rho}A)^{2}-16(\partial_{\rho}\phi)^{2}&=-16\mathcal{V}_{6}\,. \label{Eq:EOMc3}
\end{align} 
The superpotential $\mathcal{W}_{1}$ yields a system of equations that admits the solution 
$\phi=0$ and ${\cal A}=\frac{2}{3}\rho$. It can be expanded in powers of
small $\phi$: 
\beqs
\label{Eq:W3}
{\cal W}_{1}(\phi)&=&-\frac{4}{3}-2\phi^2+\frac{4}{3}\phi^3-\frac{7}{6}
\phi^4+\frac{2}{3}\phi^5-\frac{61}{180}\phi^6\,+\,{\cal O}(\phi^7)\,.
\eeqs
The quadratic term in this expansion shows that the solutions can be interpreted in terms of
the vacuum expectation value of an operator of dimension $\Delta=3$ in the dual five-dimensional 
strongly-coupled field theory~\cite{Gursoy:2002tx}.

Besides providing a useful solution-generating technique, the 
superpotential formalism also plays a role in defining an unambiguous, covariant and physically 
motivated subtraction scheme in the calculation of the free energy.   
To this purpose, we notice that the system admits a second choice of superpotential, 
that we call ${\cal W}_2$, and that 
can be written as a power-expansion for small $\f$:
\beqs
\label{Eq:W2}
{\cal W}_2(\phi)&=&-\frac{4}{3}-\frac{4}{3}\phi^2+\frac{16}{3}\phi^3+\frac{86}{3}
\phi^4+\frac{848}{3}\phi^5+\frac{988658}{315}\phi^6\,+\,{\cal O}(\phi^7)\,.
\eeqs
We are not aware of the existence of a closed form solution  to
Eq.~(\ref{Eq:super}) that satisfies this expansion.
Notice that, while encompassing the same AdS$_6$ solution of the first-order system derived from 
${\cal W}_{2}$, in this case the solutions of the first-order system
correspond to deformations of the dual field theory 
by the non-trivial coupling of the same operator of dimension $\Delta=3$.   
As we shall see, by choosing to adopt  ${\cal W}_2(\phi)$ as the form of one of the boundary-localised terms in the complete gravity action, we can provide the counter-terms in the holographic 
renormalisation procedure, and guarantee that all the divergences are cancelled for any asymptotically AdS$_6$ backgrounds.

\section{Classes of solutions}
\label{Sec:Solutions}

In this section we present the classes of solutions that we will refer to as SUSY, IR-conformal, confining, and skewed, together with their IR expansions. We also introduce a few additional, more general, singular solutions, including ones that preserve five-dimensional Poincar\'e invariance. We introduce the relevant UV expansions for the two scalars $\{\f,\c\}$ and the warp factor $A$, which are valid for all these classes of solutions. 

\subsection{UV expansions}
\label{Sec:UV}

We present here the large-$\r$ expansions for $\f$, $\chi$ and $A$ in terms of a convenient holographic coordinate defined by $z\equiv e^{-2\r/3}$. We truncate each expansion at $\mathcal{O}(z^{11})$. These expansions are used in the numerical analysis for all classes of solutions in order to extract values for the set of UV parameters $\{\f_{2},\f_{3},\chi_{5},\chi_{U},A_{U}\}$ that unambiguously identify each background, and to compute the free energy. All solutions we are interested in have the same formal UV expansion, as they all correspond to deformations of the same supersymmetric fixed point. The expansions are given by the following equations.
{\small\begin{align}
	\f(z)=\f_{2}z^2 &+\f_{3}z^3-6\f_{2}^{2} z^4 -4(\f_{2}\f_{3})z^5+\left(\frac{29 \f_{2}^3}{2}-\f_{3}^2\right)z^6+\frac{339}{20} \f_{2}^2 \f_{3}z^7\nn\\&+ \left(\frac{77 \f_{2}
		\f_{3}^2}{10}-\frac{146 \f_{2}^4}{3}\right)z^8 + \left(\frac{19 \f_{3}^3}{12}-\frac{8497 \f_{2}^3 \f_{3}}{105}\right)z^9 \\ \nn &+ \left(\frac{6752 \f_{2}^5}{35}-\frac{1986 \f_{2}^2 \f_{3}^2}{35}\right)z^{10}+ \left(\frac{4127161 \f_{2}^4 \f_{3}}{10080}-\frac{3427 \f_{2} \f_{3}^3}{180}\right)z^{11}+\mathcal{O}(z^{12})\, ,
	\end{align} }
{\small\begin{align}
	\chi(z)=\chi_{U}&-\frac{\log (z)}{3}-\frac{\f_{2}^2 z^4}{12}+\chi_{5} z^5+ \left(\frac{8 \f_{2}^3}{9}-\frac{\f_{3}^2}{12}\right)z^6+\frac{32}{21} \f_{2}^2 \f_{3}z^7+ \left(\frac{3
		\f_{2} \f_{3}^2}{4}-\frac{77 \f_{2}^4}{16}\right)z^8\nn\\
	&+ \left(-\frac{1072 \f_{2}^3 \f_{3}}{135}+\frac{25 \chi_{5} \f_{2}^2}{36}+\frac{4 \f_{3}^3}{27}\right)z^9\nn\\&+ \left(-\frac{15
		\chi_{5}^2}{64}+\frac{172 \f_{2}^5}{9}-\frac{3181 \f_{2}^2 \f_{3}^2}{600}+\frac{9 \chi_{5} \f_{2} \f_{3}}{8}\right)z^{10} \\ \nn &+\left(\frac{44776 \f_{2}^4 \f_{3}}{1155}-\frac{200
		\chi_{5} \f_{2}^3}{33}-\frac{96 \f_{2} \f_{3}^3}{55}+\frac{25 \chi_{5} \f_{3}^2}{44}\right)z^{11}+\mathcal{O}(z^{12})\, ,
	\end{align} }
{\small\begin{align}
	A(z)= A_{U}&-\frac{4 \log (z)}{3}-\frac{\f_{2}^2 z^4}{3}+ \left(\frac{\chi_{5}}{4}-\frac{3 \f_{2} \f_{3}}{5}\right)z^5+ \left(\frac{32 \f_{2}^3}{9}-\frac{\f_{3}^2}{3}\right)z^6+\frac{128}{21}
	\f_{2}^2 \f_{3}z^7\nn\\&+\left(3 \f_{2} \f_{3}^2-\frac{77 \f_{2}^4}{4}\right)z^8+\frac{1}{2160} \bigg(-69508 \f_{2}^3 \f_{3}+375 \chi_{5} \f_{2}^2+1280 \f_{3}^3\bigg)z^9 \\ \nn &+\frac{1}{3600} \bigg(-3375 \chi_{5}^2+275200 \f_{2}^5-78936 \f_{2}^2 \f_{3}^2\bigg)z^{10}\nn\\ \nn &+\frac{1}{18480} \bigg(2932864 \f_{2}^4 \f_{3}-28000 \chi_{5} \f_{2}^3-135324
	\f_{2} \f_{3}^3+2625 \chi_{5} \f_{3}^2\bigg)z^{11}+\mathcal{O}(z^{12})\, .
	\end{align} }\\
For convenience, we also write explicitly the UV expansions for the two combinations $c$ and $d$ that were introduced in Section~\ref{Sec:EOM}: 
{\small\begin{align}
c(z)=4A_{U}-\c_{U}&-5\log(z)-\frac{5\f_{2}^{2}z^4}{4}-\frac{12}{5}\f_{2}\f_{3}z^5+\left(\frac{40\f_{2}^3}{3}-\frac{5\f_{3}^2}{4}\right)z^{6}+\frac{160}{7}\f_{2}^{2}\f_{3}z^{7}\nn\\
&+\left(\frac{45\f_{2}\f_{3}^{2}}{4}-\frac{1155 \f_{2}^{4}}{16}\right)z^{8}+\left(\frac{20 \f_{3}^{3}}{9}-\frac{1087 \f_{2}^{3}\f_{3}}{9}\right)z^{9}\nn\\
&+\left(-\frac{225 \c_{5}^{2}}{64}+\frac{860 \f_{2}^5}{3}-\frac{16481 \f_{2}^2\f_{3}^2}{200}-\frac{9\c_{5}\f_{2}\f_{3}}{8}\right)z^{10} \\ \nn
&+\left(\frac{45896\f_{2}^4\f_{3}}{77}-\frac{303\f_{2}\f_{3}^3}{11}\right)z^{11}+\mathcal{O}(z^{12})\, ,
\end{align} }
{\small\begin{align}
	d(z)=A_{U}-4\c_{U}&+\left(-\frac{15\c_{5}}{4}-\frac{3 \f_{2} \f_{3}}{5}\right)z^{5}+\left(-\frac{5\f_{2}^3\f_{3}}{12}-\frac{125\c_{5}\f_{2}^2}{48}\right)z^{9}\nn\\
	&+\left(-\frac{18}{25} \f_{2}^2 \f_{3}^2-\frac{9\c_{5}\f_{2}\f_{3}}{2}\right)z^{10} \\ \nn
	&+\left(\frac{40\f_{2}^4\f_{3}}{11}+\frac{250\c_{5} \f_{2}^3}{11}-\frac{15\f_{2}\f_{3}^3}{44}-\frac{375\c_{5} \f_{3}^2}{176}\right)z^{11}+\mathcal{O}(z^{12})\,.
	\label{Eq:dd}
	\end{align} }

When computing the free energy for each class of solutions we will choose to always set $A_{U}=\c_{U}=0$.
The constraint $A=4\c\Leftrightarrow d=0$ reinstates (locally) five-dimensional Poincar\'e invariance.\ It is required for the SUSY, IR-conformal, and singular domain wall solutions, and it constrains the parameters appearing in the UV expansions. 

\subsection{SUSY solutions}
\label{Sec:Susy}

The first-order equations presented in Eqs.~(\ref{Eq:FirstOrder1}-\ref{Eq:FirstOrder3})
 of Section~\ref{Sec:Super} can be solved by performing the change of variable
$\partial_{\rho}\equiv e^{-\phi}\partial_{\tau}$, after which one obtains
\beqs
\partial_{\tau}\phi&=&-\sinh(2\phi)\,,\\
\partial_{\tau}{\cal A}&=&\frac{1}{2}\left(e^{2\phi}+\frac{1}{3}e^{-2\phi}\right)\,,
\eeqs
which are solved exactly by~\cite{Gursoy:2002tx} 
\beqs
\phi(\tau)&=&{\rm arccoth}\left(e^{2(\tau-\tau_o)}\right)\,,\label{Eq:SUSY1}\\
{\cal A}(\tau)&=&{\cal A}_o+\frac{1}{3}\log(\sinh(2(\tau-\tau_o)))+\frac{1}{6}\log(\tanh(\tau-\tau_o)))\,,\label{Eq:SUSY2}
\eeqs
where $\mathcal A_o$ and $\tau_o$ are integration constants. These supersymmetric (SUSY) solutions evolve $\phi$ monotonically from the supersymmetric fixed point towards the good singularity ($\f\to\infty$), for which 
the potential is bounded from above. We notice that the flow breaks scale invariance, and hence reduces the number of supersymmetries of the underlying theory to eight~\cite{Gursoy:2002tx}. We remind the Reader that these solutions result from the formation of a non-trivial condensate in the dual field theory.

By relating the radial coordinates $\rho$ and $\tau$, one finds that the SUSY solutions given in Eqs.~(\ref{Eq:SUSY1}-\ref{Eq:SUSY2}) have the following IR expansions:
\begin{align}
\phi(\r)&=\log(2) -\log(\r-\r_{o})+\frac{1}{80}(\r-\r_{o})^{4}+\ldots\, ,\\
\chi(\r)&=\chi_{I} + \frac{1}{3}\log(\r-\r_{o})+ \frac{1}{360}(\r-\r_{o})^{4}+\ldots\, ,\\
A(\r)&= A_{I} + \frac{4}{3}\log(\r-\r_{o})+ \frac{1}{90}(\r-\r_{o})^{4}+\ldots\, ,
\end{align}
where $\c_{I}$ and $A_{I}=4\chi_I$ are integration constants, and $\r_{o}$ is the radial position of the singularity
 in the deep IR region of the bulk.
 
\begin{figure}[t]
\begin{center}
\includegraphics[width=10.75cm]{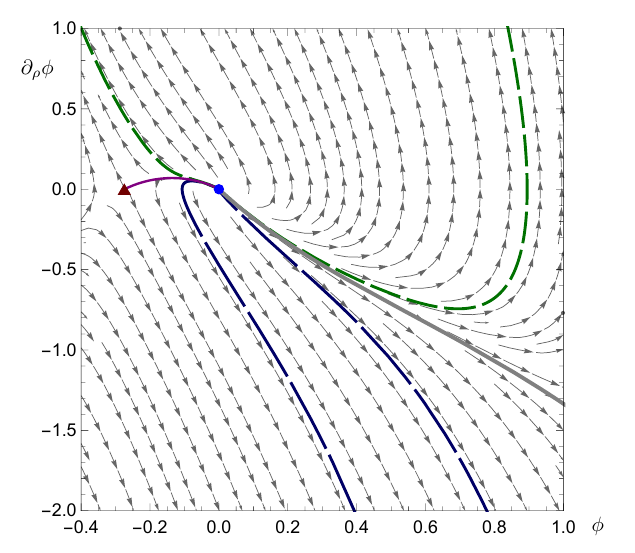}
\caption{Parametric plot of $\partial_{\r}\f$ as a function of $\f$ for solutions 
	which satisfy the domain-wall constraint $A=\frac{4}{3}{\cal A}=4\c$. 
	The blue disk and dark-red triangle respectively denote the UV and IR critical points of the
	 six-dimensional potential $\mathcal{V}_{6}$, the purple (solid) line represents the class of IR-conformal 
	 solutions introduced in Section~\ref{Sec:Conformal} with duals which flow between the two critical points, 
		 and the 
		 light-grey (solid) line represents the analytical supersymmetric solutions 
		 obtained by solving the first-order differential equations (\ref{Eq:FirstOrder1}),
		  (\ref{Eq:FirstOrder2}), and  (\ref{Eq:FirstOrder3}).\ 
		The dark-grey arrows exhibit the vector field appearing in the first order differential equation for $(\phi, \partial_\rho \phi)$.
		We also show two examples of the (good) singular solutions obeying the IR expansion
		in Section~\ref{Sec:Divergent}, for $\phi_L=-1/\sqrt{5}$ and $\phi_I=-0.3, 0.1$ (long-dashed dark-blue lines) and 
		two examples of the domain wall (badly singular) solutions from Section~\ref{Sec:SingDW}, with
		$\phi_4=-0.06, 40$ (dashed dark-green lines).}
\label{Fig:ParametricPlot}
\end{center}
\end{figure}

In Fig.~\ref{Fig:ParametricPlot}, we illustrate the space of domain-wall solutions, of which the SUSY solutions are a special case, through the following procedure. We first solve Eq.~(\ref{Eq:EOMc3}) for $\partial_{\r}A$ 
 and substitute into Eq.~(\ref{Eq:EOMc1}) to obtain a second-order differential equation in terms of $\f$ alone, then plot parametrically $(\f,\partial_{\r}\f)$, and study how the solutions flow away from the supersymmetric fixed point at the origin. We observe that the SUSY solutions (grey line) form the separatrix between numerical solutions which flow to a bad singularity ($\f\to -\infty$) and solutions which instead flow to a good singularity ($\f\to\infty$).

\subsection{IR-conformal solutions}
\label{Sec:Conformal}

A second class of solutions interpolates between the two known $\text{AdS}_{6}$ solutions of the six-dimensional model, corresponding in the boundary theory to a renormalisation group flow between two fixed points. The six-dimensional bulk geometry does not close off for any (finite) value of the holographic coordinate $\r$ and the compact dimension described by $\eta$ maintains non-zero size for all $\r$. Hence there does not exist a physical lower limit for the energy scale at which the field theories dual to this class of solutions may be probed. The IR expansions for this class of solutions are conveniently written in terms of $e^{-(5-\Delta_{IR})\frac{\r}{R_{IR}} }$, which is small in the limit $\rho \rightarrow -\infty$, and they are given by
\begin{align}
\phi(\r)&=\phi_{IR} + \big(\phi_{I}-\phi_{IR} \big)e^{-(5-\Delta_{IR})\frac{\r}{R_{IR}} }\,+\cdots\, ,\\
\chi(\r)&=\chi_I+\frac{\r}{3R_{IR}} - \frac{1}{12} \big(\phi_{I}-\phi_{IR} \big)^{2} e^{-2(5-\Delta_{IR})\frac{\r}{R_{IR}} }\,+\cdots\, ,\\
A(\r)&=A_I+\frac{4\r}{3R_{IR}} - \frac{1}{3} \big(\phi_{I}-\phi_{IR} \big)^{2} e^{-2(5-\Delta_{IR})\frac{\r}{R_{IR}} }\,+\cdots\, ,
\end{align}
where $\chi_I$ and $A_I$ are integration constants, $R^{2}_{IR}\equiv -5\big(\mathcal{V}_{6}(\phi_{IR})\big)^{-1}=\frac{5}{4}\sqrt{3}$ is the (squared) curvature radius of the $\text{AdS}_{6}$ geometry, $\Delta_{IR}=\frac{1}{2}(5+\sqrt{65})$ is the scaling dimension of the operator in the dual boundary theory that is related to the IR critical point value of the bulk scalar $\phi$, and we restrict the one free parameter $\phi_{I}\geqslant -\frac{1}{4}\log(3)$. We observe that $d\equiv A-4\c=0$ for all values of $\f_{I}$ for this class of solutions. It is also worth noting that the backgrounds defined by this class of solutions do not preserve supersymmetry. In the UV expansions, these solutions require a tuning of $\f_{3}$ against $\f_{2}$, as we will discuss in Section~\ref{Sec:IRC-FreeEnergy}. 

\subsection{Confining solutions}
\label{Sec:SolutionsConfining}

With some abuse of language, we refer to a third class of solutions as \emph{confining}.\ Here the compact dimension (described by the coordinate $\eta$) shrinks to a point at some finite value $\r_{o}$ of the holographic coordinate and the six-dimensional bulk geometry closes off smoothly.
On the boundary side of the duality this smooth tapering property of the bulk manifold is interpreted as a physical lower limit on the energy scale that may be probed in the corresponding field theory in $D=5$ dimensions. We anticipate that the IR asymptotic expansion of  these solutions is identical to those studied in Ref.~\cite{Elander:2013jqa}, and hence yield the area law expected from confinement. We will compute explicitly the spectrum in Section~IV and show that it is discrete, generalising the results of Ref.~\cite{Elander:2018aub}.

As mentioned in previous work~\cite{Wen:2004qh,Kuperstein:2004yf,
Elander:2013jqa,Elander:2018aub}, there exist exact analytical solutions of the classical equations of motion when $\f$ is constant: 
\begin{align}
\phi&=\f_{UV}=0,\\
\chi(\r)&=\chi_{I}-\frac{1}{5}\log\bigg[\cosh\bigg(\frac{{\sqrt{-5v}}}{2}(\r-\r_{o}) \bigg) \bigg]
+\frac{1}{3}\log\bigg[\sinh\bigg(\frac{{\sqrt{-5v}}}{2}(\r-\r_{o}) \bigg) \bigg],\label{Eq:ConfChi}\\
A(\r)&=A_{I}-\frac{4}{15}\log(2)+\frac{4}{15}\log\bigg[\sinh\bigg(\sqrt{-5v}(\r-\r_{o}) \bigg) \bigg]\notag\\
&\hspace{65mm}+\frac{1}{15}\log\bigg[\tanh\bigg(\frac{{\sqrt{-5v}}}{2}(\r-\r_{o}) \bigg) \bigg],\label{Eq:ConfA}
\end{align}
with $v\equiv\mathcal{V}_{6}(\f=0)$ as defined in Section~\ref{Sec:6potential}. By direct substitution of the above analytical solutions we find:
\begin{align}
\label{eq:confcd1}
e^{c(\r)}=e^{4A(\r)-\c(\r)}&=\frac{1}{2}e^{4A_{I}-\c_{I}}\sinh\left(\frac{10}{3}(\r-\r_{o})\right)\, ,\\
\label{eq:confcd2}
e^{d(\r)}=e^{A(\r)-4\c(\r)}&=e^{A_{I}-4\c_{I}}\coth\left(\frac{5}{3}(\r-\r_{o})\right)\, .
\end{align} 
 These solutions can be generalised by series expanding for small $(\r - \r_o)$ and allowing for non-trivial values of $\f$ for small $(\r - \r_o)$, to obtain expansions which may be used to construct a generalised family of numerical solutions.
  
We obtain the numerical solutions by solving the classical equations of motion, subject to boundary conditions obtained from the following IR (small $(\r - \r_o)$) expansions: 
{\small\begin{align}
\phi(\r) &= \phi_{I}-\frac{1}{12}e^{-6\phi_{I}} \left(1-4 e^{4\phi_{I}}+3 e^{8\phi_{I}}\right)(\r-\r_{o})^2 \\ \nn
&\hspace{15mm}-\frac{1}{324} e^{-12\phi_{I}}\left(4-28 e^{4\phi_{I}}+51 e^{8\phi_{I}}-27 e^{16\phi_{I}}\right)(\r-\r_{o})^4+\mathcal{O}\left((\r-\r_o)^6\right)\, ,\\
\chi(\r)&=\chi_{I}+\frac{1}{3}\log\left(\frac{5}{3}\right)+\frac{1}{3}\log (\r-\r_{o})
-\frac{1}{27} e^{-2\phi_{I}} \Big(2+\sinh\big(4\phi_{I}+\log (3)\big)\Big)(\r-\r_{o})^2\, \\ \nn
&\hspace{26mm}+\frac{5}{486}e^{-4\phi_{I}}\Big(2+\sinh\big(4\phi_{I}+\log (3)\big)\Big)^2(\r-\r_{o})^4+\mathcal{O}\left((\r-\r_o)^6\right)\, ,\\
A(\r)&= A_{I}+\frac{1}{3}\log\left(\frac{5}{3}\right)+\frac{1}{3}\log (\r-\r_{o})
-\frac{7}{324}e^{-6\phi_{I}}\left(1-12 e^{4\phi_{I}}-9 e^{8\phi_{I}}\right)(\r-\r_{o})^2\, \\ \nn
	 &\hspace{0mm}+\frac{1}{17496}\left(108-67 e^{-12\phi_{I}}+636 e^{-8\phi_{I}}-2124 e^{-4\phi_{I}}-1053 e^{4\phi_{I}}\right)(\r-\r_{o})^4+\mathcal{O}\left((\r-\r_o)^6\right)\, ,
\end{align}}%
where $\chi_{I}$ and $A_{I}$ generalise the integration constants appearing in the analytical solutions, and the third integration constant $\r_{o}$ may be chosen to fix the point at which the geometry ends. We will comment on the fourth integration constant $\phi_I$ momentarily. By using these expressions, we find that:
\begin{align}
e^{c(\r)}=e^{4A(\r)-\c(\r)}&= e^{4A_{I}-\c_{I}}\,f\big(\f_{I},\,(\r-\r_{o})\big)\, ,\\
e^{d(\r)}=e^{A(\r)-4\c(\r)}&= e^{A_{I}-4\c_{I}}\,g\big(\f_{I},\,(\r-\r_{o})\big)\, ,\label{Eq:Exp(d)conf}
\end{align}
where the functions $f$ and $g$ are known for $\phi_I=0$, and otherwise can be determined numerically.

The additive integration constant $A_{I}$ may be removed by a rescaling of the Minkowski coordinates. By contrast, because $\eta$ is a periodic coordinate with period $2\pi$, we are required to fix $\c_{I}$ to avoid a conical singularity at $\r_{o}$. In proximity of this point, the six-dimensional geometry resembles a two-dimensional space described by the following metric:
\begin{align}
 \text{d}s_{2}^{2}&= \text{d}\r^{2}+e^{6\c}\text{d}\eta^{2}\\
 &=\text{d}\r^{2}-\frac{5}{4}ve^{6\c_{I}}(\r-\r_{o})^2\text{d}\eta^{2}+\cdots\,,
\end{align}
from which we extract the required constraint:
\begin{equation}
\c_{I}=\frac{1}{6}\log\bigg(\frac{-4}{5v}\bigg)=-\frac{1}{3}\log\left(\frac{5}{3}\right)\,.
\end{equation} 
The one remaining free parameter of this system is $\phi_{I}$, which we constrain to take values $\phi_{I}\geqslant-\frac{1}{4}\log(3)$,
as we are interested only in solutions that reach back to the trivial critical point for large-$\r$ (the fact that not all possible solutions flow to the UV fixed point can be seen for the domain-wall solutions in Fig.~\ref{Fig:ParametricPlot}).

\subsection{Skewed solutions}
\label{Sec:SolutionsSingular}

There exists another class of analytical solutions with $\f=0$ for which the compact coordinate does not shrink to a point; $\chi(\r)$ is a non-monotonic function which diverges to $\infty$ at small $\r$. We refer to these solutions as \emph{skewed}. The solutions are as follows: 
\begin{align}
\phi&=\phi_{UV}=0,\\
\chi(\r)&=\c_{I}+\frac{1}{3}\log\bigg[\cosh\bigg(\frac{{\sqrt{-5v}}}{2}(\r-\r_{o}) \bigg) \bigg]
-\frac{1}{5}\log\bigg[\sinh\bigg(\frac{{\sqrt{-5v}}}{2}(\r-\r_{o}) \bigg) \bigg],\label{Eq:SkewChi}\\
A(\r)&=A_{I}-\frac{4}{15}\log(2)+\frac{4}{15}\log\bigg[\sinh\bigg(\sqrt{-5v}(\r-\r_{o}) \bigg) \bigg]\notag\\
&\hspace{65mm}-\frac{1}{15}\log\bigg[\tanh\bigg(\frac{{\sqrt{-5v}}}{2}(\r-\r_{o}) \bigg) \bigg] \,. \label{Eq:SkewA}
\end{align}
As with the confining solutions, we take note of the following two results obtained by substituting in for the skewed analytical solutions above:
\begin{align}
e^{c(\r)}=e^{4A(\r)-\c(\r)}&=\frac{1}{2}e^{4A_{I}-\c_{I}}\sinh\left(\frac{10}{3}(\r-\r_{o})\right)\, ,\\
e^{d(\r)}=e^{A(\r)-4\c(\r)}&=e^{A_{I}-4\c_{I}}\tanh\left(\frac{5}{3}(\r-\r_{o})\right)\, ,
\end{align}
which shows that these are indeed the solutions obtained from the confining ones with $\phi = 0$ (see Eqs.~\eqref{eq:confcd1} and~\eqref{eq:confcd2}) by replacing $d\rightarrow -d$, as anticipated in Section~\ref{Sec:EOM}.

Just as with the solutions that confine, we can generalise these analytical solutions to any values of $\f_I$ by series expanding for small $(\r - \r_o)$. We obtain the following IR expansions:
{\small\begin{align}
	\phi(\r) &= \phi_{I}-\frac{1}{12}e^{-6\phi_{I}} \left(1-4 e^{4\phi_{I}}+3 e^{8\phi_{I}}\right)(\r-\r_{o})^2 \\ \nn
	&\hspace{20mm}-\frac{1}{324} e^{-12\phi_{I}}\left(4-28 e^{4\phi_{I}}+51 e^{8\phi_{I}}-27 e^{16\phi_{I}}\right)(\r-\r_{o})^4+\mathcal{O}\left((\r-\r_o)^6\right)\, ,\\
	\chi(\r)&=\c_{I}-\frac{1}{5}\log\left(\frac{5}{3}\right) -\frac{1}{5}\log (\r-\r_{o})
	-\frac{1}{54} e^{-6\phi_{I}} \Big(1-12 e^{4\phi_{I}}-9 e^{8\phi_{I}}\Big)(\r-\r_{o})^2\, \\ \nn
	&\hspace{0mm}-\frac{1}{9720}e^{-12\phi_{I}}\Big[23+3e^{4\phi_{I}}
	\Big(-88+9 e^{4\phi_{I}}(38+24 e^{4\phi_{I}}+21 e^{8\phi_{I}})\Big)\Big](\r-\r_{o})^4+\mathcal{O}\left((\r-\r_o)^6\right)\, ,\\
	A(\r)&=A_{I}+\frac{1}{5}\log\left(\frac{5}{3}\right)+\frac{1}{5}\log (\r-\r_{o})
	-\frac{1}{36} e^{-6\phi_{I}} \Big(1-12 e^{4\phi_{I}}-9 e^{8\phi_{I}}\Big)(\r-\r_{o})^2\, \\ \nn
	&\hspace{-5mm}-\frac{1}{29160}e^{-12\phi_{I}}\Big[131+3e^{4\phi_{I}}
	\Big(-436+3 e^{4\phi_{I}}(508+84 e^{4\phi_{I}}+261 e^{8\phi_{I}})\Big)\Big](\r-\r_{o})^4+\mathcal{O}\left((\r-\r_o)^6\right)\, , 
	\end{align}}%
where the integration constants $\c_{I}$ and $A_{I}$ are the generalisation of the ones appearing in Eq.~(\ref{Eq:SkewChi}) and Eq.~(\ref{Eq:SkewA}), and $\f_{I}$ is the free parameter that we vary to generate the family of solutions. One can solve numerically the classical equations of motion, subject to boundary conditions derived from these IR expansions, in order to construct a class of skewed solutions.

We observe that the following relations hold:
\begin{align}
e^{c(\r)}=e^{4A(\r)-\c(\r)}&= e^{4A_{I}-\c_{I}}\,f\big(\f_{I},\,(\r-\r_{o})\big)\, ,\\
e^{d(\r)}=e^{A(\r)-4\c(\r)}&= e^{A_{I}-4\c_{I}}\,\Big[g\big(\f_{I},\,(\r-\r_{o})\big)\Big]^{-1}\, ,\label{Eq:Exp(d)skew}
\end{align}
where the functions $f$ and $g$ take exactly the same form as those in the analogous results for the confining solutions. Hence, provided that $\f_{I}^{conf}=\f_{I}^{skew}$ and $\r_{o}^{conf}=\r_{o}^{skew}$, one finds the relation
\begin{equation}
\partial_{\r}d^{conf}(\r)=-\partial_{\r}d^{skew}(\r) \,,
\end{equation}
where the $conf$ and $skew$ superscripts represent evaluation using the confining and skewed background solutions respectively. In turn, this implies that the relation
\begin{equation}
\c^{skew}(\r)-A^{skew}(\r)=-\frac{3}{5}\big(\c^{conf}(\r)+A^{conf}(\r)\big)\,
\end{equation}
is satisfied up to an additive integration constant. By comparing the UV expansions, one then finds the identifications:
\begin{align}
\label{Eq:id1}
\f_{2}^{skew}&=\f_{2}^{conf}\, ,\\
\label{Eq:id2}
\f_{3}^{skew}&=\f_{3}^{conf}\, ,\\
\label{Eq:id3}
\c_{5}^{skew}&=-\c_{5}^{conf}-\frac{8}{25}\f_{2}^{conf}\f_{3}^{conf} \,.
\end{align}

We conclude this subsection with an observation which motivates our choice of the name `skewed' for this class of solutions.\ From the six-dimensional metric in Eq.~(\ref{Eq:6Dmetric}) we can deduce the behaviour of the bulk geometry in the deep IR for these solutions. We notice by substituting for the small-$(\r-\r_{o})$ expansions that the size of the Minkowski directions scales as $(\r-\r_{o})^{\frac{2}{5}}$, while the compact dimension parametrised by $\eta$ scales as $(\r-\r_{o})^{-\frac{3}{5}}$. Hence in the $\r\to \r_{o}$ limit, the four-dimensional Minkowski volume vanishes, while the volume of the circle diverges. This contrasts with the small-$(\r-\r_{o})$ behaviour of the geometry for the confining solutions wherein the Minkowski directions maintain a fixed non-zero volume in the IR, while the circle shrinks to a point. The shrinking and expanding behaviour of the various metric components for this class of solutions motivates our choice of the name `skewed'. Appendix~\ref{Sec:CurvatureInvariants} is devoted to showing that while the confining solutions are regular, the skewed ones are singular.

\subsection{Generic (singular) solutions}
\label{Sec:Divergent}

When $\f$ diverges at the end of space, all curvature invariants diverge (see Appendix~\ref{Sec:CurvatureInvariants}).
 If $\f$ approaches $\f\rightarrow +\infty$,  we find a good singularity. These solutions are incomplete, but capture at least some salient features of the system. By contrast, in the case in which $\f\rightarrow -\infty$ at the end of space, the solutions result in a bad singularity, and we should disregard them as unphysical. Nevertheless, they play an important technical role in our study, as we anticipated in the Introduction, and as we shall see and explain in detail in Section~\ref{Sec:Free}.

We find that a broad, generic class of classical solutions can be parametrised by the  following 
expansion near the end of space at $\r=\r_o$:
\beqs
\label{Eq:series}
\phi&=&\phi_I + \phi_L \log(\rho-\rho_o) + \sum_{n=1}^{\infty} \sum_{j=0}^{2n} c_{nj} (\rho-\rho_o)^{2n+2n\,\phi_L-4j\,\phi_L}\,,
\eeqs
where the coefficients $c_{nj}$ depend on the free parameters $\phi_I$ and $\phi_L$.
Some useful details are provided in Appendix~\ref{Sec:Details}, while 
we exhibit here only the leading order terms of this expansion, 
ignoring all the power-law corrections:
\beqs
\phi(\r)&=&
\phi _I
    +\phi _L \log (\rho-\rho_o)
    \,+\cdots
\,,\\
   \chi(\r)&=&
   \chi _I+
   \frac{1}{15}
   \left(4 \zeta\sqrt{1-5 \phi _L^2}+1\right) \log (\rho-\rho_o)
   \,+\cdots
\,,\\
   A(\r)&=&
A_I+\frac{1}{15}
   \left(\zeta\sqrt{1-5 \phi _L^2}+4\right) \log (\rho-\rho_o)
   \,+\cdots
  \,.
 \eeqs
   The five integration constants  are $\phi_I$, $\chi_I$, $A_I$, $\phi_L$, and  $\r_o$,
   supplemented
   by the discrete choice $\zeta=\pm 1$. We notice that for $\phi_L=0$ and $\zeta=+1$ we recover
   the confining solutions, while for $\phi_L=0$ and $\zeta=-1$ we recover the skewed solutions.
 For $\phi_L\neq 0$ one obtains either
 solutions with a good singularity ($\phi_L<0$) or with a bad singularity ($\phi_L>0$).
 
The integration constant in front of the logarithm is constrained to take values within the range
\beqs
   -\frac{1}{\sqrt{5}} \leq \phi_L  < \frac{1}{3}\,.
   \label{Eq:bounds}
\eeqs
The lower bound $\phi_L\geq-\frac{1}{\sqrt{5}}$, arises from the requirements that both $\chi$ and $A$ be real.
For a choice that saturates this lower bound, and for $A_I=4\chi_I$, the solutions satisfy the condition $A=4\chi$ required by domain wall solutions. This parametrisation then encompasses all of the aforementioned solutions, with the exception of the IR-conformal and SUSY solutions.
  
The upper bound in Eq.~(\ref{Eq:bounds})  emerges from the requirement that all powers in Eq.~(\ref{Eq:series}) be positive. As for positive $\phi_L$ the worst power appearing at any given $n$ is $2n(1-3\phi_L)$, in order for all the powers to be positive, and that hence the IR divergence be logarithmic in $(\rho-\rho_o)$, we must require that $\phi_L<1/3$. (The same line of arguments for negative $\phi_L$ would be controlled by the $j=0$ power, in which case one would find the constraint $\phi_L>-1$.) This requirement is more stringent than requiring that $A$ and $\chi$ be real, which would yield $\phi_L \leq \frac{1}{\sqrt{5}}$. 
  
The limit $\phi_L\rightarrow \frac{1}{3}$ is such that the series expansion cannot be truncated nor resummed: at all infinitely many levels of $n$ one finds additive contributions proportional to $(\r-\r_o)^0$, to $(\r-\r_o)^{4/3}$ and so on. We discuss a related class of solutions in the next subsection.

\subsection{Badly singular domain wall solutions}
\label{Sec:SingDW}

Finally, we also found another class of singular domain-wall solutions,
for which the IR expansion is the following:

\beqs
\label{eq:IRexpBadlySingularDW}
\phi(\r)&=&
\frac{1}{6} \log \left(\frac{9}{4}\right)+\frac{\log (\r-\r_o)}{3}+ {\phi_4}(\r-\r_o)^{4/9}
+\sum_{j=2}^{\infty}f_j \,(\r-\r_o)^{\frac{4j}{9}}\,,\\
   \chi(\r)&=&
   \chi_I+\frac{1}{27}\log (\r-\r_o)
   +\frac{2}{5}{\phi_4}\, (\r-\r_o)^{4/9} 
   +\sum_{j=2}^{\infty}g_j \,(\r-\r_o)^{\frac{4j}{9}}\,.
\eeqs
Some more details about this expansion, truncated at the
order of $(\r-\r_o)^4$, are presented in Appendix~\ref{Sec:SingDWDetails}.
Together with the domain-wall constraint $A=4\chi$, this expansion identifies a class of solutions that depend on the trivial parameter
$\chi_I$, and two additional parameters: the position $\r_o$ of the end of space, and the integration constant $\phi_4$.
The coefficients $f_j$ and $g_j$ are polynomial functions of $\phi_4$. This family of solutions is the (non-trivial) limiting case of the solutions in Section~\ref{Sec:Divergent} obtained when $\phi_L\rightarrow 1/3$.
The freedom in choosing $\phi_I$ in the generic singular solutions  is replaced here by the freedom in $\phi_4$. We verified explicitly that the singularity is not removed by the lift to $D=10$ dimensions (see Appendix~\ref{sec:LiftTo10D}).

Although we cannot exclude a priori
the existence of  additional singular backgrounds with more exotic IR behaviours,
our exploration of the space of solutions that connect to the trivial ($\phi=0$) fixed point for large $\rho$, 
performed by perturbative generation of IR asymptotic expansions and evolution towards larger values of $\r$,
was confirmed by the result of scanning numerically the five-dimensional space of perturbations of the $\phi=0$ critical point,
and evolving the solutions backwards, towards small $\r$.
We did not find any indications that additional
 solutions with asymptotic UV behaviour in Section~\ref{Sec:UV}
 exist outside of the classes discussed in this section.

\subsection{Scale setting}
\label{Sec:ScaleSetting}

To facilitate comparison between all classes of solutions we choose to set $A_{U}=0$ and $\chi_{U}=0$ in all cases; the former 
assignment is permitted since the classical equations of motion are invariant 
under an additive shift of $A(\r)$, while the latter can be achieved by a rescaling
 of the radial coordinate $z\rightarrow z\,e^{3\chi_{U}}$. We are hence left with the UV parameters $\{\f_2,\f_3,\chi_5\}$.

Moreover, we find it useful to introduce a quantity that we use to set the scale
in the observables deduced from the free energy {(see later, in Section~\ref{Sec:Free})}, and that we 
conveniently
define as the time a massless particle takes to reach the end of space from the UV boundary, following Ref.~\cite{Csaki:2000cx}:
\begin{equation}
\label{Eq:Scale}
\Lambda^{-1}\equiv
\int_{r_{o}}^{\infty}\text{d}\tilde{r}\,e^{-A(\tilde{r})}=\int_{\r_{o}}^{\infty}\text{d}\tilde{\r}\,e^{\c(\tilde{\r})-A(\tilde{\r})}\, ,
\end{equation}
where $A$ and $\c$ are evaluated on the backgrounds. When a dual field theory interpretation exists, $\Lambda$ can be thought of as a characteristic energy scale, which governs among other things the mass gap of the theory.

We notice, by looking at the metric, that a trivial rigid rescaling of the coordinates $x^{\mu}\rightarrow \lambda \, x^{\mu}$ and $\eta \rightarrow \lambda \, \eta$ is equivalent to a rigid shift of $A$ and $\chi$ as $A\rightarrow A+\frac{4}{3}\log(\lambda)$ and $\chi\rightarrow \chi+\frac{1}{3}\log(\lambda)$. This is to be accompanied by a shift $\rho \rightarrow \rho - \frac{3}{2} \log(\lambda)$ such that $\chi_U$ and $A_U$ remain equal to zero. Under such a rigid shift, one can see that ${\Lambda} \rightarrow \lambda \, {\Lambda}$, and $ {\phi}_2\rightarrow \lambda^2  {\phi}_2$. It hence becomes evident that $\hat \phi_2 \equiv {\phi}_2{\Lambda}^{-2}$ is an invariant (dimensionless) quantity, which we denote by the hat. In the following, we often express our results in terms of such dimensionless quantities, by which we mean that we are measuring in units of $\Lambda$.

In order to appreciate the need for a scale setting procedure in the comparison of different classes of solutions, consider that the space of free parameters has different dimensionality for the confining and 
skewed solutions;  for the confining  solutions, 
the IR parameter $\c_I$ is fixed by the requirement of avoiding a 
conical singularity in the small $\r$ region of the bulk geometry, 
but no such constraint exists for the skewed solutions, in which the space 
does not smoothly shrink to a point.\footnote{In constructing the skewed solutions numerically, we exploit the fact that, as discussed in Section~\ref{Sec:SolutionsSingular}, they can be obtained (up to a trivial additive integration constant) from the confining solutions by making the substitution $d \rightarrow -d$.} To ensure that we can properly compare these 
two classes of solutions when plotting the free energy,
we measure all quantities in units of $\Lambda$, effectively reducing by one the dimension 
of the parameter space for the skewed solutions.  
We apply the same procedure to all other solutions as well, thus enabling us to compare the different branches of solutions in a consistent manner.

\section{Mass spectra, a tachyon, and a dilaton}
\label{Sec:Spectra}

Applying the dictionary of gauge-gravity dualities, the spectrum of small fluctuations around an asymptotically-AdS supergravity background can be interpreted in terms of the spectrum of bound states of the strongly-coupled dual field theory. All classes of solutions that we introduced in Section~\ref{Sec:Solutions} have the same asymptotically-AdS expansion for large $\r$, but only the third class of geometries (those which we referred to as \emph{confining}) have a regular end of space. We hence restrict our 
 attention to this class of solutions in this section, as they are the only candidates for
 admitting an interpretation in terms of confining field theories.
 
We devote this section of the paper to two calculations. We first compute the mass spectra for the full set of bosonic field excitations of the dimensionally-reduced model presented in Section~\ref{Sec:Reduction}. We then repeat the computation for the scalar excitations implementing the \emph{probe approximation}, according to the prescription described in Ref.~\cite{Elander:2020csd}. The former exercise will reveal the existence of a tachyonic spin-0 state in a certain region of parameter space for the class of confining solutions. The latter will show that, in proximity of this region, one scalar state is not only parametrically light, but also an (approximate) dilaton.       
 
\subsection{Mass spectra}
\label{Sec:GaugeInvariant}
 
We present in this subsection the mass spectra of fluctuations of the various bosonic supergravity fields of the sigma model coupled to five-dimensional gravity. We interpret the states as spin-$0,1,2$ glueballs of the dual confining field theory in four dimensions. In order to conduct this numerical analysis we employ the convenient gauge-invariant formalism developed in Refs.~\cite{Bianchi:2003ug,Berg:2005pd,Berg:2006xy,Elander:2009bm,Elander:2010wd}. The equations satisfied by the scalar fluctuations $\mathfrak{a}^{a}=\mathfrak{a}^{a}(M,\r)$ are given by
\be
\label{eq:scalarflucs1}
0=\left[\frac{}{}e^{\chi}\mathcal{D}_{\rho}(e^{-\chi}\mathcal{D}_{\rho}) + (4\partial_{\rho}A)\mathcal{D}_{\rho}
+e^{2\chi-2A}M^{2}\right]\mathfrak{a}^{a} - e^{2\chi} \mathcal X^{a}_{\ c}
\mathfrak{a}^{c}\,,
\ee
where $M$ is the mass of the composite states in the dual theory, and where
\begin{align}
\label{eq:scalarflucs2}
\mathcal X^{a}_{\ c}=
&-e^{-2\chi}\mathcal{R}^{a}_{\ bcd}
\partial_{\rho} \Phi^{b}
\partial_{\rho} \Phi^{d} 
+ D_{c}\bigg(G^{ab}\frac{\partial \mathcal{V}}{\partial  \Phi^{b}}\bigg)\, \notag\\
&+\frac{4}{3\partial_{\rho} A}
\bigg[\partial_{\rho} \Phi^{a}\frac{\partial \mathcal{V}}{\partial  \Phi^{c}}+G^{ab}\frac{\partial \mathcal{V}}{\partial  \Phi^{b}}\partial_{\rho}\Phi^{d}G_{dc}\bigg]
+\frac{16\mathcal{V}}{9(\partial_{\rho}A)^{2}}\partial_{\rho} \Phi^{a}\partial_{\rho} \Phi^{b}G_{bc}\, .
\end{align}
In all these expressions the quantities $\c$, $A$, ${\Phi}$, and $\mathcal V$ are evaluated on the background. Moreover, given a field $X^a$, we defined the sigma-model covariant and background covariant derivatives by $D_b X^a = \partial_b X^a + \mathcal G^a{}_{bc} X^c$ and $\mathcal D_\rho X^a = \partial_\rho X^a + \mathcal G^a{}_{bc} \partial_\rho \Phi^b X^c$ with the connection $\mathcal G^a{}_{bc} = \frac{1}{2} G^{ad} \left( \partial_b G_{cd} + \partial_ c G_{db} - \partial_d G_{bc} \right)$, while the sigma-model Riemann tensor is given by $\mathcal R^a{}_{bcd} = \partial_c \mathcal G^a{}_{bd} - \partial_d \mathcal G^a{}_{bc} + \mathcal G^a{}_{ce} \mathcal G^e{}_{bd} - \mathcal G^a{}_{de} \mathcal G^e{}_{bc}$. We impose the following  boundary conditions:\footnote{In practice, the equivalent form of the boundary condtions given in Eq.~(14) of Ref.~\cite{Elander:2014ola} turns out to be especially convenient in the numerical implementation.}
{\small
	\be
	\label{eq:scalarflucs3}
	\left.\frac{}{}e^{-2\chi}\partial_{\rho} \Phi^{c}
	\partial_{\rho} \Phi^{d}G_{db}\mathcal{D}_{\rho}\mathfrak{a}^{b}\right|_{\rho_{i}}
	=-\left.\left[\frac{}{}\frac{3\partial_{\rho}A}{2}e^{-2A}M^{2}\delta^{c}_{\ b}
	-\partial_{\rho} \Phi^{c}\bigg(\frac{4\mathcal{V}}{3\partial_{\rho}A}\partial_{\rho} \Phi^{d}G_{db} + \frac{\partial\mathcal{V}}{\partial\Phi^{b}} \bigg) \right]\mathfrak{a}^{b}\right|_{\rho_{i}}\, .
	\ee}

To compute numerically the mass spectra for the fluctuations of the fields, it is necessary to introduce regulators in the form of radial coordinate cutoffs; $\r_{1}$ is a (non-physical) infrared regulator chosen so that $\r_{o}<\r_{1}$, and $\r_{2}$ is chosen as the endpoint of the backgrounds in the far UV at large $\r$. The physical results are obtained by removing the two holographic regulators, i.e. by taking the limits $\r_1 \rightarrow \r_o$ and $\r_2 \rightarrow +\infty$. For a comprehensive explanation of this procedure (and our notation and conventions), and details not immediately important for the purposes of this paper, see Refs.~\cite{Elander:2010wd,Elander:2013jqa,Elander:2018aub}. In our numerical study of the spectrum, we chose $\r_1$ ($\r_2$) sufficiently close to the end of space (boundary), that cutoff effects are negligible---we estimate that the numerical precision is accurate to within a few percent.

The fluctuations of the pseudo-scalars $\pi^{i}$ satisfy the same equation as the scalar fluctuations above, with $G_{\pi\pi}=e^{-6\c-2\f}$, while the equations of motion for all the other fluctuations are the following~\cite{Elander:2018aub}:
\begin{align}
0&=\left[\frac{}{}\partial^{2}_{\r}+(4\partial_{\r}A-\partial_{\r}\c)\partial_{\r}+e^{2\c-2A}M^{2}\right]\mathfrak{e}^{\mu}_{\ \nu}\, ,\\
0&= P^{\mu\nu}\left[\frac{}{}e^{-\c}\partial_{\r}\left(\frac{}{}e^{2A+7\c}\partial_{\r}V_{\nu}\right)+M^{2}e^{8\c} V_{\nu}\right]\, ,\\
0&= P^{\mu\nu}\left[\frac{}{}e^{-\c}\partial_{\r}\left(\frac{}{}e^{2A+\c-2\f}\partial_{\r}A^{i}_{\nu}\right)+M^{2}e^{2\c-2\f} A^{i}_{\nu}\right]\, ,\\
0&=\partial^{2}_{\r}X+\left(\frac{}{}5\partial_{\r}\c-2\partial_{\r}A+2\partial_{\r}\f\right)\partial_{\r}X
+\left(M^{2}e^{-2A+2\c}-\frac{8}{9}e^{-6\f}\right)X\, ,\\
0&=
\left[\frac{}{}M^{2}+e^{3\c-4\f}
\partial_{\r}(e^{2A-5\c+4\f}\partial_{\r})-\frac{8}{9}e^{2A-2\c-6\f}\right]P^{\mu\nu}B_{6\nu}\, ,\\
0&=P^{\mu\nu}\big[\partial_{\r}
\big(e^{-\c}\partial_{\r}X_{\nu} \big)
-e^{-\c}(2\partial_{\r}\chi-2\partial_{\r}\phi)\partial_{\r}X_{\nu}
+e^{\c}(e^{-2A}M^{2}-\frac{8}{9}e^{-2\c-6\f} )X_{\nu} \big]\, ,\\
0&=P^{\mu\r}P^{\nu\s}\big[M^{2}e^{-2A}+e^{-5\c-4\f}\partial_{\r}\big(e^{3\c+4\f}\partial_{\r} \big) 
-\frac{8}{9}e^{-2\c-6\f} \big]B_{\r\s}
\,,
\end{align}
where $P^{\mu\nu} \equiv \eta^{\mu\nu} - \frac{q^\mu q^\nu}{q^2}$. The fluctuations $X$ and $X_{\mu}$ obey generalised boundary conditions, that reduce to Dirichlet in the limit of interest to this paper (see Eqs.~(B41) and~(B.42) of Ref.~\cite{Elander:2018aub} for detailed technical explanations). All other fluctuations obey Neumann boundary conditions.

The confining solutions are characterised by the constant $\phi_I \geqslant -\frac{1}{4}\log(3)$, where the lower bound would correspond to the IR fixed point of the dual five-dimensional QFT (in the sense that it results in a constant solution for the scalar field $\phi = \phi_{IR}$). Conversely, $\phi_{I} = \phi_{UV} = 0$ corresponds to the UV fixed point of the dual five-dimensional field theory. While the background solutions and spectra for $-\frac{1}{4}\log(3)\leqslant\phi_{I}\leqslant0$ have been presented in Ref.~\cite{Elander:2018aub}, the results for $\f_I>0$ are new to this work.

We show the results for the computation of the mass spectrum in Figs.~\ref{Fig:Spectra1} and \ref{Fig:Spectra2}. In Appendix~\ref{sec:SpectraLambda} we also show the same numerical results, but normalised with the scale setting parameter $\Lambda$ defined in Eq.~\eqref{Eq:Scale}. For the region $-\frac{1}{4}\log(3)\leqslant\phi_{I}\leqslant0$, each plot is in agreement with our previous computation in Ref.~\cite{Elander:2018aub}; of more interest are the observations that for large enough $\f_{I}$ one of the states in the scalar spectrum becomes tachyonic, and that the lightest massive states in two of the other towers ($B_{6\mu}$ and $B_{\mu\nu}$) appear to become massless in the limit of large $\phi_{I}$.

\begin{figure}[t]
\includegraphics[width=17.8cm]{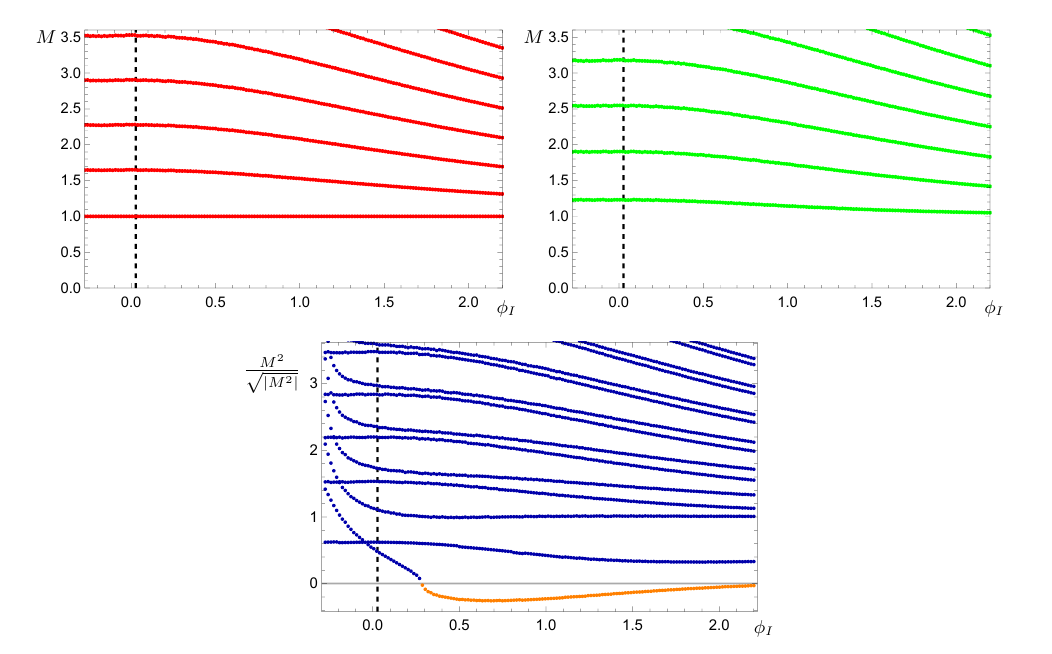}
\caption{The spectra of masses $M$, as a function of the one free parameter $\phi_I$ 
	characterising the confining solutions, 
	and normalised in units of the lightest tensor mass, computed with $\r_1 = 10^{-4}$ and $\r_2 = 12$. From top to bottom, left to right, 
	the spectra of fluctuations of the tensors $\mathfrak{e}^{\mu}_{\ \nu}$ (red), 
	the gravi-photon $V_{\mu}$ (green) and the two scalars 
	$\phi$ and $\chi$ (blue). The orange points in the plot of the scalar 
	mass spectrum represent values of $M^{2}<0$ and hence denote a tachyonic state. 
	We also show by means of the vertical dashed lines the case $\phi_I= \phi_I^c>0$, the critical value
	  that is introduced and discussed in Section~\ref{Sec:Free}.}
\label{Fig:Spectra1}
\end{figure}

\begin{figure}[t]
\includegraphics[width=17cm]{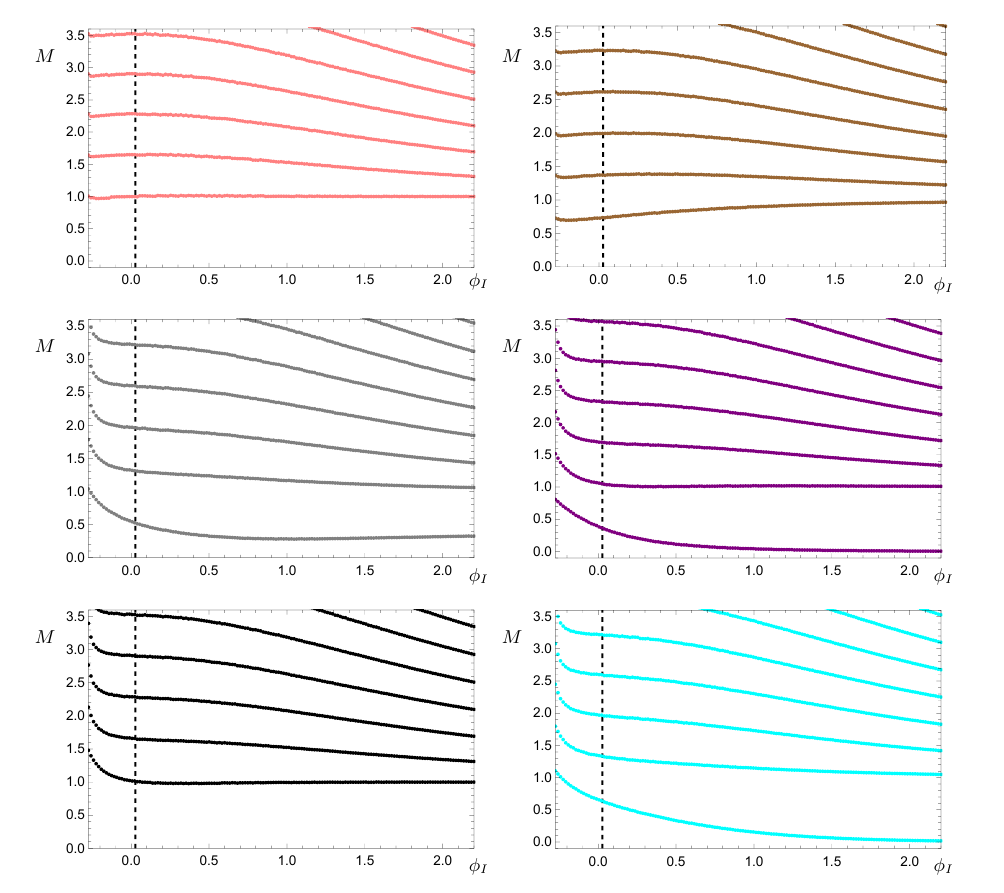}
\caption{The spectra of masses $M$ as a function of the scale parameter $\phi_I$, 
	normalised in units of the lightest tensor. From top to bottom, left to right, the spectra of fluctuations 
	of the pseudo-scalars $\pi^{i}$ forming a triplet \textbf{3} of $SU(2)$
	 (pink), vectors  $A^{i}_{\mu}$ forming a triplet \textbf{3} of $SU(2)$ (brown), 
	 $U(1)$ pseudo-scalar $X$ (grey), 
	 $U(1)$ transverse vector $B_{6\mu}$ (purple), 
	 $U(1)$ transverse vector $X_{\mu}$ (black) and 
	 the massive U(1) 2-form $B_{\mu\nu}$ (cyan). 
	 The spectrum was computed using the regulators $\r_1 = 10^{-4}$ and $\r_2 = 12$ with the exception of the $U(1)$ pseudo-scalar $X$ for which we used $\r_1 = 10^{-7}$ in order to minimize the cutoff effects present for the very lightest state at large values of $\phi_I$.
	We also show by means of the vertical dashed lines the case $\phi_I= \phi_I^c>0$, the critical value
	  that is introduced and discussed in Section~\ref{Sec:Free}.}
\label{Fig:Spectra2}
\end{figure}

\subsection{Probe scalars and dilaton mixing}
\label{Sec:Probe}
 
This is one of the central subsections to the paper. We analyse the composition of the scalar particles  in the spectrum in terms of fluctuations of the background fields, in order to establish whether any of them can, at least approximately, be identified with the dilaton. The magnitude of the condensates  in the underlying dynamics, as evinced from the  parameters $\phi_3$ and $\chi_5$ (see Appendix~\ref{Sec:ParametricPlots}), changes along the branch of confining solutions, providing a natural interpretation for the emergence of a dilaton and its properties. We will further return to this point, later in the paper.
 
The spin-0 mass spectrum presented in the previous subsection represents the solutions to the scalar fluctuation equation for the gauge-invariant combination (see Refs.~\cite{Bianchi:2003ug,Berg:2005pd,Berg:2006xy,Elander:2009bm,Elander:2010wd,Elander:2020csd}) given by
\be
\label{Eq:GothicA}
\mathfrak{a}^{a}(M,\rho)=\varphi^{a}(M,\rho)-\frac{\partial_{\rho}{\Phi}^{a}(\rho)}{6\partial_{\rho}A(\rho)}h(M,\rho)\, ,
\ee
where $\varphi^{a}(M,\r)$ are the first-order fluctuations of the scalar fields about their respective background solutions ${\Phi}^{a}(\r)$,
 while $h(M,\r)$ describes small perturbations of the trace of the four-dimensional tensor component 
 of the ADM-decomposed five-dimensional metric tensor.
  In terms of the dual field theory, $\varphi^{a}$ are associated with generic scalar operators 
  that define the theory, while $h$ is associated to the dilatation operator.

We are interested in determining to what extent any of the  scalar particles is a dilaton, i.e. whether mixing effects between 
$\varphi^{a}$ and $h$ are important. To this end, in this subsection we repeat the computation of the mass spectrum in the
 spin-0 sector, by using the \emph{probe approximation}: we neglect the contribution of the metric 
 perturbation $h$ in Eq.~(\ref{Eq:GothicA}), effectively removing any back-reaction
  the scalar fluctuations may have on the bulk geometry (for details, see \cite{Elander:2020csd}). We then check how well the resulting 
  spectrum computed with $\mathfrak{a}^{a}\rvert_{h=0}$ agrees with the correct computation making use of $\mathfrak{a}^{a}$. 
  If we find that the two calculations yield results that are in good agreement
   then we may infer that the contribution of the metric perturbation is negligible and hence 
   the spin-0 state is not a dilaton; if, by contrast, the two results disagree, then this is a clear indication of the fact that the 
   metric perturbation affects significantly the spectrum and hence the scalar state has a significant dilaton component.

\begin{figure}[t]
\begin{center}
\includegraphics[width=14.5cm]{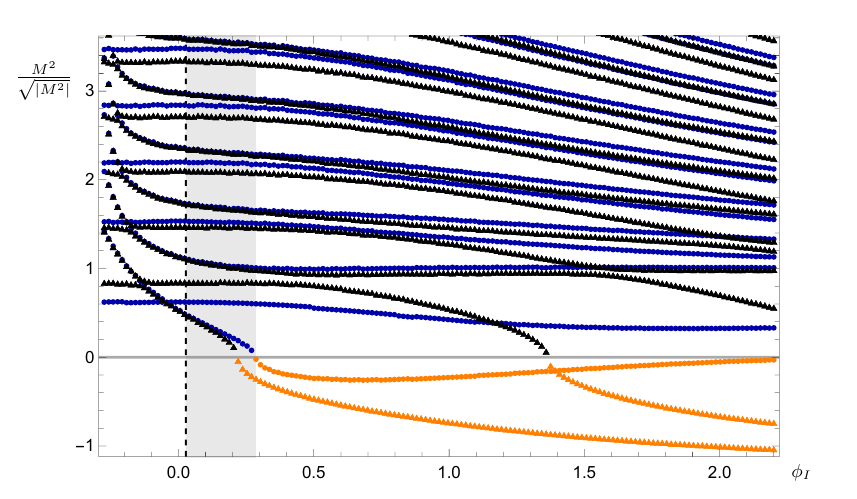}
\caption{The spectra of scalar masses $M$ as a function of the parameter $\phi_{I}$
	along the confining branch of solutions, normalised in units of the lightest tensor mass, computed with $\r_1 = 10^{-4}$ and $\r_2 = 12$. As in Fig.~\ref{Fig:Spectra1},
	 the blue disks represent the two scalars of the model $\phi$ and $\chi$, while the orange disks denote the tachyon. 
	 We additionally include the results of our mass spectrum computation using the probe approximation for $M^{2}>0$
	  (black triangles) and $M^{2}<0$ (orange triangles); note that these do not represent additional states.
	We also show by means of the vertical dashed line the case $\phi_I= \phi_I^c>0$, the critical value
	  that is introduced and discussed in Section~\ref{Sec:Free}. The shaded gray region indicates the metastable region of parameter space.}
\label{Fig:FullScalar}
\end{center}
\end{figure}

\begin{figure}[t]
\begin{center}
\includegraphics[width=14.5cm]{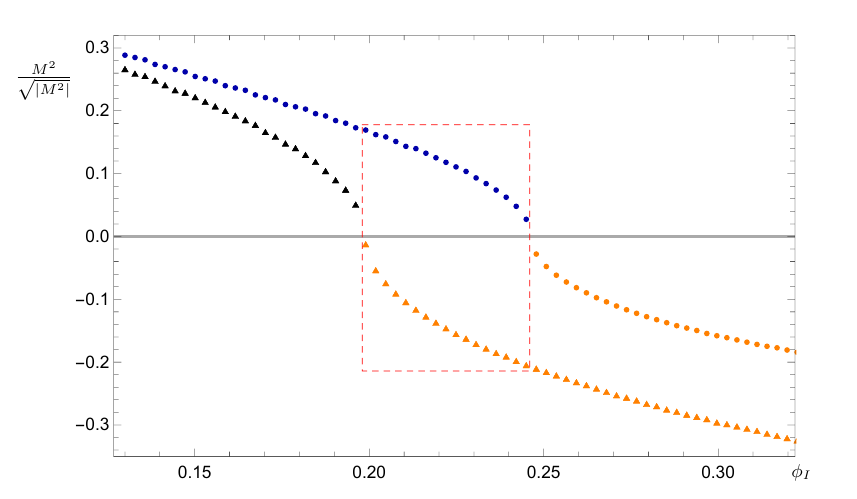}
\caption{A magnification of the plot shown in Fig.~\ref{Fig:FullScalar}. We normalised the masses $M$ in units of the lightest tensor mass, and computed with $\rho_1 = 10^{-9}$ and $\rho_2 = 15$. We focus in particular on the lightest state of the spectrum, in the plot region where the tachyonic states first appear. The dashed red box is intended to enclose an important feature of the full spectrum in Fig.~\ref{Fig:FullScalar}, namely a region of $\f_{I}$ parameter space wherein the probe approximation completely disagrees with the full gauge-invariant scalar computation. We see that there exists a finite range of values of the IR parameter $\f_{I}$ for which the squared masses $M^{2}$ of the physical scalars $\mathfrak{a}^{b}$ and the probes $\mathfrak{a}^{b}\rvert_{h=0}$ differ by a minus sign, and hence the probe approximation unambiguously fails.}
	\label{Fig:FullScalarZoom}
\end{center}
\end{figure}

As can be seen in Figs.~\ref{Fig:FullScalar} and \ref{Fig:FullScalarZoom}, the probe approximation is not accurate, and for all values of $\phi_I$ at least one of the lightest states is not well captured. This state is the lightest scalar for large, negative $\phi_I$, and becomes the next to lightest state for $\phi_I$ close to zero. 
This is a  mixed state that has a significant overlap with the dilatation operator.
We have already discussed the case $\f\leq 0$ elsewhere~\cite{Elander:2020csd}, and we will not return 
to the details of that discussion here. Interestingly,
for $\f_I\sim 0.25$, starting from the region in close proximity of (but before) the appearance of the tachyon,
the discrepancy between the probe approximation and the mass of the lightest physical scalar
becomes much more pronounced
(see Fig.~\ref{Fig:FullScalarZoom}). 
In this region of parameter space the lightest scalar particle can be rendered 
parametrically light with respect to all other states, and it is legitimate to interpret it as an approximate dilaton.
It is to be noticed that the next-to-lightest state is still not well captured by the probe approximation, due to mixing effects.

Let us now discuss the confining solutions with large values of $\phi_I$. In the limit $\phi_I \rightarrow +\infty$, the plots in Appendix~\ref{Sec:ParametricPlots} show that $\hat \phi_2 \rightarrow 0$ and $\hat \phi_3 \rightarrow +\infty$, as in the SUSY solution. Since $\hat \phi_2$ is connected to the explicit breaking of scale invariance, while $\hat \phi_3$ encodes its spontaneous breaking, in this limit one expects the emergence of an exact dilaton. This is confirmed by the fact that mass squared of the tachyon approaches zero from below, as can be seen in Fig.~\ref{Fig:FullScalar}. While these solutions are unphysical, this observation nevertheless provides a non-trivial check of our analysis. We further note that as $\phi_I$ is increased, the probe approximation results in additional heavier states becoming lighter and eventually tachyonic. This reinforces the fact that it is not only the tachyon and the lightest scalar that mix with the dilaton, but some of the heavier states as well. We finally notice that, besides a small number of light, discrete states, the spectrum of heavy particles in four dimensions becomes densely packed, eventually degenerating into a gapped continuum. Early evidence of this phenomenon can be seen in all the mass spectra, in Figs.~\ref{Fig:Spectra1}~and~\ref{Fig:Spectra2}. This final observation is reminiscent of the features that emerge in proximity of the CVMN solution \cite{Chamseddine:1997nm,Maldacena:2000yy} along the baryonic branch of the Klebanov-Strassler system~\cite{Elander:2017cle,Elander:2017hyr}---see also Refs.~\cite{Brandhuber:2000fr,Brandhuber:2002rx} that study the gravity dual of the Coulomb branch of $\mathcal N = 4$ super Yang-Mills.

We conclude this section by summarising our results for the spectrum and interpreting them in terms of 
the dual field theory. We consider only the  regular (confining) solutions, and we start from the region of parameter space in proximity of the backgrounds with $\phi=0$. The dual field theory is given by a supersymmetric fixed point in $D=5$ dimensions, that admits two deformations. One corresponds to the insertion of an operator of dimension $\Delta=3$, the source for which is encoded in the boundary value of the field $\phi$, via the coefficient $\phi_2$, and the response function, which is related to the coefficient $\phi_3$.
 The other is the compactification on a circle of one of the space-like dimensions, which is encoded in the gravity theory by the marginal deformation corresponding to $\chi$---by the coefficient $\chi_5$. The gravity solutions all correspond to dual theories that confine, in the usual sense typical of strongly-coupled gauge theories in four dimensions.
 
 Scale symmetry is both spontaneously and explicitly broken. The spectrum of bound states in proximity of 
 $\phi=0$ contains two almost degenerate scalar bound states: the lightest of them is well captured by the 
 probe approximation, and it corresponds to fluctuations sourced by the operator of dimension $\Delta=3$.
Its overlap with the dilaton is negligible.
The other state, conversely, can be identified with an approximate dilaton (in the sense that it would couple to the dilatation operator as a dilaton does), and its dynamical origin is the unsuppressed vacuum 
expectation value of the marginal operator.
We highlight the fact that $\phi_3$ vanishes when $\phi_2\rightarrow 0$, but this is not the case for $\chi_5$ (see the plots in Appendix~\ref{Sec:ParametricPlots}). This region of parameter space resembles
 generic Yang-Mills theories: there is no sense in which the explicit breaking of scale invariance is parametrically small compared to the scale of spontaneous breaking, and hence while one of the scalar bound states inherits some of the properties of an approximate dilaton, it is not parametrically light.
We further discuss the regime in which $\phi_3$ is large in Section~\ref{Sec:phasetransition},
where we return to the results of the exercise performed in the current subsection.

\section{Free energy and a phase transition}
\label{Sec:Free}

In this section we discuss the stability of backgrounds belonging to all the distinct classes of solutions
introduced earlier on. We do so by computing the free energy density of the system, with a prescription that
allows us to compare unambiguously to one another solutions belonging to different classes.

\subsection{General action and formalism}
\label{Sec:FreeGeneral}

Our first step  is to derive the free energy of the solutions from
the  truncated action of the  scalar field $\f$ coupled to gravity in $D=6$ dimensions---while setting
equal to zero all other fields.
We  include a boundary at $\r=\r_2$ as a regulator, with the understanding that the physical field theory
results will be recovered at the end of the calculations by taking the limit $\r_2\rightarrow +\infty$.
We also need to introduce a regulator in the IR: despite the fact that some of the solutions we consider are
completely smooth, the physical space is bounded
by $\r_1<\r<\r_2$. It is understood that eventually
we will  take $\r_1\rightarrow \r_o$, with $\r_o$ the physical end of the geometry.
The presence of boundaries requires on general grounds adding to the action the Gibbons-Hawking-York terms $S_{GHY,i}$ and boundary-localised potentials $S_{pot,i}$, for $i=1,2$. We hence write the action as follows.
\beqs
\label{Eq:GenAction}
\mathcal S&=& \mathcal S_{bulk}+\sum_{i=1,2}\big(\mathcal S_{GHY,i}+ \mathcal S_{pot,i}\big)\nonumber\\
&=&\int\text{d}^{4}x\,\text{d}\eta\,\text{d}\r \sqrt{-\hat{g}_{6}}\bigg(\frac{\mathcal{R}_{6}}{4}-\hat{g}^{\hat{M}\hat{N}}\partial_{\hat{M}}\f\partial_{\hat{N}}\f-\mathcal{V}_{6}(\f)\bigg)\nonumber\\
&&\left.+ \sum_{i=1,2} (-)^i\int\text{d}^{4}x\,\text{d}\eta\sqrt{-\tilde{\hat g}}
\bigg(\frac{K}{2}+\l_{i} \bigg)\right|_{\r=\r_{i}} \,,
\eeqs
where $\hat{g}_{\hat{M}\hat{N}}$ is the metric tensor for the six-dimensional line
element in Eq.~(\ref{Eq:6Dmetric}) for $V_M=0$, $\hat{g}_{6}$ is its determinant,
$\mathcal{R}_{6}$ is the corresponding Ricci scalar, and $\tilde{\hat g}_{\hat{M}\hat{N}}$ is the metric induced on each boundary.

In order to define the induced metric, we introduce the six-vector $n_{\hat{M}}=(0,0,0,0,1,0)$,
that satisfies the defining relations:
\begin{align}
	1&=\hat{g}_{\hat{M}\hat{N}}n^{\hat{M}}n^{\hat{N}}=n^{\hat{M}}n_{\hat{M}}\, ,\\
	0&=n^{\hat{M}}(\hat{g}_{\hat{M}\hat{N}}-n_{\hat{M}}n_{\hat{N}})\,.
\end{align}
The covariant derivative is written in terms of the connection as
\begin{align}
	\nabla_{\hat{M}}f_{\hat{N}}&\equiv\partial_{\hat{M}}f_{\hat{N}}-\Gamma^{\hat{Q}}_{\hat{M}\hat{N}}f_{\hat{Q}}\, ,\\
	\Gamma^{\hat{P}}_{\hat{M}\hat{N}}&\equiv\frac{1}{2}\hat{g}^{\hat{P}\hat{Q}}\big(\partial_{\hat{M}}\hat{g}_{\hat{N}\hat{Q}}+\partial_{\hat{N}}\hat{g}_{\hat{Q}\hat{M}}-\partial_{\hat{Q}}\hat{g}_{\hat{M}\hat{N}}\big)\, .
\end{align}
We can now define the induced metric tensor $\tilde{\hat g}_{\hat{M}\hat{N}}$ and the extrinsic curvature $K$ as follows:
\begin{align}
	\tilde{\hat g}_{\hat{M}\hat{N}}&\equiv \hat{g}_{\hat{M}\hat{N}}-n_{\hat{M}}n_{\hat{N}}\, ,\\
	K&\equiv \hat g^{\hat{M}\hat{N}}K_{\hat{M}\hat{N}}=\hat g^{\hat{M}\hat{N}}\nabla_{\hat{M}}n_{\hat{N}}\, ,
\end{align}
so that with our conventions we find that
\begin{equation}
	K=-\hat g^{\hat{M}\hat{N}}\Gamma^{5}_{\hat{M}\hat{N}}=4\partial_{\r}A-\partial_{\r}\chi\, = \partial_\r c \,.
\end{equation}

In order to calculate the free energy, one needs to evaluate the action on-shell. The bulk part of the action then has two components: one proportional to the equations of motion themselves, that hence vanishes when evaluated on any classical background solution,
and a second part that reduces to a total derivative. We can use Eq.~(\ref{Eq:EOMd}) from Section~\ref{Sec:EOM} to rewrite the bulk action as:
\begin{equation}
	\mathcal S_{bulk} = \mathcal S_{bulk,1} + \mathcal S_{bulk,2} = -\frac{3}{8}\int_{\r_{1}}^{\r_{2}}\text{d}^{4}x\text{d}\eta\text{d}\r\,\partial_{\r}\big(e^{4A-\chi}\partial_{\r}A\big)\, .
\end{equation}
Explicit evaluation shows that the boundary-localised contributions, evaluated on-shell, yield
\beqs
{\cal S}_{GHY,1}&=&
-\int\text{d}^{4}x\text{d}\eta\, e^{4A-\chi}\Big(2\partial_{\r}A-\frac{1}{2}\partial_{\r}\chi\Big)\Big\rvert_{\r=\r_{1}}
\,,\\
{\cal S}_{pot,1}&=&
-\int\text{d}^{4}x\text{d}\eta\, e^{4A-\chi}\Big(\lambda_{1}\Big)\Big\rvert_{\r=\r_{1}}\,,\\
{\cal S}_{GHY,2}&=&
\int\text{d}^{4}x\text{d}\eta\, e^{4A-\chi}\Big(2\partial_{\r}A-\frac{1}{2}\partial_{\r}\chi\Big)\Big\rvert_{\r=\r_{2}}
\,,\\
{\cal S}_{pot,2}&=&
\int\text{d}^{4}x\text{d}\eta\, e^{4A-\chi}\Big(\lambda_{2}\Big)\Big\rvert_{\r=\r_{2}}\label{Eq:Spot}\,.
\eeqs
The free energy $F$ and the free-energy density ${\cal F}$ are defined as
\begin{equation}
	F\equiv - \lim_{\r_1\rightarrow \r_o}\lim_{\r_2\rightarrow +\infty}{\cal S}\equiv\int\text{d}^{4}x\text{d}\eta\,\cf\, ,
\end{equation}
which yields the general result
\begin{align}
\label{Eq:F}
{\cal F}&= \lim_{\r_1\rightarrow \r_o}\frac{1}{8}e^{4A-\chi}\Big(13\partial_{\r}A-4\partial_{\r}\chi+8\lambda_{1}\Big)\Big|_{\r_1}\notag\\
&-\lim_{\r_2\rightarrow +\infty}\frac{1}{8}e^{4A-\chi}\Big(13\partial_{\r}A-4\partial_{\r}\chi+8\lambda_{2}\Big)\Big|_{\r_2}\,.
\end{align}
In the body of the calculations, we adopt the following prescription. We choose $\lambda_1=-\frac{3}{2}\partial_{\rho}A$ and $\lambda_2 = {\cal W}_2$ (it is sufficient to know the form of ${\cal W}_2$ up to quadratic order in $\phi$ in order to extract the divergent and finite parts), and
as a result the free energy density is
\begin{equation}
\label{Eq:Flambda}
{\cal F}=\lim_{\r_1\rightarrow \r_o}\frac{1}{8}e^{4A-\chi}\Big(\partial_{\r}A-4\partial_{\r}\c\Big)\Big|_{\r_1}
-\lim_{\r_2\rightarrow +\infty}\frac{1}{8}e^{4A-\chi}\Big(13\partial_{\r}A-4\partial_{\r}\chi+8\mathcal{W}_{2}\Big)\Big|_{\r_2}\,.
\end{equation}
\indent The choice of $\lambda_1$ is dictated by the  requirement that the variational principle be well defined, 
and the variation of the bulk action supplemented by the IR boundary action yields the bulk equations of motion
and boundary conditions at $\r=\r_1$.\footnote{This leads to the requirement that $\lambda_1 |_{\rho_1}=-\frac{3}{2}\partial_{\rho}A |_{\rho_1}$ evaluated at the IR boundary, hence explaining the aforementioned choice. Note, however, that we do not need to know the explicit functional dependence of $\lambda_1$ on $\phi$ in order to perform our calculation of the free energy.}
We find that with this choice
\begin{equation}
\label{eq:FreeEnergyIRplusUV}
{\cal S}_{GHY,1} + {\cal S}_{pot,1}=-\frac{1}{2} \int \di^4 x \di \eta\left(\frac{}{} 
e^{4A-\c}\left(\partial_{\r}A-\partial_{\r }\c\right)\right)\Big|_{\r_1}\,,
\end{equation} 
and by looking at the IR expansions of the regular solutions we find that the boundary-localised action
does not contribute to their free energy in the $\r_1\rightarrow \r_o$ limit. 
Hence, in the case in which the geometry closes smoothly in the IR, 
the presence of the regulator is unnecessary and has no physical effect.
We are now in a position to apply this prescription to all other solutions as well.

The choice of $\lambda_2$ is dictated by covariance, locality, and the requirement that 
all divergences cancel \cite{Bianchi:2001kw,Skenderis:2002wp,Papadimitriou:2004ap}. In the case at hand, in general one expects two types of UV divergences: 
one driven by the bulk cosmological constant, and one by the (square of the)
mass deformation $\f_2^2$. Because of these two divergences, ${\cal F}$ and its 
second derivative with respect to the source $\f_2$ are scheme-dependent. 
This is a generic feature, commonly  appearing in many holographic free energy calculations,
and has been observed in other contexts (see for instance the discussions in Ref.~\cite{Bobev:2013cja}).
For our purposes, it has one important implication: the classical statistical mechanics 
concavity theorems do not trivially apply to our 
results for the free energy, the minima of which will not exhibit a concavity with definite sign.
With our choice of $\lambda_2$, dictated by holographic renormalisation,
and by making use of the UV expansions and of the relation $\partial_{\r}=-\frac{2}{3}z\partial_z$, we find that
\beqs
{\cal S}_{GHY,2}&=&\int\di^4 x\di \eta \left. \frac{e^{4A_U-\chi_U}}{z^5}\,\left(\frac{5}{3}-\frac{5}{12}\phi_2^2z^4+ 0 \times z^5 + \cdots \right)\right|_{\r_2}\,,\\
{\cal S}_{pot,2}&=&\int\di^4 x\di \eta \left. \frac{e^{4A_U-\chi_U}}{z^5}\left(-\frac{4}{3}+\frac{1}{3}\phi_2^2z^4+\frac{8}{15}\phi_2\phi_3z^5 + \cdots \right)\right|_{\r_2}\, ,\\
{\cal S}_{bulk,2}&=&\int\di^4 x\di \eta \left. \frac{e^{4A_U-\chi_U}}{z^5}\left(-\frac{1}{3}+\frac{1}{12}\phi_2^2z^4+ \frac{1}{80} \left(4\phi_2\phi_3+25\chi_5 \right)z^5 + \cdots \right)\right|_{\r_2}\,.
\eeqs
The divergences exactly cancel, leaving a finite contribution to the free energy.

We observe that the contribution to the free energy coming from evaluation at the IR boundary $\r_1$ in Eq.~\eqref{eq:FreeEnergyIRplusUV} happens to be proportional to the combination appearing in Eq.~(\ref{Eq:C}).
This contribution hence coincides with a conserved quantity, that we can evaluate at any value 
of the coordinate $\r$. It is convenient to evaluate it at the UV boundary, where we notice that (as expected) it
gives a finite contribution. 
By substituting the general UV expansions, we hence obtain the following final result for the free energy density:
\begin{align}
\label{Eq:Fquasiquasifinal}
\cf&=\frac{1}{16}e^{4A_{U}-\chi_{U}}\big(4\phi_{2}\phi_{3}+25\chi_{5}\big)
-\frac{1}{48}e^{4A_{U}-\chi_{U}}\big(28\phi_{2}\phi_{3}+15\chi_{5}\big)\\
\label{Eq:Fquasifinal}
&=-\lim_{\r_2\rightarrow +\infty}e^{4A-\chi}\Big(\frac{3}{2}\partial_{\r}A+\mathcal{W}_{2}\Big)\Big|_{\r_2}\,\\
\label{Eq:Ffinal}
&=-\frac{1}{12}e^{4A_{U}-\chi_{U}}\big(4\phi_{2}\phi_{3}-15\chi_{5}\big)\, ,
\end{align}
where in the first line 
the first term comes from the $\r_{1}\rightarrow+\infty$ limit evaluation of the first term 
in Eq.~(\ref{Eq:Flambda}), and the second from the $\r_{2}$ limit evaluation. 
The second line is a general combination of all the contributions.
The third line is our main result, and we will return to it when we discuss each individual class of solutions, in the subsections to follow. For completeness, and to elucidate some subtle differences, we repeat this calculation in the five-dimensional language, in Appendix~\ref{sec:FreeEnergyin5D}, with identical results.

\subsection{Domain wall solutions}
\label{Sec:IRC-FreeEnergy}

If we impose the (domain-wall) constraint $A=4\chi$, this  introduces two additional 
constraints on the five UV parameters:
\begin{align}
A_{U}&=4\chi_{U}\, ,\\
\chi_{5}&=-\frac{4}{25}\f_{2}\f_{3}\, .
\label{Eq:DWc}
\end{align}  
From these two relations we may deduce the values of $\chi_{5}$ and $\chi_{U}$ given the other three parameters. 
We notice that the above constraint on $\c_{5}$ causes the first term of Eq.~(\ref{Eq:Fquasiquasifinal}) to vanish exactly, and we hence obtain the following expression for the free energy of the domain-wall (DW) solutions, 
which include, among others, the SUSY as well as the IR-conformal  solutions:
\begin{equation}
\cf^{(DW)}=-\frac{8}{15}e^{4A_{U}-\chi_{U}}\phi_{2}\phi_{3}\,.
\label{Eq:DWf}
\end{equation}

In the case of the IR-conformal solutions (IRC), one numerical background may be used 
 to generate any other by an additive shift of the holographic coordinate. 
The following ratio is an invariant:
\be
\kappa \equiv \frac{\lvert\f_{3}\rvert}{\lvert\f_{2}\rvert^{\frac{3}{2}}}\,.
\ee
We find numerically that $\kappa\simeq 2.87979$, so that the final result for the free energy is
\begin{equation}
\cf^{(IRC)}=-\frac{8}{15}\kappa\,\phi_{2}\,\lvert\phi_{2}\rvert^{\frac{3}{2}}\simeq
-\frac{8}{15}(2.87979)\,\phi_{2}\,\lvert\phi_{2}\rvert^{\frac{3}{2}}\, .
\end{equation}

\begin{table}[t]
	\begin{center}
		\begin{tabular}{|c|c|c|c|c|c|c|}
			\hline\hline
			Class &  $A_U$ &  $\chi_U$ & $\phi_2$  & $\phi_3$  & $\chi_5$ & Scale setting \cr
			\hline\hline
			SUSY  &  $0$ &  $0$ & $0$  & Free  & $A=4\chi$ & None \cr
			IR-conformal  &  $0$ &  $0$ & $<0$  & $\phi_3=\kappa \phi_2|\phi_2|^{1/2}$  & 
			$A=4\chi$ & None  \cr
			Confining  &  $0$ &  $0$ & Free   & Curvature sing.   & Conical sing. & $\Lambda$ \cr
			Skewed  &  $0$ &  $0$ & Free  & $c^{skew}=c^{conf}$  &  $d^{skew}=-d^{conf}$ & $\Lambda$ \cr
			Good Singular  &  $0$ &  $0$ & Free   & Free   & Free  & $\Lambda$ \cr
			Bad Singular  &  $0$ &  $0$ & Free   & Free   & Free  & $\Lambda$ \cr
			DW Singular &  $0$ &  $0$ & Free  & Free  & $A=4\chi$ & $\Lambda$ \cr
			\hline\hline
		\end{tabular}
	\end{center}
	\caption{Parametrisation, constraints and scale setting procedure of each class of solutions considered in the text,
	and in Figs.~\ref{Fig:Plot} and~\ref{Fig:AnotherPlot}.
	The scale $\Lambda$ is defined in Eq.~(\ref{Eq:Scale}), and has been used to restore physical units in ${\cal F}$ and  
		$\phi_2$ in the  energetics. 
		In the case of the IR-conformal solutions and of the SUSY solutions, no scale setting is
		used, because ${\cal F}=0$ in the former, and ${\cal F}=-\frac{8}{15}\kappa \phi_2|\phi_2|^{3/2}$ in the latter,
		but $\Lambda$ is not defined.
		The SUSY singular solutions are represented by a point in Figs.~\ref{Fig:Plot} and~\ref{Fig:AnotherPlot}, 
		the IR conformal, confining, skewed and singular DW solutions are represented by lines, and finally the generic (good as well as bad) singular solutions
		 cover a two-dimensional portion of the $(\hat{\phi}_2,\hat{\cal F})$ plane (see Figs.~\ref{Fig:Plot}
		  and~\ref{Fig:AnotherPlot}).
		}
	\label{Fig:summarytable}
\end{table}

\subsection{Numerical implementation}
\label{sec:numimp}

The general result for the free-energy density for all solutions  is in Eq.~(\ref{Eq:Ffinal}):
\begin{equation}\nonumber
\cf=-\frac{1}{12}e^{4A_{U}-\chi_{U}}\big(4\phi_{2}\phi_{3}-15\chi_{5}
\big)\,.
\end{equation}

All the classes of solutions we discuss are known numerically, and are obtained by exploiting the 
IR expansions we reported in Section~\ref{Sec:Solutions}. We implement a numerical routine to extract a table of UV parameter values $\{ \phi_2, \phi_3,\chi_5 \}$ for solutions of  each class, having set $A_{U}=\chi_{U}=0$. To this end, we do the following.
\begin{enumerate}
	\item For each given choice of IR expansion, we numerically solve the background equations of motion for $\f(\r)$, $\chi(\r)$, and $A(\r)$, having chosen the end-of-space to be at $\rho_o = 0$ with the boundary conditions set up at a small $\rho$.
	\item Starting from these solutions, we generate new ones by shifting the radial coordinate together with $\chi$ and $A$ such that the combined effect is to set $A_{U}=\chi_{U}=0$ as required.
	\item We match each numerical solution and its derivatives with the UV expansions, and extract $\f_{2}$, $\f_{3}$, and $\c_{5}$.   
\end{enumerate}
In the third step, one needs to choose a value of the radial coordinate $\rho = \rho_m$ at which to do the matching. This choice is dictated by the requirement to minimise the effect of the numerical noise, while at the same time ensuring that $\rho_m$ is large enough that the solutions have reached the region in proximity of the $\phi=0$ critical point. We do not report the details of this laborious process, but only our main results.

We checked that the numerical determination of the UV parameters can be used to set up the boundary conditions in the UV, and by solving again the equations of motion towards small $\rho$, we recover the original backgrounds. The Reader should be alerted of the fact that the non-linear nature of the equations is such that this second process does not allow one to reproduce accurately the region of the geometry in proximity of the end of space at small $\rho$, a region that is essential in the calculation of the scale-setting parameter $\Lambda$. Indeed, this is the reason why, for the purpose of numerical studies, it is preferable to construct the solutions by choosing the boundary conditions close to the end of space in the geometry, and evolving the differential equations towards large values of the holographic coordinate $\r$. We estimate the numerical precision of our calculation of the free energy and of the parameters relevant to the energetics study to be accurate within a few percent.

Singular solutions are treated in exactly the same way as the confining solutions, thanks to the
introduction of the regulator at $\r_1$ and to the prescription we discussed earlier in this 
section. A practical simplification of the procedure is given by the observation that 
the free energy of the skewed solutions is formally identical to that for the confining solutions,
except for the replacements in Eqs.~(\ref{Eq:id1}),~(\ref{Eq:id2}), and~(\ref{Eq:id3}).
 
 In Table~\ref{Fig:summarytable} we summarise some basic properties of the
  various classes of solutions relevant to the analysis that follows.
We repeat here some important and subtle points.
The scale-setting procedure for the SUSY  and  IR-conformal solutions is treated in a different way,
for specific reasons that we describe in the next subsection.
In the case of confining solutions, $\phi_3$ and $\chi_5$ are constrained
 by the requirement of eliminating curvature and conical singularities,
respectively. For the skewed solutions, these requirements are replaced by the fact that skewed solutions can be obtained from 
confining solutions by changing the sign of the background function $d$.

From here on, we find it convenient to define the following notation. We rescale all the physical quantities by the 
appropriate power of the scale $\Lambda$ defined in Eq.~(\ref{Eq:Scale}) as
\beqs
\hat{\cal F}&\equiv& {\cal F}\Lambda^{-5}\,,\\
\hat{\phi}_2&\equiv& {\phi}_2\Lambda^{-2}\,,\\
\hat{\phi}_3&\equiv& {\phi}_3\Lambda^{-3}\,,
\eeqs
and so on for all possible physical quantities. By doing so, as we will show explicitly, we can legitimately compare solutions belonging to any of the different classes described in this paper.

\subsection{Free energy density and the phase structure}
\label{Sec:FreeAnalysis}

\begin{figure}[t]
\begin{center}
\includegraphics[width=12.5cm]{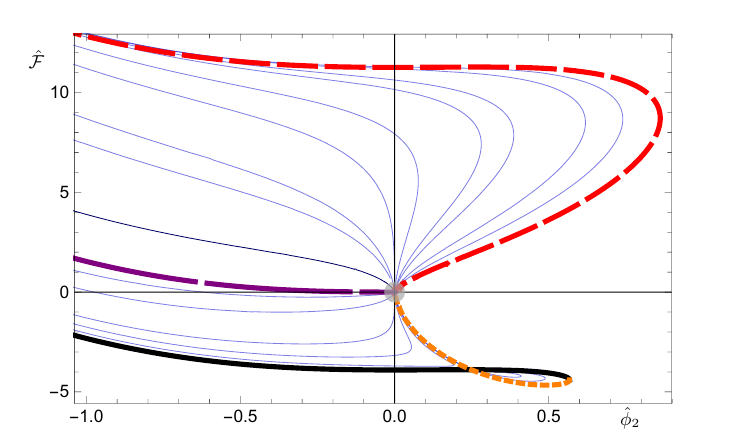}
\caption{The free energy density $\hat{\cal F}$ as a function of the deformation parameter $\hat\phi_2$ for the IR-conformal solutions (longest-dashed purple line), 
	the confining solutions (solid black line), and the skewed solutions (dashed red line), compared 
	to a few representative choices of (good) singular solutions (thin blue lines). For the latter, we generated the numerical solutions from the IR
	expansions, by setting $(\phi_L,\zeta)=(-0.02,-1)$, $(-0.04,-1)$, $(-0.08,-1)$, $(-0.15,-1)$, $(-0.2, -1)$, $(-0.25,-1)$, $(-0.3,-1)$, $(-0.35,-1)$, $(-0.35,1)$, $(-0.3,1)$, $(-0.25,1)$, $(-0.2,1)$, $( -0.15,1)$, $(-0.04,1)$, $(-0.02,1)$, respectively (top blue line to bottom blue line), and varied the value of $\phi_I$. The darker blue line, separating the cases $\zeta = \pm 1$, corresponds to the domain wall solutions obtained with $\phi_L = -1/\sqrt{5}$ and varying $\phi_I$. 
	 The SUSY solutions are represented by a grey point at the origin. 
	 The short-dashed orange line shows the region along the branch of confining solutions in which the tachyonic state appears in the scalar mass spectrum. (Note that the very top thin blue line crosses the dashed red one for large negative values of $\hat \phi_2$. We expect this to be a purely numerical artifact that could be removed with higher numerical precision.)}
\label{Fig:Plot}
\end{center}
\end{figure}

In order to investigate the energetics along all the branches of solutions, we employed a numerical 
routine to compute their free energy by extracting physical values of the 
five UV parameters; we present here the results of this numerical analysis.
In particular, we show how the free energy density ${\cal F}$ behaves as a function of  $\phi_2$,
the deformation of the theory corresponding to the aforementioned operator of dimension $\Delta=3$.
We repeat that we normalised the two quantities by the appropriate power of the scale
 $\Lambda$, in order to be able to compare different solutions.
As the plots are rather busy, showing a large amount of information, we
first devote some space to explaining how to read them, and then we analyse the physical results,
by treating separately the $\hat\phi_2<0$ and $\hat\phi_2>0$ cases.

In Fig.~\ref{Fig:Plot} we show five of the seven classes of solutions listed in Table~\ref{Fig:summarytable}.
\begin{itemize}

\item The SUSY solutions all have $\phi_2=0$, and because they satisfy the domain-wall constraint $A=4\chi$, 
by virtue of Eq.~(\ref{Eq:DWf}), which descends from Eq.~(\ref{Eq:DWc}), also ${\cal F}=0$. The integral 
defining $\Lambda$ in Eq.~(\ref{Eq:Scale}) diverges ($\Lambda\rightarrow 0$).
These solutions are
represented by the grey disk at the origin. 

\item The IR-conformal solutions exist only for $\phi_2< 0$. The integral defining $\Lambda$ in Eq.~(\ref{Eq:Scale}) diverges also in this class of solutions
($\Lambda\rightarrow 0$).
Yet, because of scale invariance, we find that ${\cal F}$ scales as a power of $\phi_2$, and we
represent these solutions with the longest-dashed purple line in Fig.~\ref{Fig:Plot}.
This line represents what would be the result of using any other possible scale-setting process for the IR-conformal solutions.

\item The confining solutions are rendered in solid black and short-dashed orange. They form a line, as we generate the solutions 
by varying the parameter $\phi_I$. We notice the existence of a maximum value of $\hat\phi_2$.
For graphical illustration, we rendered in short-dashed orange the part of the curve obtained with confining solutions
for which one of the scalar states has a negative mass squared (see Figs.~\ref{Fig:FullScalar} and~\ref{Fig:FullScalarZoom}).
Part of this tachyonic portion of the branch of solutions has free energy $\hat{\cal F}$
lower than the solutions with the same value of 
$\hat{\phi}_2$ located along the regular portion of the confining branch, and the short-dashed orange and solid black curves cross non-trivially.
This observation by itself would be proof that a phase transition takes place, were it not for the undesirable
feature that the tachyonic backgrounds would be minimising the free energy over a portion of parameter space.

\item The skewed solutions are rendered in dashed red. We obtained these solutions by changing the sign of $d\rightarrow -d$ from 
the confining solutions,
which implies the relations in Eqs.~(\ref{Eq:id1}), (\ref{Eq:id2}) and~(\ref{Eq:id3}). Also in this case there exists a maximum value
of $\hat \phi_2$.

\item The generic solutions with good singularity are depicted by thin blue lines. We choose a number of representative values 
for the parameter $\phi_L<0$, and discuss both choices of $\zeta=\pm 1$ (see Section~\ref{Sec:Divergent}). 
For $\phi_L\rightarrow 0$ the thin blue lines approximate the confining (for $\zeta=+1$) and skewed (for $\zeta=-1$) solutions, as expected.
For $\phi_L\rightarrow -\frac{1}{\sqrt{5}}$, one finds the special case of domain-wall solutions with good singularity (in this case the choice $\zeta=\pm 1$ is immaterial),
and we denote this line, which appears just above the longest-dashed purple one, with a darker shade of blue.

\end{itemize}

\begin{figure}[t]
\begin{center}
\includegraphics[width=12.5cm]{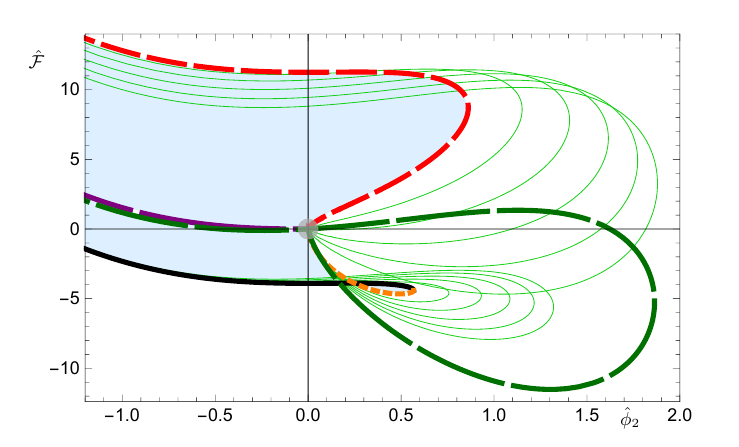}
\caption{The free energy density $\hat{\cal F}$ as a function of the deformation parameter $\hat\phi_2$ for the IR-conformal solutions (longest-dashed purple line), 
	the confining solutions (solid black line), and the skewed solutions (dashed red line), compared 
	to a few representative choices of (badly) singular solutions.
	 For the latter, we generated the numerical solutions from the IR
	expansions, by setting $(\phi_L,\zeta) = (0.05,-1)$, $( 0.1,-1)$, $( 0.15,-1)$, $( 0.2, -1)$, $(0.25, -1)$, 
         $( 0.25,1)$, $( 0.2,1)$, $( 0.15,1)$, 
	 $( 0.1,1)$, $(  0.05,1)$, 
	respectively (thin light-green lines), and varied the value of $\phi_I$.
	The long-dashed dark-green line represents the domain-wall (badly) singular solutions, obtained by varying the parameter $\phi_4$
	in the IR expansion in Section~\ref{Sec:SingDW}.
	 The SUSY solutions are represented by a grey point at the origin. 
	 The short-dashed orange line shows the region along the branch of confining solutions in which the tachyonic state appears in the scalar mass spectrum.
	 We shaded in light blue the region covered by the good singular solutions (see Fig.~\ref{Fig:Plot}).
}
\label{Fig:AnotherPlot}
\end{center}
\end{figure}

We notice one very important fact: thanks to the rescaling that defines $\hat{\cal F}$ and $\hat{\phi}_2$,
all branches of solutions depicted in Fig.~\ref{Fig:Plot} (and this holds true also in the subsequent Fig.~\ref{Fig:AnotherPlot})
connect to the origin of the diagram, with $\hat{\cal F}=0$ and $\hat{\phi}_2=0$.
This observation makes it explicitly clear that, despite the semi-classical nature of the calculations we performed, the free energy density $\hat{\cal F}$ is defined in a consistent way that allows for the comparison of all possible solutions along all the branches we identified, given that effectively they all share one common point.

All the thin blue lines are entirely contained within the region of the plot delimited by the solid black, short-dashed orange, and dashed red lines. Varying within this class of solutions, for all available choices of parameters, the confining solutions minimise $\hat{\cal F}$, while the skewed solutions maximise it.
The solutions with good singularity do not resolve either of the two problematic features of the confining class: they do not extend the plot
beyond the maximum value for $\hat \phi_2$, nor do they give us solutions with energy lower than the tachyonic sub-branch of the confining solutions.
Finally, we highlight how not only are the solutions fully contained 
inside the region delimited by the solid black, short-dashed orange, and dashed red curves, but also that, by varying $\phi_L$, we can span the entirety of this region.

In Fig.~\ref{Fig:AnotherPlot}, we add to the set of solutions on display several 
representative choices of badly singular solutions (in thin light green),
chosen by varying $\phi_L$ and $\zeta$, as well as
 the domain wall ones discussed in Section~\ref{Sec:SingDW} (in long-dashed dark green). 
 We replace the solutions with a good singularity by shading 
in light blue the whole region of the plane $(\hat\phi_2\,,\,\hat{\cal F})$ delimited by the confining and skewed solutions.
We notice two important features: for some choices of parameters, badly singular solutions exist that 
exceed the upper bound on $\hat\phi_2$
that we identified when discussing  the confining solutions, and furthermore there are domain-wall,
 badly singular solutions with free energy lower
that those along the tachyonic portion of the confining branch of solutions.

\begin{figure}[t]
\begin{center}
\includegraphics[width=12.5cm]{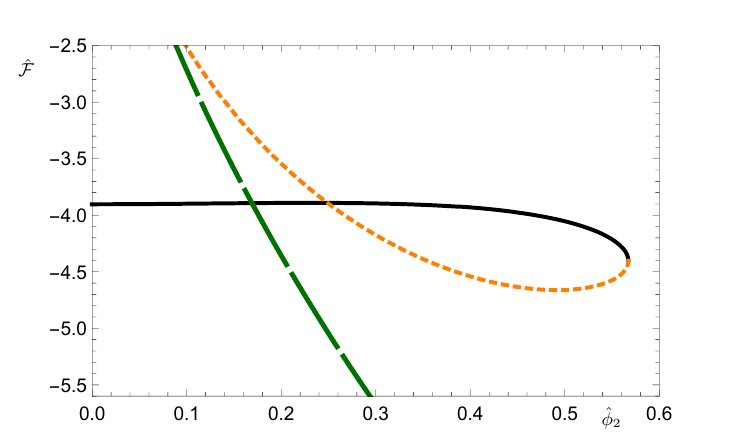}
\caption{The free energy density $\hat{\cal F}$ as a function of the deformation parameter $\hat\phi_2$ for the confining solutions (solid black and short-dashed orange lines), and
	 the domain-wall (badly) singular solutions, obtained by varying the parameter $\phi_4$
	in the IR expansion in Section~\ref{Sec:SingDW} (long-dashed dark-green line).}
\label{Fig:YetAnotherPlot}
\end{center}
\end{figure}

The plot in Fig.~\ref{Fig:AnotherPlot} clearly displays the features expected in the presence 
of a phase transition, and we will return to it shortly.
Fig.~\ref{Fig:YetAnotherPlot} is a detail of  Fig.~\ref{Fig:AnotherPlot}, in which we retained only
the confining solutions (solid black and short-dashed orange lines) and the badly  singular domain-wall solutions (in long-dashed dark green).
We highlight the region in proximity of the intersection between the two lines,
which identifies a critical value $\hat \phi_2^c$ of the deformation parameter $\hat \phi_2$.
The minimum of the free energy density is given by confining solutions for $\hat \phi_2 < \hat\phi_2^c$,
and by badly singular domain-wall solutions for $\hat\phi_2 > \hat \phi_2^c$. The tachyonic section of the 
confining branch is never a minimum of the free energy at fixed $\hat \phi_2$.

We  now discuss the physics lessons we learn from the combination of Figs.~\ref{Fig:Plot}, 
\ref{Fig:AnotherPlot} and~\ref{Fig:YetAnotherPlot}.
For negative values of $\hat \phi_2$, we find that all classical solutions identified in the body of the paper 
have finite free energy density $\hat{\cal F}$, and this is bounded from below by the confining solutions,
and from above by the skewed solutions. All other solutions have free energy somewhere in between---they
 include the SUSY solutions, the IR-conformal solutions, and all the generic 
 solutions discussed in Section~\ref{Sec:Divergent} and Section~\ref{Sec:SingDW}.
If we were to restrict attention to $\phi\leq 0$ (as done for example in Refs.~\cite{Elander:2013jqa} and~\cite{Elander:2018aub}), 
there would be no benefit from the study of solutions other
than the confining ones, which already minimise the free energy, have no curvature nor conical singularities, 
admit a sensible field theory interpretation, and the spectrum of the small fluctuations around these background solutions 
can be interpreted in terms of the discrete mass spectrum of bound states of the dual field theory.

When we analyse the region with $\hat\phi_2>0$, we find the existence of a critical choice $\hat\phi_2^c$
 for which a phase transition takes place, with the physically realised background minimising
the free energy density being given by confining solutions when $\hat\phi_2 < \hat\phi_2^c$,
and singular domain wall solutions for $\hat\phi_2 > \hat\phi_2^c$.
Interestingly, while the spontaneous compactification of one of the space-time dimensions of the theory
is energetically favoured in the confined phase, beyond the critical point the theory prefers to preserve (locally)
the full five-dimensional Poincar\'e invariance.
The critical parameters at the transition are extracted from the numerical study, and
we find:
\beqs
\hat\phi_2^c &\simeq & 0.169\,,\\
\phi_I^c&\simeq& 0.027 \,,\\
\phi_4^c &\simeq&  98.9\,,\\
{\hat{\cal F}}^c &\simeq& -3.893\,.
\eeqs
We also find that the UV parameters in the gravity analysis show a sharp discontinuity in the values assumed in the phase with a shrinking circle (denoted by the subscript $_<$) and in the domain-wall phase (denoted by the subscript $_>$):
\beqs
\hat\phi_{3\,<}^c &\simeq & -0.092\,,~~~~~~~~\hat\phi_{3\,>}^c \,\simeq \,43.2 \,,\\
\hat\chi_{5\,<}^c &\simeq & -3.12\,,~~~~~~~~\hat\chi_{5\,>}^c \,\simeq \,-1.17 \,.
\eeqs
In particular, we notice the  enhancement of  $\hat\phi_{3\,>}^c$.

\subsection{Properties of the phase transition}
\label{Sec:phasetransition}

Having established the existence of a first order phase transition, we devote this subsection to characterising it. We also return to its relation with the physical spectrum of the bound states of the dual theory along the confining branch.

As repeatedly stated, two non-trivial operators are present in the dual field theory. We identify the source for the operator of dimension $\Delta = 3$ with the leading-order coefficient $\phi_2$ in the UV expansion exhibited at the beginning of Section~\ref{Sec:Solutions}. We can express this statement by adopting the following definition:
 \beqs
 \phi_2&\equiv&
 \lim_{\rho_2\rightarrow +\infty} e^{2A-2\chi}\phi(\r_2)\,,
 \label{Eq:phi2}
 \eeqs
which is manifestly consistent with the UV expansion. In the study we performed of the free energy density ${\cal F}$ we kept the source of the other non-trivial operator fixed (we set $A_U=0=\chi_U$), and studied how ${\cal F}$ varies as a function of the source $\phi_2$. Moreover, in order to facillitate the comparison of different branches, we implemented a scale-setting procedure by defining the energy scale $\Lambda$, allowing us to compare dimensionless quantities.

We now define two dynamical quantities that play a role similar to that of order parameters, and study them as we cross from one side to the other of the phase transition. In analogy with the magnetization of a system in thermodynamics, the first such parameter is defined as the variation of the free energy density with respect to the source $\phi_2$ (holding $A_U=0=\chi_U$ and $\Lambda$ fixed) measured in units of $\Lambda$:
\beq
	\hat{\mathcal M} \equiv \Lambda^{-3} \frac{\partial}{\partial \phi_2} \mathcal F(\phi_2,\Lambda) = \frac{\partial}{\partial \hat\phi_2} \hat{\mathcal F}(\hat \phi_2) \,.
\eeq
We cannot write this in closed form, as it requires expressing explicitly the coefficients $\hat\phi_3(\hat\phi_2)$ and $\hat\chi_5(\hat\phi_2)$, appearing in the expression for the free energy density, in terms of $\hat\phi_2$. But we can evaluate the derivative numerically. From Figs.~\ref{Fig:Plot} and~\ref{Fig:AnotherPlot}, we see that $\hat{\mathcal M}$ is a well-defined quantity for the confining, skewed, IR-conformal, and singular domain-wall solutions. In the more general singular solutions (represented by the thin blue and lighter green lines in Figs.~\ref{Fig:Plot} and~\ref{Fig:AnotherPlot}), an additional parameter remains undetermined in terms of $\hat \phi_2$ (see Table~\ref{Fig:summarytable}), and therefore the variation with respect to $\hat \phi_2$ is ambiguous.

\begin{figure}[t]
\includegraphics[width=16cm]{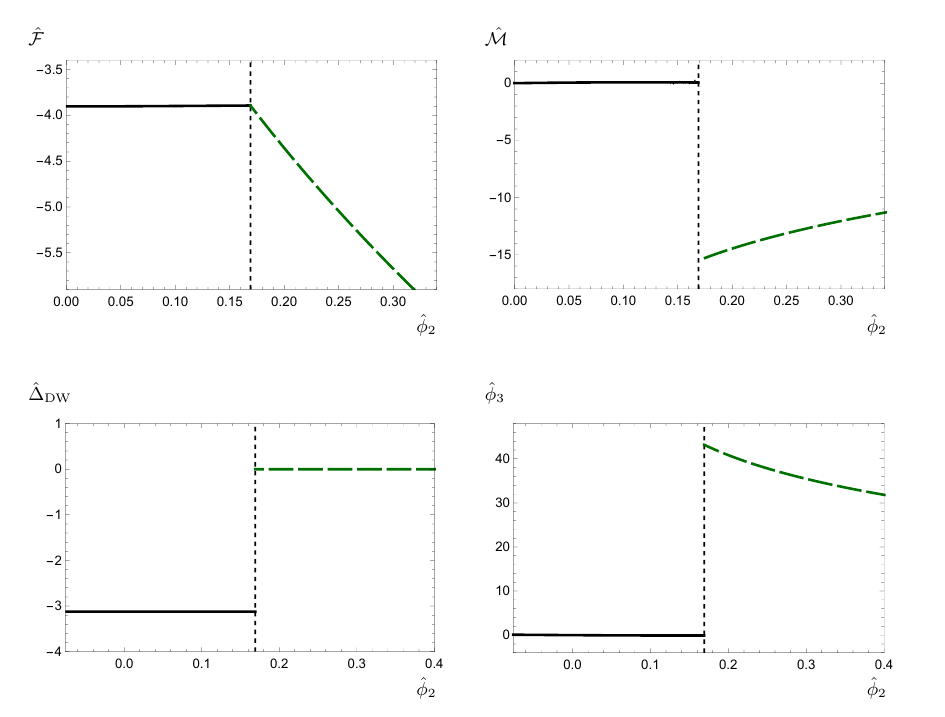}
\caption{The free energy density ${\hat{\cal F}}$ (top-left) and its derivative $\hat{\mathcal M} \equiv \frac{\partial {\hat{\cal F}}}{\partial \hat\phi_2}$ (top-right), as a function of the deformation parameter $\hat\phi_2$, for solutions within the confining (solid black), and singular domain-wall (long-dashed dark-green) classes. The bottom panels show the order parameter $\hat \Delta_{\rm DW}$ (bottom-left) and $\hat \phi_3$ (bottom-right), for the same solutions.}
\label{Fig:discontinuity}
\end{figure}

The second parameter that we define measures how much  Poincar\'e invariance in
$D=5$ dimensions is broken, and is given by
\beq
\label{Eq:DWparameter}
	\hat \Delta_{\rm DW} \equiv \hat \chi_5 + \frac{4}{25} \hat\phi_2 \hat\phi_3 \,.
\eeq
As can be seen from the leading-order parameter appearing in the UV-expansion of the combination $d=A-4\chi$ in Eq.~(\ref{Eq:dd}), $\hat \Delta_{\rm DW}$ vanishes for the domain-wall background solutions, for which $d=0$.

In Fig.~\ref{Fig:discontinuity} we show a detail of the functions $\hat{\cal F}$ and $\hat{\mathcal M} \equiv \frac{\partial {\hat{\cal F}}}{\partial \hat\phi_2}$ in the vicinity of the phase transition. The derivative has been evaluated numerically. The plots show clear evidence of a strong first-order phase transition: while the free energy is continuous, its derivative with respect to the deformation parameter is not. The two bottom panels of Fig.~\ref{Fig:discontinuity} show that  in the physical phase in which the confining solutions are realised, the order parameter $\hat \Delta_{\rm DW}$ is large, while $\hat\phi_3$ is negligible, and vice versa $\hat\phi_3$ is large along the singular domain-wall solutions for which $\hat \Delta_{\rm DW}=0$.

\begin{figure}[t]
\begin{center}
\includegraphics[width=10cm]{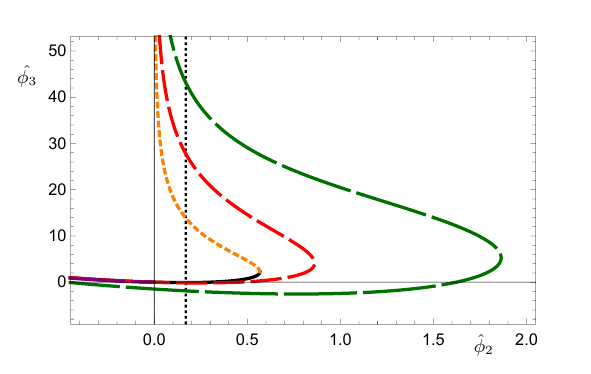}
\caption{The UV parameter ${\hat\phi_3}$ as a function of the deformation parameter $\hat\phi_2$, for the confining (solid black and short-dashed orange), skewed (dashed red), IR-conformal (longest-dashed purple), and singular domain-wall (long-dashed dark-green) classes of background solutions.}
\label{Fig:YetAnotherPlot2}
\end{center}
\end{figure}

Along the branch of confining solutions, the dynamics captured by the gravity theory favours the shrinking  to zero size of the compact dimension  spanned by $\eta$, which in field theory terms corresponds to confinement of the dimensionally-reduced dual theory. Conversely, along the branch of singular domain-wall solutions, the theory is preserving (locally) the higher-dimensional Poincar\'e invariance, with the formation of a condensate for the dimension-3 operator $\mathcal O_3$ associated with $\phi$, whose magnitude is related to the coefficient $\phi_3$ of the subleading term in the UV expansion of $\phi$. In Fig.~\ref{Fig:YetAnotherPlot2}, we show $\hat\phi_3$ as a function of $\hat\phi_2$ for a few of the branches of solutions. The confining, skewed, and singular domain-wall branches all share the feature that $\hat \phi_3$ diverges as $\hat \phi_2 \rightarrow 0$. This reflects the fact that in this limit, they all approach the solution we called SUSY, in which  both $\phi_2$ and $\chi_5 $ vanish, but the combination $\hat \phi_3 = \phi_3\Lambda^{-3}$ diverges. The regions in parameter space for which $\hat \phi_3$ diverges are never energetically favoured. Moreover, while $\hat\phi_3 \gg 1$ on the singular domain-wall branch close to the phase transition, the singular nature of this class of solutions makes a field theory interpretation problematic, and it is unknown whether this feature would remain in a more complete treatment of the gravity description.

We can now return to the discussion of the spectrum of bound states along the branch of confining solutions,
that we started in Section~\ref{Sec:Probe}.
The behaviour of $\phi_3$ and $\chi_5$
is related to the nature of the approximate dilaton state.
In particular, the region of parameter space in which $\phi_3$ is large compared to the 
 dynamical scale $\Lambda$ of the theory
is the region of large and positive $\phi_I$,
for which we see in the spectrum the appearance first of a parametrically light (approximate) dilaton state that eventually becomes tachyonic. This region of parameter space is not physically realised, as the confining solutions are energetically disfavoured, compared to the singular domain wall solutions. Some of the metastable configurations leading to a very light dilaton might be long lived, but exploring this possibility would require a detailed study of the bubble rate of the phase transition, which goes far beyond our current purposes (see Ref.~\cite{Bigazzi:2020phm} for a recent study in this direction). Nevertheless, it is reassuring to notice how the (failure of the) probe approximation captures correctly the existence of a region of parameter space in which the condensate $\langle \mathcal O_3\rangle$ is parametrically enhanced.
 
Some degree of complication in interpreting the spectrum of scalar bound states along the physical, confining branch
of solutions arises because of the interplay between the two possible operators developing 
vacuum expectation values (VEVs).
This is particularly subtle in the region with positive, large values of $\phi_I$. 
As can be seen in Figs.~\ref{Fig:phi2}, \ref{Fig:phi3},  \ref{Fig:chi5}, and \ref{Fig:chi5phi2(shift)} in Appendix~\ref{Sec:ParametricPlots}, by following the solid black and short-dashed orange lines, in the limit 
in which $\phi_I\rightarrow +\infty$ one ultimately drives towards suppressing
 the explicit symmetry breaking parameter 
$\hat \phi_2 \rightarrow 0$. One of the condensates vanishes in this unphysical limit, 
as  $\hat \chi_5\rightarrow 0$, 
yet  scale invariance is broken spontaneously by the divergence of the other condensate, signalled by the fact that $\hat\phi_3\rightarrow +\infty$. Albeit unphysical (because of the tachyon, and of the presence of a phase transition) the analysis of this region of parameter space is quite interesting as a way to test our theoretical tools. The reason why the mass of the lightest scalar fluctuation in the system shows significant discrepancy with the probe approximation is the emergence of this divergently large condensate. At finite, small values of $\phi_I$, the effects of explicit symmetry breaking are not small, and the mixing effects between the two scalar particles sourced by both the operators developing VEVs are not negligible either.

We finally notice that the critical value of $\phi_I^c$ that sets the upper limit to the reach of the field-theory interpretation of the confining solutions is comparatively small with respect to the value at which the tachyon emerges. By examining Fig.~\ref{Fig:FullScalar}, one sees that in immediate proximity of this value of $\phi_I$  the lightest scalar is not a dilaton, and neither is it appreciably much lighter than the other states of the system. The next-to-lightest state, though, shows significant discrepancy with the probe approximation, and it should be interpreted as an approximate (not so light) dilaton, which exists because $\hat \Delta_{\rm DW} \neq 0$, signalling the presence of a condensate.

Furthermore, our estimate of $\hat{\phi}_2^c$ is likely an overestimate: the domain-wall, badly singular solutions cannot be
the ones realised physically, and in a more complete gravity theory other solutions must take over the dynamics beyond a new critical point  $\hat{\phi}_2^{cc}\leq\hat{\phi}_2^c$.
Potentially, this might happen at $\hat{\phi}_2^{cc}=0$ (see Section~\ref{Sec:freeagain} for a complementary discussion).
This might make the phase transition even stronger, but we do not have the quantitative elements to support this suggestion.

We must close this discussion by repeating the observation that two pathologies are still present: we could not find any solutions, neither regular nor singular, corresponding to arbitrarily large values of $\hat \phi_2$,
and furthermore the phase transition we identified seems to indicate that the energetically favoured solutions 
for $\hat\phi_2>\hat\phi_2^c$ are singular, and hence do not admit a sensible physical interpretation in terms of dual field theory quantities. Our interpretation of these results is that there is an upper bound to the choice of $\hat\phi_2<\hat\phi_2^c$ for which the gravity description at our disposal admits a holographic field theory interpretation. The other phase exists only as a phase of the gravity theory, regulated by putting boundaries 
$\r_1<\r<\r_2$ on the radial (holographic) direction. This unusual feature resembles what happens in the presence of bulk phase transitions  in the study of lattice field theories. We will explore this observation further in Section~\ref{Sec:freeagain}.

We are forced to conclude that large (positive) deformations of the field theory due to the dimension-3 operator
$\mathcal O_3$ dual to the scalar $\phi$ cannot be captured by this gravity model. Whether or not extensions of the gravity theory can overcome this limitation is unknown: given that Romans supergravity does not contain other scalar fields, such extensions either might involve allowing for non-trivial behaviours of the fields removed by the reduction on $S^4$ of massive type-IIA, or might require the inclusion of extended objects that are not captured by the supergravity approximation. We leave this challenging problem open to future exploration.

\subsection{An alternative approach to  the free energy density}
\label{Sec:freeagain}

In the previous subsections, we introduced appropriate regulators $\r_1$ and $\r_2$, as well as  a suitably defined set of boundary-localised terms, chosen according to a prescription that allows one to remove all divergences and to compare to one another the free energy density ${\cal F}$ of different, independent background configurations. In particular, we derived Eq.~(\ref{Eq:Fquasifinal}), which we reproduce here for convenience
\beqs
\cF&=-\lim_{\r_2\rightarrow +\infty}e^{4A-\chi}\Big(\frac{3}{2}\partial_{\r}A+\mathcal{W}_{2}\Big)\Big|_{\r_2}\,.\nonumber
\eeqs
For the same purpose, we repeat the definition of the scale $\Lambda$, taken from Eq.~(\ref{Eq:Scale}):
\begin{equation}
\Lambda^{-1}\equiv\lim_{\r_2\rightarrow +\infty}\int_{\r_{o}}^{\rho_2}\text{d}\tilde{\r}\,e^{\c(\tilde{\r})-A(\tilde{\r})}\,.
\nonumber
\end{equation}
We also studied the energetics as a function of the leading-order coefficient $\phi_2$ in the UV
 expansion exhibited at the beginning of Section~\ref{Sec:Solutions}, and that we can write 
 by copying Eq.~(\ref{Eq:phi2}):
 \beqs
 \phi_2&\equiv&
 \lim_{\rho_2\rightarrow +\infty} e^{2A-2\chi}\phi(\r_2)\,.\nonumber
 \eeqs
 
 The strategy we followed in the previous subsections consisted of first taking the $\rho_2\rightarrow +\infty$ 
 limit in these three expressions, and then studying the resulting phase structure for the theory. There is another possible way to perform this study, and we explore it in this section. We can first hold fixed $\Delta\rho \equiv \rho_2 - \rho_o$ and study the phase structure encoded in the dependence of $\hat{\cal F}\equiv {\cal F} \Lambda^{-5}$ on $\hat{\phi_2}\equiv\phi_2 \Lambda^{-2}$, and only afterwards take the limit $\rho_2\rightarrow +\infty$ by looking at how the phase structure evolves in the limit in which the boundary of the gravity theory is removed. We hence introduce the following quantities:
\beqs
 \tilde{\cF}(\rho_2)&\equiv&-e^{4A-\chi}\Big(\frac{3}{2}\partial_{\r}A+\mathcal{W}_{2}^f \Big)\Big|_{\r_2} \,, \hspace{0.5cm}
 {\mathcal W}_2^f \equiv -\frac{4}{3}-\frac{4}{3}\phi^2 \,, \\
 \tilde{\Lambda}^{-1}(\rho_2)&\equiv&\int_{\r_{o}}^{\rho_2}\text{d}\tilde{\r}\,e^{\c(\tilde{\r})-A(\tilde{\r})} \,, \hspace{0.5cm}
 \tilde{\phi}_2(\rho_2)\equiv
 e^{2A-2\chi}\phi(\r_2) \,,
\eeqs
which are the finite-$\r_2$ analogues of their  infinite-$\r_2$ limits. We will study them at finite $\Delta\rho$, perform the minimisation of  $\tilde{\cal F} \tilde{\Lambda}^{-5}$, and identify possible phase transitions, and only afterwards take $\rho_2\rightarrow +\infty$. Notice that in defining ${\mathcal W}_2^f$ we chose to retain only the terms of ${\mathcal W}_2$ that give divergent and finite order contributions to the free energy in the $\rho_2 \rightarrow +\infty$ limit; this corresponds to a particular choice of subtraction scheme.

\begin{figure}[t]
\includegraphics[width=15cm]{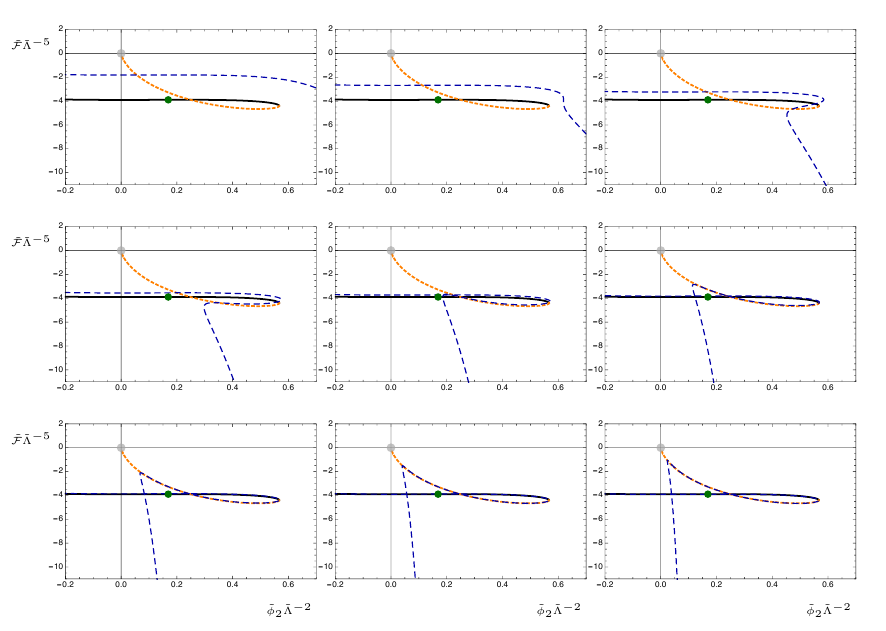}
\caption{The regulated free energy $\tilde{\cal F}\tilde{\Lambda}^{-5}$ of the confining solutions,
	 as a function of the regulated
	deformation  $\tilde{\phi}_2\tilde{\Lambda}^{-2}$ (dashed blue lines), for various choices of $\Delta \r=3,4,5,6,7,8,9,10,11$ (top to bottom, left to right). The solid black and short-dashed orange lines depict the renormalised free energy 
	$\hat{\cal F}$ in Figs.~\ref{Fig:Plot}, \ref{Fig:AnotherPlot}, and~\ref{Fig:YetAnotherPlot}. The green dot marks the location of crossing between the branch of regular solutions and the branch of singular domain wall solutions. The grey dot denotes the SUSY solutions.}
\label{Fig:AppendixF}
\end{figure}

This alternative approach is closely related to what is normally done on the lattice, where one first performs a rough scan of the lattice parameter space, to identify possible artificial phase transitions of the lattice theory, and then restricts the lattice studies to the region connected with the field theory, before taking the continuum limit, in this way avoiding completely unphysical regions of parameter space. In this section we perform this study, restricting our attention to the confining solutions. We will show that the procedure yields the same results as those discussed in the bulk of the paper, in the physical region. The existence of spurious phase-transitions in the gravity side of the gauge-gravity correspondence has been observed before, though in a different context, dealing with the treatment in the probe approximation of extended object embedded in curved backgorunds~\cite{Faedo:2014naa}. We stress that the phase transition is not a feature of the field theory, but rather of the regulated gravity dual (although it should be possible to interpret our results in terms of a finite cutoff in the field theory).

The results of this analysis are shown in Fig.~\ref{Fig:AppendixF}. We display the results for the confining solutions only, by comparing the regulated free energy $\tilde{\cal F}\tilde{\Lambda}^{-5}$, for various choices of $ 3 \leq \Delta \rho \leq 11$, to the result of the renormalised analysis. For $\Delta \rho \gsim 5$ the signature of a phase-transition appears, and moreover there is no maximum allowed value of $\tilde{\phi}_2\tilde{\Lambda}^{-2}$.

The branch that takes over the dynamics at large $\Delta \rho$, at least for large positive values of
 $\tilde{\phi}_2\tilde{\Lambda}^{-2}$, has no genuine field theory dual interpretation, 
 as it exists only when we retain the finite  UV cutoff $\rho_2$
 when minimising the free energy.
 We notice that this rather rough analysis seems to suggest that the phase transition takes place at
  smaller  (positive) values of 
 the deformation parameter $\tilde \phi_2 \tilde \Lambda^{-2}$, when compared with the analysis conducted in the bulk of the  paper.
 We also notice that the comparison is not rigorous, as it is affected by the presence of arbitrary 
 scheme dependences.
 This dependence on the order of limits, on the scheme, and the fact that the energetics of the dominant solution
 is dominated by spurious cut-off effects, are typical of what in the lattice literature are called 
 bulk phase transitions.

\section{Summary}
\label{Sec:Summary}

We presented a first realisation, within top-down holography, of one particular strategy for building a dilaton scenario, which is inspired by the ideas in Refs.~\cite{Kaplan:2009kr} and~\cite{Pomarol:2019aae}. In this scenario, a parametrically light dilaton would emerge as a light scalar particle for choices of the parameters that bring the theory in close proximity of a dynamical instability (and in the presence of enhanced condensates). However, we also found direct evidence of a phase transition, effectively preventing the dynamics from approaching arbitrarily close to the aforementioned instability, and hence none of the scalar particles can be made arbitrarily light along the physical branch of solutions. Nevertheless, the lightest particle can be dialled to have arbitrarily small mass along the metastable solutions of the same branch.

This approach represents an appealing, alternative search strategy for dynamical realisations of the dilaton, in contrast to starting by establishing first the existence  of a moduli space in the field theory.\footnote{Top-down holographic realisations of the latter approach already exist, though for limited and quite non-trivial systems, for example
along the baryonic branch of the Klebanov-Strassler 
system~\cite{Elander:2017cle,Elander:2017hyr}.}
This study complements the work done by other authors,
either guided by considerations emerging from lower-dimensional statistical-mechanics 
systems~\cite{Gorbenko:2018dtm,Gorbenko:2018ncu}
or by holographic models built within the bottom-up approach to holography~\cite{Pomarol:2019aae}.
The primary difference is that we proposed and studied a calculable model built within 
top-down holography. 

The example we considered is the
six-dimensional half-maximal supergravity written by Romans~\cite{Romans:1985tw}, dimensionally reduced on a circle.
The lift of the solutions to $D = 10$ massive Type-IIA is known~\cite{Romans:1985tz,Brandhuber:1999np,Cvetic:1999un}
(alternative lifts in Type IIB exist as well~\cite{Hong:2018amk,Jeong:2013jfc}).
The equations of motion admit a special solution with  AdS$_6$ geometry and trivial $\phi=0$. 
This solution can be interpreted as the dual of a strongly coupled fixed point
in the large-$N$ limit of a class of supersymmetric
field theories in $D=5$ dimensions that has been studied extensively in the literature~\cite{Seiberg:1996bd,Morrison:1996xf,Intriligator:1997pq,Douglas:1996xp,Ganor:1996pc} 
(see also Refs.~\cite{Aharony:1997ju,Aharony:1997bh,DeWolfe:1999hj,BenettiGenolini:2019zth} and 
references therein, and the discussion in Ref.~\cite{Brandhuber:1999np}).

The scalar $\phi$ in the gravity theory corresponds
to an operator of dimension $\Delta=3$  in the dual five-dimensional theory. 
Its coupling and condensate are related to the coefficients
$\phi_2$ and $\phi_3$ in the asymptotic expansion of background gravity solutions.
It is known that by tuning $\phi_2$ and $\phi_3$ one can build the gravity dual of 
the field-theory renormalization group flow towards what can be interpreted as a second, non-supersymmetric,
perturbatively stable fixed point~\cite{Gursoy:2002tx}---although it 
is not known that this fixed point exists in the dual field theory.

The field theory admits compactification of one spatial direction of the 
geometry on a circle, hence  breaking five-dimensional
Poincar\'e invariance. The size of the circle in the gravity theory is a function of $\rho$, controlled by the field $\c$, and in particular by the coefficient $\c_5$ appearing in its asymptotic (UV) expansion.
When the circle shrinks, the resulting strongly-coupled four-dimensional dual theory confines.
The gravity description hence provides a comparatively simple description of 
confinement in four dimensions, along the lines  suggested by Witten~\cite{Witten:1998zw},
but in a somewhat simpler environment~\cite{Wen:2004qh,Kuperstein:2004yf}.

The simultaneous combination of these two deformations had been studied so far
only for values of $\phi \leq 0$~\cite{Elander:2013jqa,Elander:2018aub}.
In this paper, we extended our study by first  allowing $\phi>0$, and secondly by 
complementing the calculation of the spectrum with the study of the free energy density ${\cal F}$.
Furthermore, we considered several new general classes of background gravity solutions,
all of which approach the aforementioned AdS$_6$ 
geometry for large values of the radial direction $\r$.
Some are regular, and are the 
main subject of our attention, some have a good singularity, in the sense defined by Gubser~\cite{Gubser:2000nd},
and some have a bad singularity.
\begin{itemize}
	\item We called {\bf SUSY} the solutions that satisfy the  first-order equations for the system in $D=6$ dimensions.
	These solutions are supersymmetric, exhibit a good singularity, and
	preserve five-dimensional Poincar\'e invariance---in the gravity 
	language this last property corresponds to the constraint $d=A-4\chi=0$, with $A$ the warp factor in the metric, 
	as discussed 
	in the main body of the paper.
	\item The {\bf IR-conformal} solutions
	correspond to the aforementioned flows between the two fixed points. They preserve
	five-dimensional  Poincar\'e invariance, but break supersymmetry. These solutions are regular.
	\item With some abuse of language, we called {\bf confining} solutions the regular ones in which 
	Poincar\'e invariance is reduced to four dimensions, and in which the compact circle shrinks smoothly to zero size
	at a finite value of the radial direction $\r \rightarrow \r_o$. The holographic interpretation 
	of such backgrounds involves both the compactification of the dual five-dimensional theory on a circle, and then 
	linear confinement of the resulting dimensionally-reduced four-dimensional strongly coupled theory.
	\item A related class of gravity solutions can be obtained 
	from the confining ones by  changing the sign of the function $d=A-4\chi$.
	These solutions have the same symmetries as the confining ones, but the size of the circle diverges  
	for $\r\rightarrow \r_o$, and as a result the geometry has a (good) naked singularity. We called these
	solutions {\bf skewed}.
	\item We also included in our survey three other classes of singular solutions. We found that they can either result in {\bf good singular solutions} or in  {\bf bad singular solutions}.
	(The constraint $A=4\chi$ yields the subclass of {\bf singular domain-wall solutions}.)
	While not representative of dual field theory configurations, we found that 
	the badly singular domain-wall solutions play an important role in the energetics of the gravity theory.
\end{itemize}

We summarised in Table~\ref{Fig:summarytable} all these classes of solutions,
and how we chose to parametrise them. We introduced a scale setting procedure via the function $\Lambda$ defined in Eq.~(\ref{Eq:Scale}), and showed the dimensionality of the resulting space of solutions.
We plotted the free energy in Figs.~\ref{Fig:Plot} and~\ref{Fig:AnotherPlot}.

Our first new finding is that the regular, confining solutions exist also for positive values of $\phi>0$.
We hence extended the one-parameter family of solutions studied in 
earlier publications~\cite{Elander:2013jqa,Elander:2018aub}.
We computed the spectrum of fluctuations of all
the 32 bosonic degrees of freedom of the five-dimensional
theory obtained by dimensional reduction on the circle.
Our second new result is the mass spectrum, that can be seen in Figs.~\ref{Fig:Spectra1}--\ref{Fig:FullScalarZoom}.

The salient feature of the mass spectrum is what brings this work in contact with the 
line of arguments in Refs.~\cite{Kaplan:2009kr,Gorbenko:2018dtm,Gorbenko:2018ncu,Pomarol:2019aae}.
While  the confining solutions are regular, 
by moving along the one-parameter class  labelled by $\phi_I$,
the mass squared of the lightest scalar glueball becomes progressively smaller (in units of the mass of the tensor,
which we use to set the scale in the spectrum), until it becomes tachyonic at some finite,
positive value of $\phi_I$.
This instability is our third new result.
The reason why this is interesting is that, if interpreted naively, this system would 
yield an example of  a theory in which by tuning the parameter  $\phi_I$ one could dynamically 
produce a hierarchy of scales between the mass of the lightest scalar particle and the rest
of the spectrum.  By making use of the probe approximation (as suggested in Ref.~\cite{Elander:2020csd}),
we also showed that in the region of parameter space in which the lightest scalar has a parametrically suppressed mass---in proximity to the region in which a  tachyon emerges---
the associated particle is indeed an approximate dilaton (see Fig.~\ref{Fig:FullScalarZoom}),
which is our fourth original result. In connection with this, we also noted the divergent behaviour of the parameter $\hat \phi_3$ in the limit of $\phi_I\rightarrow +\infty$.

Unfortunately though, the naive interpretation contained in the previous paragraph has to be used with caution.
To show why, we studied the energetics of the classical solutions, and found another additional result.
The tachyonic instability appears for values of the deforming parameter $\hat \phi_2$ for 
which the solution has  free energy ${\hat{\cal F}}$ that is higher than that of other solutions.
This is the typical feature expected in the presence of
a first-order phase transition.
It is hence not possible to dial the boundary parameter to approach arbitrarily close to the massless case, as this would require exploring a branch of metastable and unstable solutions, well past a phase transition.

We could only identify two branches of solutions within the confining class, by varying $\phi_2$.
Furthermore, a maximum value 
of the parameter $\hat \phi_2$ emerged, further confirming the incompleteness of the energetics discussion 
when restricted to the confining solutions only.
The picture became more clear once we included in the discussion also singular solutions.
For arbitrarily large $\hat\phi_2>0$, we could not find a ground state solution (within the restrictions defining our ansatz for the background metric)---free of gravity singularities---that admits a trustable field theory interpretation. Yet, for values of $\hat\phi_2>{\hat\phi}_2^c$, we showed that there exist singular solutions with free energy lower than that of the regular, confining solutions. Conversely, for negative $\hat\phi_2<0$, the singular solutions have free energy higher than the confining ones. Hence, the phase transition takes place at $\hat\phi_2={\hat\phi}_2^{cc}$ (with $0\leq \hat\phi_2^{cc} \leq \hat\phi_2^c$), and all the solutions with $\hat\phi_2 > {\hat\phi}_2^{cc}$ along the confining branch are either metastable or unstable.
In particular, there is not a parametrically  light dilaton near the transition: although the lightest
bound state is a scalar, and its mass is slightly smaller than in other regions of the 
physical portion of parameter space, it does not show the properties expected by an approximate dilaton,
and its mass is not parametrically, nor numerically, small. The next-to-lightest state, though, is at least approximately a dilaton, but it is heavier, and its mass does not show any special features in the region of parameter space immediately adjacent to the phase transition.

We repeat again that the phase transition we find is not a field theory feature, but rather it exists only in the gravity theory. As discussed in Section~\ref{Sec:freeagain}, this is not contradictory, as gauge/gravity dualities relate only physical objects in the physically related phase of the theory, and the gravity theory (with finite radial direction $\r_1 < \r < \r_2$) may exhibit a more general phase structure. Nevertheless, it is interesting to notice how the physical properties of the bound states in the region of parameter space that admits a field-theory interpretation are influenced by the phenomena taking place past the phase transition.

\section{Conclusion and outlook}
\label{Sec:Conclusions}

Along a new branch of regular solutions of Romans supergravity,
we found a tachyonic  instability  by studying the mass spectrum
of the fluctuations of the sigma-model coupled to gravity.
By approaching this instability in the space of parameters,
we found that the lightest scalar state  in the spectrum turns into
a tunably light approximate dilaton, which could be realised in a metastable configuration of the system. A condensate is enhanced when moving along this branch of solutions, spontaneously breaking (approximate) scale invariance.
But we also found that the instability
is hidden away by a strong first-order phase transition, so that the lightest scalar
state along the stable phase
is not parametrically light, and  it is the next-to-lightest scalar state that behaves as an approximate dilaton
(in association with an enhancement of one of the condensates).
We hence uncovered a concrete realisation within top-down holography of arguments
closely resembling those of Ref.~\cite{Pomarol:2019aae,Gorbenko:2018ncu,Gorbenko:2018dtm},
although in a generalised form.

Our study admits a clear (though not simple) interpretation, and our action is taken from the established catalogue of rigorously  defined supergravity theories. We also tested the formal tools that would be needed to perform this type of analysis in other supergravity theories. This paper establishes the basis for the development of a systematic future research programme, encompassing the exploration of the vast catalogue of known supergravity theories---possibly encompassing the technically more challenging 
cases in which one does not recover an AdS geometry asymptotically far in the UV.

While we found a strong first-order phase transition, there may be other models realising this mechanism, and it is not known a priori how strong the first-order phase transitions should be in general. They might be very weak. There are well known examples in physics of systems in which first-order phase transitions sit along critical lines (in parameter space) that have an end point.
If one could identify a supergravity theory 
realising this type of critical behaviour, then it would be interesting to 
repeat our analysis in more detail within such a system.
A direct calculation could  establish whether the phase transition takes place in the proximity of
the end point of the critical line. We might find that the whole spectrum scales without producing a hierarchy, 
and hence asymptotically reproduces the scaling behaviours expected in the presence of
explicitly broken scale invariance. Conversely, one might discover that the spectrum still behaves as in Figs.~\ref{Fig:FullScalar} 
and~\ref{Fig:FullScalarZoom} and a light dilaton emerges.
If so, its existence would be connected to the enhancement of non-trivial condensates in the vacuum, which can be checked explicitly. This possibility, if realised, would have important theoretical and phenomenological implications, and hence motivates us to further pursue our programme in the future.

\begin{acknowledgments}
We thank A.~Pomarol for useful discussions, and C.~N\'u\~nez and D.C.~Thompson for comments on an earlier version of the manuscript. The work of MP has been supported in part by the STFC Consolidated Grants ST/L000369/1 and ST/P00055X/1. MP has also received funding from the European Research Council (ERC) under the European Union's Horizon 2020 research and innovation programme under grant agreement No 813942. JR has been supported by STFC, through the studentship ST/R505158/1. DE was supported by the OCEVU Labex (ANR-11-LABX-0060) and the A*MIDEX project (ANR-11-IDEX-0001-02) funded by the ``Investissements d'Aveni'' French government program managed by the ANR.
\end{acknowledgments}

\appendix

\section{A few gravitational (curvature) invariants}
\label{Sec:CurvatureInvariants}

In this appendix we find it useful to present and discuss the results for some of the 
curvature invariants of the theory in $D=6$ dimensions---the Ricci scalar $R = \mathcal R_6$, the
 Ricci tensor squared $R^2_2\equiv R_{\hat{M}\hat{N}}R^{\hat{M}\hat{N}}$, 
and the Riemann tensor squared  $R^2_{4}\equiv R_{\hat{M}\hat{N}\hat{R}\hat{S}}R^{\hat{M}\hat{N}\hat{R}\hat{S}}$.
We adopt the six-dimensional metric ansatz in
Eq.~\eqref{Eq:6Dmetric}. After using the equations of motion presented in Section~\ref{Sec:EOM}, we find that
\beqs
\label{eq:curvatureinvariants}
R &=& 6\mathcal{V}_{6}+4\big(\partial_{\r}\f\big)^{2} \,, \nonumber \\
R^{2}_{2} &=& 6\mathcal{V}_{6}^{2}+8\mathcal{V}_{6}\big(\partial_{\rho}\phi\big)^2+16\big(\partial_{\r}\f\big)^{4} \,, \\
R^{2}_{4}
&=&\frac{1}{250} \Bigg(32(\partial_{\r}d)^2 \Big(4\partial_{\r}d \sqrt{36(\partial_{\r}d)^2+15\sqrt{5}
	\sqrt{6 R^{2}_{ 2}-R^{2}}-30
	R}+24(\partial_{\r}d)^2 \nonumber \\ \nonumber
&&\hspace{30mm}+5 \sqrt{5} \sqrt{6 R^{2}_{ 2}-R^{2}}-10 R\Big)-25
\left(R^{2}-10 R^{2}_{ 2}\right)\Bigg) \,,
\eeqs
where in deriving the expression for $R_4^2$, we made use of the fact that for our solutions $\partial_\rho c > 0$ (see Eq.~\eqref{eq:RGinequality}).

\begin{figure}[t]
\begin{center}
\includegraphics[width=15.5cm]{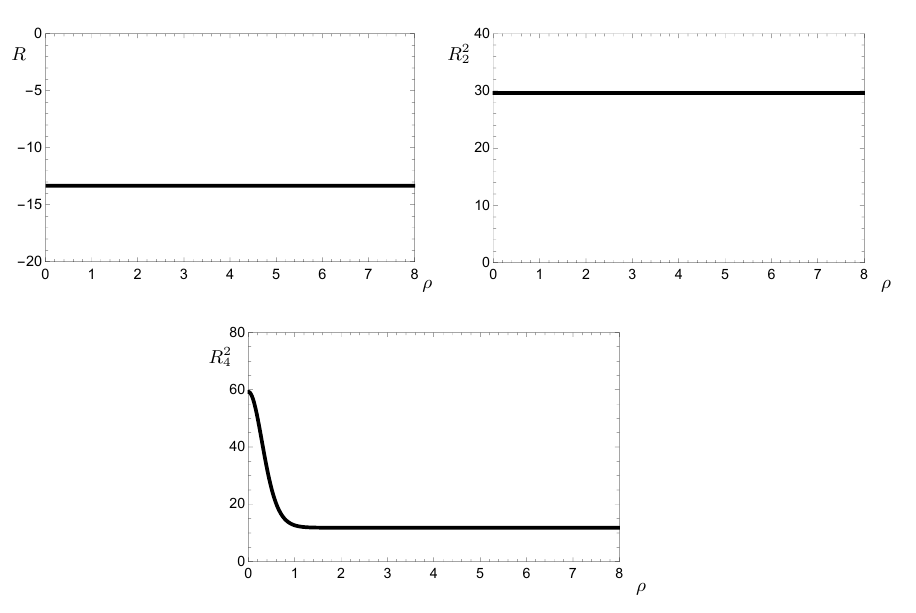}
\caption{Gravitational invariants computed using the analytic confining solutions with $\phi=0$, shown in Eqs.~(\ref{Eq:ConfChi}, \ref{Eq:ConfA}). 
	From top to bottom, left to right: $R\equiv g^{\hat{M}\hat{N}}R_{\hat{M}\hat{N}}$, $R_2^2\equiv 
	R_{\hat{M}\hat{N}}R^{\hat{M}\hat{N}}$, and $R_4^2\equiv
	R_{\hat{M}\hat{N}\hat{R}\hat{S}}R^{\hat{M}\hat{N}\hat{R}\hat{S}}$. }
\label{Fig:ConfInvariants}
\includegraphics[width=15.5cm]{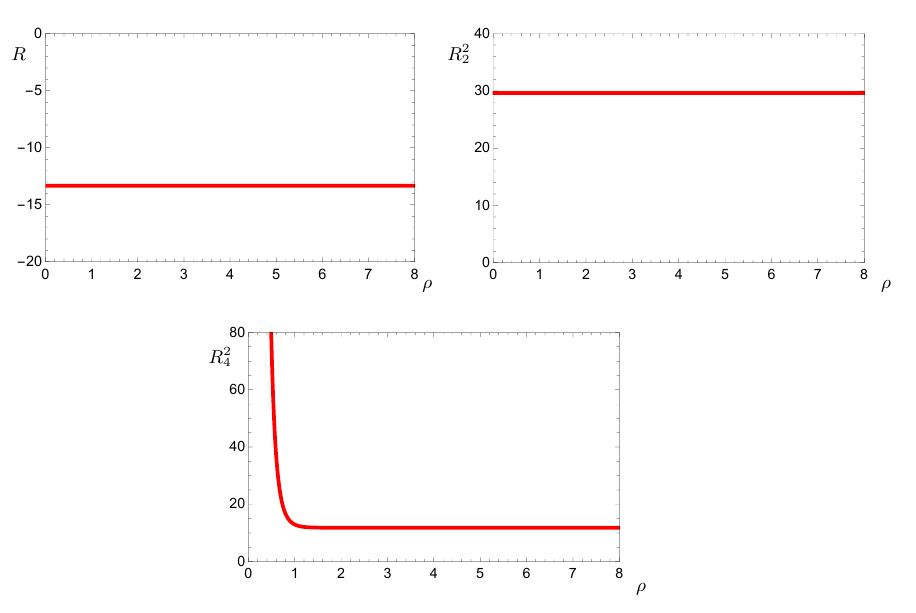}
\caption{Gravitational invariants computed using the analytic skewed solutions with $\phi=0$, shown in Eqs.~(\ref{Eq:ConfChi}, \ref{Eq:ConfA}). 
	From top to bottom, left to right: $R\equiv g^{\hat{M}\hat{N}}R_{\hat{M}\hat{N}}$, $R_2^2\equiv 
	R_{\hat{M}\hat{N}}R^{\hat{M}\hat{N}}$, and $R_4^2\equiv
	R_{\hat{M}\hat{N}\hat{R}\hat{S}}R^{\hat{M}\hat{N}\hat{R}\hat{S}}$. }
\label{Fig:SkewInvariants}
\end{center}
\end{figure}

From these expressions, and from the knowledge of the smooth potential ${\cal V}_6$ in Eq.~(\ref{Eq:V6}),
 one sees that as long as $\f$ does not diverge, both the Ricci scalar and the square of the Ricci tensor remain finite.
This is the case for the regular solutions that we called {\it confining}, but it also 
holds true for the {\it skewed} solutions, for which 
the singularity manifests itself only at the level of the square of the Riemann tensor.
In Figs.~\ref{Fig:ConfInvariants} and~\ref{Fig:SkewInvariants}, 
 we plot the curvature invariants for 
representative examples of solutions belonging to these two classes, which we chose to be the
analytical backgrounds with $\phi=0$, corresponding to the confining and skewed solutions, respectively.

\section{IR expansions of the generic singular solutions in Section~\ref{Sec:Divergent}}
\label{Sec:Details}

This appendix complements the discussion in
Section~\ref{Sec:Divergent}.
Explicit evaluation of the first terms in the series expansion
performed near the end of  the geometry, which would correspond to the deep IR 
of the field theory, including all terms with $n\leq 2$, yields the following:

\beqs
\phi(\r)&=&
\phi _I
    +\phi _L \log (\rho-\rho_o)-\frac{e^{2 \phi _I} \left(7 \phi
   _L+3\right)}{4 \left(\phi _L+1\right){}^2 \left(2 \phi
   _L+3\right)} (\rho-\rho_o)^{2 \phi _L+2}
   \nonumber+\\
   &&+
\frac{e^{-2 \phi _I} \left(7 \phi _L-3\right) }{3 \left(\phi
   _L-1\right){}^2 \left(2 \phi _L-3\right)}(\rho-\rho_o)^{2-2 \phi _L}  \nonumber+\\
   &&+
  \frac{e^{-6 \phi _I} \left(9-23 \phi _L\right)}{108 \left(1-3 \phi
   _L\right){}^2 \left(2 \phi _L-1\right)}  (\rho-\rho_o)^{2-6 \phi _L}+
\nonumber\\
&&   
   +\frac{e^{4 \phi _I} \left(\phi _L \left(\phi _L \left(79 \phi _L+197\right)+144\right)+30\right)}{8
   \left(\phi _L+1\right){}^4 \left(2 \phi _L+3\right){}^2 \left(4 \phi _L+5\right)} (\rho-\rho_o)^{4 \phi _L+4}
\nonumber+\\
&&
-\frac{\phi _L \left(12 \phi _L^4+19 \phi _L^2+9\right)}{6 \left(\phi _L^2-1\right){}^2 \left(4 \phi _L^2-9\right)}\,(\rho-\rho_o)^4+\\
&&
+\frac{e^{-4 \phi _I}(\rho-\rho_o)^{4-4\phi_L} }{108 \left(1-3 \phi_L\right){}^2 \left(3-2 
   \phi_L\right){}^2 \left(\phi _L-1\right){}^4 \left(\phi_L+1\right){}^2 \left(2 \phi _L-1\right) \left(2 \phi _L+3\right) \left(4 \phi _L-5\right)}\times\nonumber\\
    &&\times \left\{\frac{}{}\phi _L \left(\phi _L \left(\phi _L \left(\phi_L \left(\phi _L \left(\phi _L \left(\phi _L \left(\frac{}{}-78428 \phi _L^2+76312 \phi_L+75857\frac{}{}\right)
\right.\right.\right.\right.\right.\right.\right.\nonumber+\\
&&+\left.\left.\left.\left.\left.\left.\left.\frac{}{}
147745\right)-431341\right)+213007\right)+67591\right)
   -85329\right)+24561\right)-2295\frac{}{}\right\}
\nonumber+\\
&&
+\frac{e^{-8 \phi _I} \left(4958 \phi _L^4-8837 \phi _L^3+6004 \phi _L^2-1831 \phi _L+210\right)}{162 \left(16 \phi _L^2-34 \phi _L+15\right) 
\left(6 \phi _L^3-11
   \phi _L^2+6 \phi _L-1\right){}^2}
(\rho-\rho_o)^{4-8\phi_L}\nonumber+\\
&&
+\frac{e^{-12 \phi _I} \left(\phi _L \left(-5391 \phi _L^2+6591 \phi _L-2674\right)+360\right)}
{5832 \left(1-3 \phi _L\right){}^4 \left(1-2 \phi _L\right){}^2   \left(12 \phi _L-5\right)}
(\rho-\rho_o)^{4-12\phi_L}\nonumber
\,+\cdots
\,, \nonumber
\eeqs

\beqs
   \chi(\r)&=&
   \chi _I+
   \frac{1}{15}
   \left(4 \zeta\sqrt{1-5 \phi _L^2}+1\right) \log (\rho-\rho_o)\nonumber+\\
&&   +\frac{e^{2 \phi _I} \left(\phi _L-2 \zeta\sqrt{1-5 \phi _L^2}+1\right)}{6 \left(\phi _L+1\right){}^2 \left(2 \phi _L+3\right)} 
   (\rho-\rho_o)^{2
   \phi _L+2}\nonumber+\\
   &&
-\frac{2 e^{-2 \phi _I} \left(-2\zeta\sqrt{{1}-5{\phi _L^2}}-\phi_L+1\right) }{9 \left(\phi _L-1\right){}^2 \left(2 \phi
   _L-3\right)}(\rho-\rho_o)^{2-2 \phi _L}+\nonumber\\
&&   
 +  \frac{e^{-6 \phi _I} \left(-2 \zeta\sqrt{{1}-5{\phi _L^2}}-3\phi_L+1\right) }{162 \left(1-3 \phi _L\right){}^2 \left(2 \phi
   _L-1\right)}  (\rho-\rho_o)^{2-6 \phi _L}  +
   \nonumber\\
&&   
+\frac{e^{4 \phi _I} (\rho-\rho_o)^{4 \phi _L+4}}{24 \left(\phi _L+1\right){}^4 \left(2 \phi _L+3\right){}^2 \left(4 \phi _L+5\right)}
\left\{\phi _L \left(\phi _L \left(-24 \phi _L+44 \zeta\sqrt{1-5 \phi _L^2}
\right.\right.\right.\nonumber+\\
&&\left.\left.\left.
-71\right)
+92 \zeta\sqrt{1-5 \phi _L^2}-66\right)+44
   \zeta\sqrt{1-5 \phi _L^2}-19\right\}\,+ \nonumber\\
   &&
+   \frac{(\rho-\rho_o)^4}{45 \left(\phi _L^2-1\right){}^2   \left(4 \phi _L^2-9\right)}
\left\{-24 \left(\zeta\sqrt{1-5 \phi _L^2}-1\right) \phi _L^4\right.+\nonumber\\
&&
\left.+5 \left(2 \zeta\sqrt{1-5 \phi
   _L^2}+3\right) \phi _L^2-26 \zeta\sqrt{1-5 \phi _L^2}+1\right\}\,
   \nonumber+\\
&&
+\frac{e^{-4 \phi _I} (\rho-\rho_o)^{4-4\phi_L}}{324 \left(1-3 \phi _L\right){}^2 \left(3-2 \phi _L\right){}^2 \left(\phi
   _L-1\right){}^4 \left(\phi _L+1\right){}^2 \left(2 \phi _L-1\right) \left(2 \phi _L+3\right) \left(4 \phi _L-5\right)}\times\nonumber\\
   &&\times
\left\{\frac{}{}9 \left(344 \zeta\sqrt{1-5 \phi _L^2}-169\right)\frac{}{}
+\phi _L \left(\phi _L \left(5 \left(8468 \zeta\sqrt{1-5 \phi _L^2}-7123\right)\right.\right.\right.
+\nonumber\\
    &&\left.\left.\left.\left.
+\phi _L \left(88
   \left(433 \zeta\sqrt{1-5 \phi _L^2}+137\right)\right.\right.\right.\right.\right.\left.\left.\left.
+\,\phi _L \left(\phi _L \left(\phi _L \left(4 \phi _L \left(\phi _L \times\frac{}{}
\right.\right.\right.\right.\right.\right.\right.\nonumber\\
&&\left.\left.\left.\left.\left.\left.\left.
\times \left(-5864 \phi _L-10124 \zeta\sqrt{1-5 \phi
   _L^2}+7547\right)+928 \zeta\sqrt{1-5 \phi _L^2}+8595\right)
   \right.\right.\right.\right.\right.\right.\frac{}{}+\nonumber\\
    &&\left.\left.\left.\left.\left.\left.\left.
   +107972 \zeta\sqrt{1-5 \phi _L^2}-1353\right)+27964 \zeta\sqrt{1-5 \phi _L^2}-122406\right)+\right.\right.\right.\right.\right.
   \nonumber\\
   &&\left.\left.\left.\left.\left.-152272 \zeta\sqrt{1-5 \phi
   _L^2}+94701\right)\right)\right)-22740 \zeta\sqrt{1-5 \phi _L^2}+13026\right)\frac{}{}\right\}
\nonumber+\\
&&
+\frac{e^{-8 \phi _I}(\rho-\rho_o)^{4-8\phi_L} }{243 \left(1-3 \phi _L\right){}^2 
   \left(1-2 \phi _L\right){}^2 \left(\phi _L-1\right){}^2
   \left(2 \phi _L-3\right) \left(8 \phi _L-5\right)}\times\\
   &&\times
\left\{\phi _L \left(\phi _L \left(2 \phi _L \left(340 \phi _L+230 \zeta\sqrt{1-5 \phi _L^2}-617\right)
\right.\right.\right.\nonumber+\\
&&\left.\left.\left.
-682 \zeta\sqrt{1-5 \phi _L^2}+879\right)+356
   \zeta\sqrt{1-5 \phi _L^2}-290\right)-62 \zeta\sqrt{1-5 \phi _L^2}+37\right\}
\nonumber+\\
&&
+\frac{e^{-12 \phi _I}(\rho-\rho_o)^{4-12\phi_L}}{17496 \left(1-3 \phi _L\right){}^4 \left(1-2 \phi _L\right){}^2 \left(12 \phi _L-5\right)}\times\nonumber\\
&&\times  \left\{-3 \phi _L \left(3 \phi _L \left(168 \phi _L+76 \zeta\sqrt{1-5 \phi _L^2}-199\right)-188 \zeta\sqrt{1-5 \phi _L^2}+234\right)+\nonumber\right.\\
&&
\left.
-116 \zeta\sqrt{1-5 \phi
   _L^2}+91\right\}
\nonumber
\,+\cdots
\,,
\eeqs

\beqs
   A(\r)&=&
a_I+\frac{1}{15}
   \left(\zeta\sqrt{1-5 \phi _L^2}+4\right) \log (\rho-\rho_o)+
   \nonumber\\
&&+   \frac{e^{2 \phi _I} \left(8 \phi _L-\zeta\sqrt{1-5 \phi _L^2}+8\right) (\rho-\rho_o)^{2
   \phi _L+2}}{12 \left(\phi _L+1\right){}^2 \left(2 \phi _L+3\right)}\nonumber+\\
   &&
   -\frac{e^{-2 \phi _I} \left(-\zeta\sqrt{{1}-5{\phi _L^2}}-8 \phi_L+8\right) (\rho-\rho_o)^{2-2 \phi _L}}{9 \left(\phi _L-1\right){}^2 \left(2 \phi
   _L-3\right)}+\nonumber\\
   &&
   +\frac{e^{-6 \phi _I} \left(-\zeta\sqrt{{1}-5{\phi _L^2}}-24 \phi
   _L+8\right) (\rho-\rho_o)^{2-6 \phi _L}}{324 \left(1-3 \phi _L\right){}^2 \left(2 \phi
   _L-1\right)}
     +
     \nonumber\\
&&        
+\frac{e^{4 \phi _I} (\rho-\rho_o)^{4 \phi _L+4}}{24 \left(\phi _L+1\right){}^4 \left(2 \phi _L+3\right){}^2 \left(4 \phi _L+5\right)}
%\times\nonumber\\
%&&\times
\left\{ \left(\phi _L \left(\phi _L \left(-96 \phi _L+11 \zeta\sqrt{1-5 \phi _L^2}
\nonumber+\right.\right.\right.\right.\\
&&\left.\left.\left.\left.
-284\right)
+23 \zeta\sqrt{1-5 \phi _L^2}-264\right)+11
   \zeta\sqrt{1-5 \phi _L^2}-76\right)\right\}+
\nonumber\\
&&
+  \frac{(\rho-\rho_o)^4}{90 \left(\phi _L^2-1\right){}^2
   \left(4 \phi _L^2-9\right)}
   %\times\nonumber\\
   %&&\times
\left\{
-12 \left(\zeta\sqrt{1-5 \phi _L^2}-16\right) \phi _L^4
\right.+\\
&& \left.
+5 \left(\zeta\sqrt{1-5 \phi _L^2}+24\right) \phi _L^2-13 \zeta\sqrt{1-5 \phi _L^2}+8\right\}\,
   \nonumber+\\
&&
+\frac{e^{-4 \phi _I}(\rho-\rho_o)^{4-4\phi_L}}{324 \left(1-3 \phi _L\right){}^2 
   \left(3-2 \phi _L\right){}^2 \left(\phi _L-1\right){}^4 \left(\phi
   _L+1\right){}^2 \left(2 \phi _L-1\right) \left(2 \phi _L+3\right) \left(4 \phi _L-5\right)}\times\nonumber\\
   &&\times
 \left\{\phi _L \left(\phi _L \left(5 \left(2117 \zeta\sqrt{1-5 \phi _L^2}-28492\right)
 %+
%\right.\right.\right.\nonumber\\ 
%&&\left.\left.\left.
 +\phi _L 
\left(\phi _L \left(\phi _L \left(\phi _L \left(4 \phi
   _L \left(\phi _L \,\times\nonumber\frac{}{}
   \right.  \right.  \right.  \right.  \right.  \right. \right. \right.\\
   &&\left.\left.\left.\left.\left.\left.\left.\left.
   \times\left(-23456 \phi _L-2531 \zeta\sqrt{1-5 \phi _L^2}+30188\right)+4 \left(58 \zeta\sqrt{1-5 \phi _L^2}
   +8595\right)\right)+
   \right.\right.\right.\right.\right.\right.\right.\nonumber\\ 
&&\left.\left.\left.\left.\left.\left.\left.
   +26993 \zeta\sqrt{1-5 \phi
   _L^2}-5412\right)+6991 \zeta\sqrt{1-5 \phi _L^2}-489624\right)
      +\right.\right.\right.\right.\right.\nonumber\\
&&\left.\left.\left.\left.\left.\frac{}{}   
   -38068 \zeta\sqrt{1-5 \phi _L^2}+378804\right)
   +9526 \zeta\sqrt{1-5 \phi _L^2}+48224\right)\right)%+
      \right.\right.\nonumber+\\ 
&&\left.\left.
   -5685 \zeta\sqrt{1-5
   \phi _L^2}+52104\right)+774 \zeta\sqrt{1-5 \phi _L^2}-6084\right\}
\nonumber+\\
&&
+\frac{e^{-8 \phi _I}   (\rho-\rho_o)^{4-8\phi_L}}{486 \left(1-3 \phi _L\right){}^2 \left(1-2 \phi _L\right){}^2 \left(\phi _L-1\right){}^2
   \left(2 \phi _L-3\right) \left(8 \phi _L-5\right)}
     \times\nonumber\\
  &&\times 
 \left\{\phi _L \left(\phi _L \left(2 \phi _L \left(2720 \phi _L+115 \zeta\sqrt{1-5 \phi _L^2}-4936\right)
 \right.\right.\right.\nonumber+\\
 &&\left.\left.\left.
 -341 \zeta\sqrt{1-5 \phi _L^2}+7032\right)+178
   \zeta\sqrt{1-5 \phi _L^2}-2320\right)-31 \zeta\sqrt{1-5 \phi _L^2}+296\right\}
\nonumber+\\
&&
+\frac{e^{-12 \phi _I} (\rho-\rho_o)^{4-12\phi_L}}{17496 \left(1-3 \phi _L\right){}^4 \left(1-2 \phi _L\right){}^2 \left(12 \phi _L-5\right)}
%\times\nonumber\\
%&&\times
\left\{-3 \phi _L \left(3 \phi _L \left(672 \phi _L+19 \zeta\sqrt{1-5 \phi _L^2}
+\right.\right.\right.\nonumber\\
&&\left.\left.\left.
-796\frac{}{}\right)
-47 \zeta\sqrt{1-5 \phi _L^2}+936\right)-29 \zeta\sqrt{1-5 \phi
   _L^2}+364\right\}
%\nonumber+\\
%&&
\,+\cdots\nonumber
  \,.
\eeqs

In the numerical studies included in the body of the paper (e.g. in the calculations illustrated by Fig.~\ref{Fig:Plot}), we retained a few additional higher-order terms in these expressions in order to minimise noise and improve convergence of the numerical studies.

\section{IR expansions of the singular domain wall solutions in Section~\ref{Sec:SingDW}}
\label{Sec:SingDWDetails}

In this appendix, we show explicitly some of the terms in the
series expansion around the end-of-space of the geometry, 
for the solutions discussed in Section~\ref{Sec:SingDW}.
For convenience, we truncate the expansion at the order ${\cal O}((\r-\r_o)^4)$,
although we retained also a few additional higher-order terms 
 in some of the numerical calculations described in the main body of the paper.

\beqs
\phi(\r)&=&
\frac{1}{6} \log \left(\frac{9}{4}\right)+\frac{\log (\r-\r_o)}{3}+ {\phi_4}(\r-\r_o)^{4/9}-\frac{69}{85}  {\phi_4}^2\,(\r-\r_o)^{8/9}+\nonumber\\
&&
- \left(\frac{3
   \sqrt[3]{\frac{3}{2}}}{14}+\frac{6046 {\phi_4}^3}{1785}\right)\,(\r-\r_o)^{4/3}+\\
&&
+\frac{
   \left(6047325\times 2^{2/3} \sqrt[3]{3}\,{\phi_4}+241437236 {\phi_4}^4\right)}{7586250}   \,(\r-\r_o)^{16/9}\,+\nonumber\\
&&
   +\frac{ \left(-711782775\times 2^{2/3} \sqrt[3]{3}{\phi_4}^2-26218571272 {\phi_4}^5\right)}{146667500}\,(\r-\r_o)^{20/9} +\nonumber\\
&&
+\frac{ (\r-\r_o)^{8/3} }{11519265450000}\left\{\frac{}{}3^{2/3}276494428125 \sqrt[3]{2}  +\right.\nonumber\\
&&\left.+296581753795800\ 2^{2/3} \sqrt[3]{3} {\phi_4}^3+  9755979537544064 {\phi_4}^6\nonumber
 \frac{}{} \right\}\,+\,\\
  &&+(\r-\r_0)^{28/9} \left\{
    -\frac{5428197 \left(\frac{3}{2}\right)^{2/3} \phi_4}{1994300}
  -\frac{2360772721213 \sqrt[3]{\frac{3}{2}} \phi_4^4}{9496720625}
 \,+ \right.\nonumber\\
&&\left.\frac{}{}  
  -\frac{26498325334552196 \phi_4^7}{7264991278125}\nonumber
  \right\}\,+
  \\
 && +\frac{(\r-\r_o)^{32/9}}{773620668463671875}
   \left\{\frac{}{} 6151561447643611875 \sqrt[3]{2} 3^{2/3} \phi_4^2
+
   \nonumber\right.\\
   &&\left.\frac{}{}+
   432480088524305869575\times 2^{2/3} \sqrt[3]{3}\phi_4^5 \nonumber
  +   11365219568455804037008\phi_4^8
   \right\}\,+
   \\
   && +   \frac{(\r-\r_o)^4}{155961926762276250000} \left\{\frac{}{}  -2413524250949484375\,+
   \right.\nonumber\\
&&\left.   -7074613549111472685000 \sqrt[3]{2}\, 3^{2/3}  \phi_4^3 
 \right.\nonumber+\\
  &&\left.\frac{}{} 
 -371140233877399872460300\times   2^{2/3} \sqrt[3]{3}\phi_4^6
  \right.\nonumber+\\
  &&\left.\frac{}{} 
  -8759105677773799489866912\phi_4^9\nonumber
\right\}\,+\,\cdots\,,
\eeqs

\beqs
   \chi(\r)&=&
   \chi_I+\frac{1}{27}\log (\r-\r_o)
   +\frac{2}{5}{\phi_4}\, (\r-\r_o)^{4/9} \nonumber
   -\frac{244}{255}{\phi_4}^2\, (\r-\r_o)^{8/9}+\\
&&   
   + \left(\frac{1}{7} \sqrt[3]{\frac{3}{2}}+\frac{14668 {\phi_4}^3}{5355}\right)(\r-\r_o)^{4/3}+ \\
&&
   -\frac{3
   \left(166175\times 2^{2/3} \sqrt[3]{3} {\phi_4}+6954744 {\phi_4}^4\right)}{2528750}\, (\r-\r_o)^{16/9}+\nonumber\\
&&
+\frac{\left(2318168025\times 2^{2/3} \sqrt[3]{3} {\phi_4}^2+84249694712 {\phi_4}^5\right)}{3300018750}\,(\r-\r_o)^{20/9} +\nonumber\\
&&
+\frac{(\r-\r_o)^{8/3}}{17278898175000} \left\{-  3^{2/3}108193471875 \sqrt[3]{2}+\nonumber\frac{}{}\right.\\
&&\left.\frac{}{}-45915207465600\times 2^{2/3} \sqrt[3]{3} {\phi_4}^3- 1414212386352128 {\phi_4}^6
\right\}\,\,+
 \nonumber \\
&&      
   +\frac{(\r-\r_o)^{28/9} }{319659616237500}
   \left\{\frac{}{}
      58007569300875  \sqrt[3]{2} \,3^{2/3}  \phi_4
      \,+\right.\nonumber\\
&& \left.\frac{}{}
   +3409626494789460\times 2^{2/3} \sqrt[3]{3} \,\phi_4^4
   +89150405935813024 \phi_4^7
   \frac{}{}\right\}
\nonumber \\
&&       
   +\frac{(\r-\r_o)^{32/9}}{41775516097038281250}
   \left\{\frac{}{}
      -38576134638208906875 \sqrt[3]{2}\, 3^{2/3} \phi_4^2
      \,+\right.\nonumber\\
&& \left.\frac{}{}
         -1921139818985605118175\times 2^{2/3}   \sqrt[3]{3}  \phi_4^5
   -43318844111573477101952  \phi_4^8
   \frac{}{}\right\}\,+
\nonumber \\
&&    
+\frac{(\r-\r_o)^4}{3509143352151215625000} \left\{\frac{}{}
14718312628622578125
   \,+\right.\nonumber\\
&& \left.\frac{}{}
   +18585357769872307035000 \sqrt[3]{2} \,3^{2/3} \phi_4^3
   \,+\right.\nonumber\\
&& \left.\frac{}{}
      +744000468587964720553300\times   2^{2/3} \sqrt[3]{3} \phi_4^6
       \right.\nonumber+\\
  &&\left.\frac{}{} 
+14846531094788772880019552  \phi_4^9
   \frac{}{}\right\} \,+\,\cdots\,.\nonumber
\eeqs

The domain wall condition  $A=4\chi$ restores (locally) Poincar\'e invariance in $D=5$ dimensions.

\section{Singular domain wall solutions: lift to $D=10$ dimensions}
\label{sec:LiftTo10D}

This appendix discusses the lift of the solutions to massive type-IIA supergravity in $D=10$ dimensions. We focus on the ten-dimensional metric, which in the Einstein frame is given by
\beq
	\dd s_{10}^2 = (\sin(\xi))^{1/12} X^{1/8} \Delta^{3/8} \left( \dd s_6^2 + \dd \tilde \Omega_4^2 \right) \,,
\eeq
where
\beqs
	X &=& e^{\phi} \,, \\
	\Delta &=& X^{-3} \sin^2(\xi) + X \cos^2(\xi) \,, \\
	\dd \tilde \Omega_4^2 &=& X^2 \dd \xi^2 + X^{-1} \Delta^{-1} \cos^2(\xi) \frac{1}{4} \left[ \dd \theta^2 + \sin^2(\theta) \dd \varphi^2 + (\dd \psi + \cos(\theta) \dd \varphi)^2 \right] \,,
\eeqs
and the ranges of the angles, describing the internal four-sphere, are
\beq
	0 \leq \theta \leq \pi , \hspace{1cm} 0 \leq \varphi < 2\pi , \hspace{1cm} 0 \leq \psi < 4\pi , \hspace{1cm} - \frac{\pi}{2} \leq \xi \leq \frac{\pi}{2} \,.
\eeq
The detailed expressions for the remaining non-zero background fields, the dilaton and the Ramond-Ramond four-form, can be found in Refs.~\cite{Cvetic:1999un,Jeong:2013jfc}.\footnote{Our conventions are such that they agree with Section~3.1.3 of Ref.~\cite{Elander:2013jqa} putting $g = 1$.}

Because of the factor $\sin(\xi)^{1/12}$ in the ten-dimensional metric, all the solutions we consider are singular at $\xi = 0$. For non-zero values of $\xi$, the behaviour of the curvature invariants differs depending on the different classes considered in the body of this paper, as we shall now see. For simplicity, consider the ten-dimensional Ricci scalar evaluated at $\xi = \pi/2$, given by
\beq
	R^{(10)} \big|_{\xi = \frac{\pi}{2}} = 6 e^{-\phi} + 9 e^{3\phi} +\frac{1}{2} e^{\phi} \left( 12 \mathcal V_6(\phi) + \partial_\phi \mathcal V_6(\phi) + 4 (\partial_\rho \phi)^2 \right) \,.
\eeq
As can be seen, $R^{(10)} \big|_{\xi = \frac{\pi}{2}}$ remains finite as long as $\phi$ remains finite as a function of $\rho$. This is the case for the confining and skewed solutions. The class of badly singular domain-wall solutions introduced in Section~\ref{Sec:SingDW} plays a prominent role in our analysis, being the energetically favoured branch for $\hat \phi_2 > \hat \phi_2^c$. Using the IR expansion given in Eq.~\eqref{eq:IRexpBadlySingularDW}, we obtain that
\beq
	R^{(10)} \big|_{\xi = \frac{\pi}{2}} = \frac{5 \left(\frac{2}{3}\right)^{2/3}}{9} \rho^{-5/3} + \frac{\left(\frac{2}{3}\right)^{2/3} \phi_4}{9} \rho^{-11/9} + \mathcal O(\rho^{-7/9})\,,
\eeq
confirming the singular nature of these solutions also in $D = 10$ dimensions (even away from $\xi = 0$).

\section{Mass spectra in units of $\Lambda$}
\label{sec:SpectraLambda}

\begin{figure}[t]
\includegraphics[width=16cm]{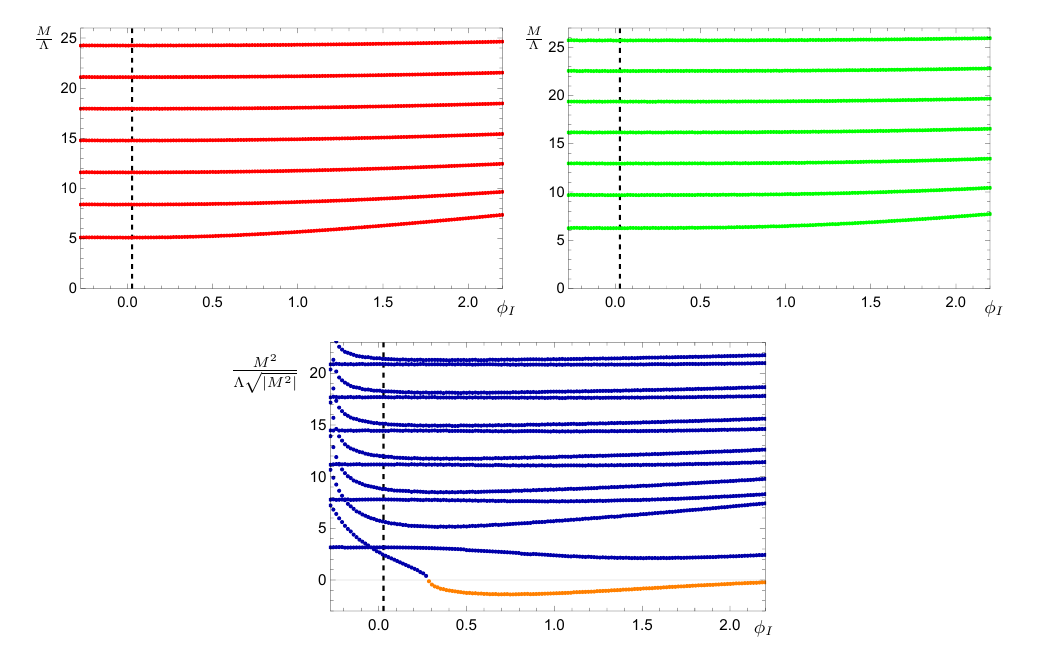}
\caption{The spectra of masses $M$, as a function of the one free parameter $\phi_I$ 
	characterising the confining solutions, 
	and normalised in units of $\Lambda$, computed with $\r_1 = 10^{-4}$ and $\r_2 = 12$. From top to bottom, left to right, 
	the spectra of fluctuations of the tensors $\mathfrak{e}^{\mu}_{\ \nu}$ (red), 
	the gravi-photon $V_{\mu}$ (green) and the two scalars 
	$\phi$ and $\chi$ (blue). The orange points in the plot of the scalar 
	mass spectrum represent values of $M^{2}<0$ and hence denote a tachyonic state. 
	We also show by means of the vertical dashed lines the case $\phi_I= \phi_I^c>0$, the critical value
	  that is introduced and discussed in Section~\ref{Sec:Free}.}
\label{Fig:SpectrumLambda1}
\end{figure}

\begin{figure}[t]
\includegraphics[width=16cm]{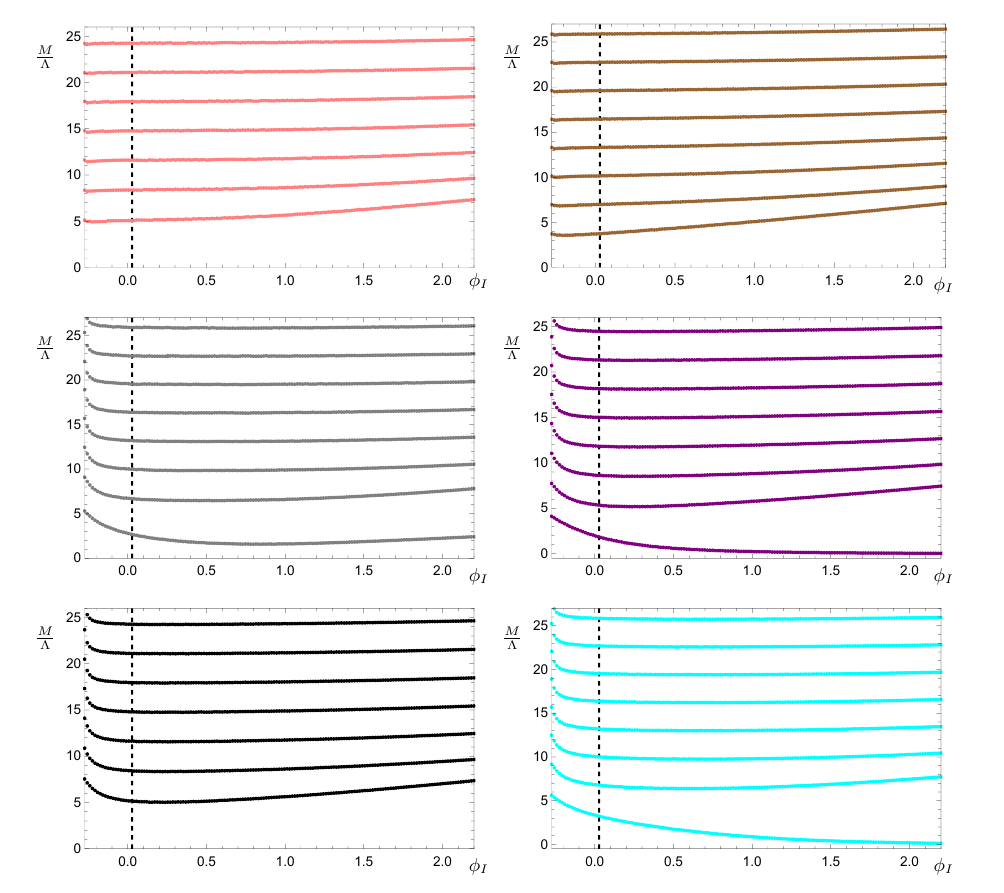}
\caption{The spectra of masses $M$ as a function of the scale parameter $\phi_I$, 
	normalised in units of $\Lambda$. From top to bottom, left to right, the spectra of fluctuations 
	of the pseudo-scalars $\pi^{i}$ forming a triplet \textbf{3} of $SU(2)$
	 (pink), vectors  $A^{i}_{\mu}$ forming a triplet \textbf{3} of $SU(2)$ (brown), 
	 $U(1)$ pseudo-scalar $X$ (grey), 
	 $U(1)$ transverse vector $B_{6\mu}$ (purple), 
	 $U(1)$ transverse vector $X_{\mu}$ (black) and 
	 the massive U(1) 2-form $B_{\mu\nu}$ (cyan). 
	 The spectrum was computed using the regulators $\r_1 = 10^{-4}$ and $\r_2 = 12$ with the exception of the $U(1)$ pseudo-scalar $X$ for which we used $\r_1 = 10^{-7}$ in order to minimize the cutoff effects present for the very lightest state at large values of $\phi_I$.
	We also show by means of the vertical dashed lines the case $\phi_I= \phi_I^c>0$, the critical value
	  that is introduced and discussed in Section~\ref{Sec:Free}.}
\label{Fig:SpectrumLambda2}
\end{figure}

In this appendix, we show the mass spectra normalised in units of the scale $\Lambda$, in order to facilitate the comparison with the results of Section~\ref{Sec:Free}. The results are depicted in Figs.~\ref{Fig:SpectrumLambda1} and~\ref{Fig:SpectrumLambda2}. The only purpose of these plots is to allow the Reader to easily relate the scale setting procedures we used in the calculaton of the spectrum and of the phase structure.

\section{A few parameteric plots}
\label{Sec:ParametricPlots}

\begin{figure}[t]
\includegraphics[width=16cm]{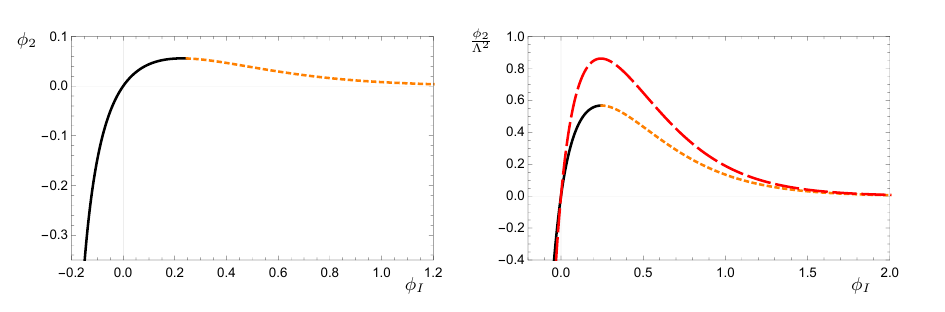}
\caption{Plots showing the relationship between the UV expansion parameter 
	$\f_{2}$ and the IR parameter $\f_{I}$, in the solutions we called {\it confining} (solid black and short-dashed orange lines) and {\it skewed} (dashed red line).
	The left plot shows the bare parameters extracted: the solid black and dashed red lines agree, as
	 with $\f_{2}^{conf}=\f_{2}^{skew}$, $\f_{I}^{conf}=\f_{I}^{skew}$. The right plot shows the same parameters
	  after rescaling with the appropriate powers of $\L$.}
\label{Fig:phi2}
\includegraphics[width=16cm]{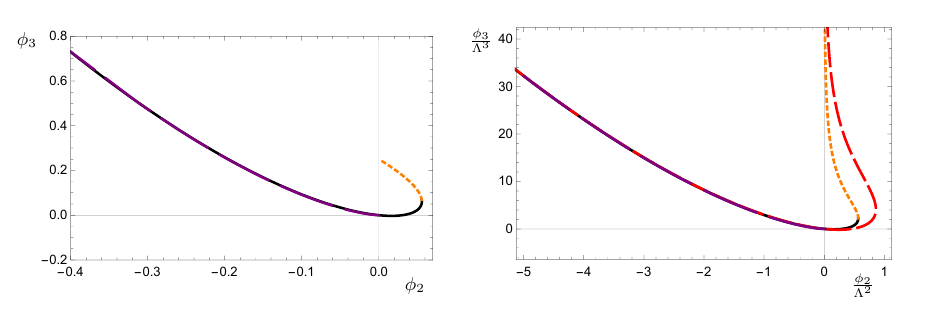}
\caption{Plots showing the relationship between the two UV expansion parameters $\f_{2}$ and $\f_{3}$ for solutions belonging to the confining (solid black and short-dashed orange lines), skewed (dashed red line), and IR-conformal (longest-dashed purple line) classes. 
	The left plot shows the base parameters extracted by matching to the UV expansions, 
	with $\f_{2}^{conf}=\f_{2}^{skew}$, $\f_{3}^{conf}=\f_{3}^{skew}$. The right panel shows the same parameters
	 after rescaling with the appropriate powers of $\L$. (For $\phi_2 \leq 0$, although the confining, skewed, and IR-conformal classes are not in complete agreement, they are close enough that in these plots the solid black and dashed red lines are hidden behind the longest-dashed purple one.)}
\label{Fig:phi3}
\end{figure}

\begin{figure}[t]
\includegraphics[width=16cm]{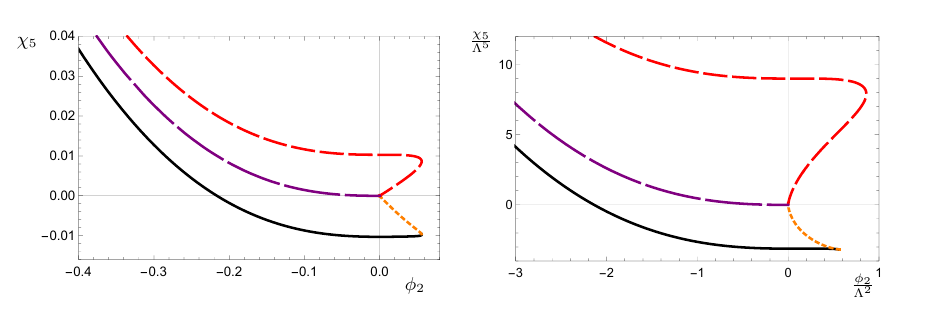}
\caption{Plots showing the relationship between the two UV expansion parameters $\f_{2}$ and $\c_{5}$ for  solutions within the confining (solid black and short-dashed orange lines), skewed (dashed red line), 
	and IR-conformal (longest-dashed purple line) classes. The left plot shows the parameters extracted by matching to the UV expansions, 
	with $\f_{2}^{conf}=\f_{2}^{skew}$, $\f_{3}^{conf}=\f_{3}^{skew}$, and $\c_{5}^{skew}=-\c_{5}^{conf}-\frac{8}{25}\f_{2}^{conf}\f_{3}^{conf}$. The right panel shows the same parameters after rescaling with the appropriate powers of $\L$.}
\label{Fig:chi5}
\includegraphics[width=16cm]{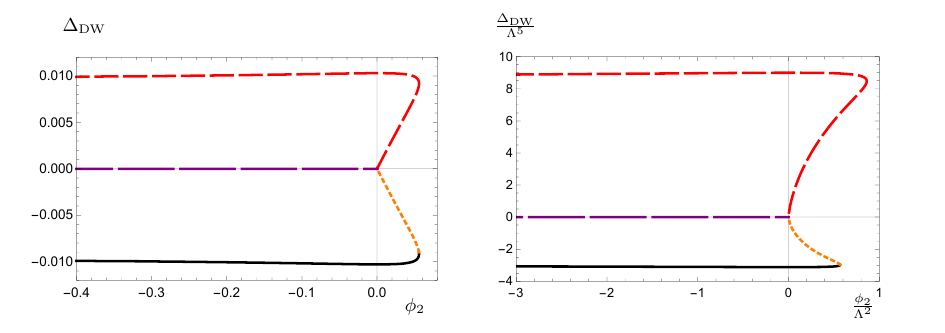}
\caption{The coefficient $\c_{5}+\frac{4}{25}\phi_2\phi_3 \equiv \Delta_{\rm DW}$ appearing in the expansion of the function $d$ in Eq.~(\ref{Eq:dd}), for solutions within the confining (solid black and short-dashed orange lines), skewed (dashed red line), and IR-conformal (longest-dashed purple line) classes. The left plot shows the parameters extracted by matching to the UV expansions. The right panel shows the same parameters after rescaling with the appropriate powers of $\L$.}
\label{Fig:chi5phi2(shift)}
\end{figure}

In this appendix we show some additional details of the numerical results we obtained 
by studying the confining and skewed solutions, and their approach to the trivial 
critical point for large values $\r$ of the radial direction.
In the main body of the text, we focused most of our attention on the values of the parameter $\phi_2$,
and on the free-energy density ${\cal F}$ along the various branches of solutions.
We show here how the other parameters, $\phi_3$, $\chi_5$ and $\phi_I$ evolve along the 
two special branches of solutions that we called {\it confining}, {\it skewed}, and {\it IR-conformal}. These parameters are extracted by following the procedure outlined in Section~\ref{sec:numimp}, and correspond to the values obtained in step~3.~of the list describing the numerical implementation.

The main qualitative features that emerge from Figs.~\ref{Fig:phi2}, ~\ref{Fig:phi3}, ~\ref{Fig:chi5}, and~\ref{Fig:chi5phi2(shift)}
are similar to what we have already described in the main text.
We notice that when studying $\phi_2$, $\phi_3$ and $\chi_5$ as a function of $\phi_I$,
two different regimes emerge.
For negative values of $\phi_I$, all the physically interesting UV parameters show a monotonic,
unbounded dependence on $\phi_I$ itself.
When $\phi_I>0$, the fact that a maximum value of $\phi_2$ is reached at finite $\phi_I$ gives rise to the 
non-trivial shape of the curves shown in the three figures. We find it useful to show also the results along the IR-conformal branch of solutions where appropriate.

\section{Formulation  of the free energy in $D=5$ dimensions}
\label{sec:FreeEnergyin5D}

In this appendix we rewrite the same system of Section~\ref{Sec:FreeGeneral} in the language of a sigma-model of two scalars $\phi$ and $\chi$
in $D=5$ dimensions. We remind the Reader that this is derived by assuming that none of the background fields depend on $\eta$, and then performing
dimensional reduction of the system. As detailed elsewhere~\cite{Elander:2018aub},
for the bulk action one finds that
\beqs
{\cal S}_{bulk}&=&\int \di \eta \left\{ \tilde{{\cal S}}_{bulk}
+\frac{1}{2}\int \di^4 x \di r \partial_M\left(\sqrt{-g_5} g^{MN}\partial_N \chi\right)\right\}\,,
\eeqs
with
\beqs
\tilde{{\cal S}}_{bulk}&=&\int \di^4 x \di r \sqrt{-g_5}\left(\frac{{\cal R}_5}{4}-\frac{1}{2}G_{ab}
g^{MN}\partial_M\Phi^a\partial_N \Phi^b -{\cal V}(\phi,\chi)\right)\,,
\eeqs
where the sigma-model metric is $G_{ab}={\rm diag} (2,6)$ in the basis $\{\phi,\chi\}$,
and the potential is ${\cal V}=e^{-2\chi}{\cal V}_6$.

Hence, by just replacing the equations of motion we find
\beqs
\tilde{{\cal S}}_{bulk}&=&-\frac{3}{8}\int_{\r_1}^{\r_2} \di^4 x \di \rho \partial_{\rho}\left(e^{4A-\chi} \partial_{\rho} A \right)
-\frac{1}{2}\int_{\r_1}^{\r_2} \di^4 x \di \rho \partial_{\rho}\left(e^{4A-\chi}\partial_{\rho} \chi\right)\,.
\eeqs
The boundary-localised GHY term at $\r=\r_2$ now reads\footnote{The sign of the term proportional to $\partial_{\rho} A$ is the opposite of that which is stated just after Eq.~(2.23) of Ref.~\cite{Elander:2018aub}.}
\beqs
\tilde{{\cal S}}_{GHY,2}&=&\left.\int \di^4x\sqrt{-\tilde{g}}\frac{\tilde{K}}{2}\right|_{\r_2}\,=\,
\left.2 \int \di^4x e^{4A-\chi}\partial_{\rho} A\right|_{\r_2}\,.
\eeqs
The boundary-localised potential term at $\r=\rho_2$ reads
\beqs
\tilde{{\cal S}}_{pot,2}&=&\left.\int\di^4 x \sqrt{-\tilde{g}}\left(\frac{}{}\tilde{\lambda}_2\frac{}{}\right)\right|_{\r_2}
\,=\,\left.\int\di^4 x e^{4A}\left(\frac{}{}\tilde{\lambda}_2\frac{}{}\right)\right|_{\r_2}\,,
\eeqs
which by comparing to Eq.~(\ref{Eq:Spot}) implies that we must
choose $\tilde{\lambda}_2\equiv e^{-\chi}\lambda_2$.

In the five-dimensional language, even regular solutions in six dimensions may be singular---in the sense that the curvature singularity in $D=5$ dimensions is resolved by the lift to $D=6$ dimensions, which makes it more transparent to understand
why we need to introduce the boundary at $\r=\r_1$.
The resulting contributions to the action are
\beqs
\tilde{{\cal S}}_{GHY,1}&=&-\left.\int \di^4x\sqrt{-\tilde{g}}\frac{\tilde{K}}{2}\right|_{\r_1}\,=\,
-\left.2 \int \di^4x e^{4A-\chi}\partial_{\rho} A\right|_{\r_1}\,,\\
\tilde{{\cal S}}_{pot,1}&=&-\left.\int\di^4 x \sqrt{-\tilde{g}}\left(\frac{}{}\tilde{\lambda}_1\frac{}{}\right)\right|_{\r_1}
\,=\,-\left.\int\di^4 x e^{4A}\left(\frac{}{}\tilde{\lambda}_1\frac{}{}\right)\right|_{\r_1}\,,
\eeqs
which again implies that $\tilde{\lambda}_1\equiv e^{-\chi} \lambda_1$.

We notice how the GHY terms in the description in $D=5$ dimensions
combine with the total derivative distinguishing ${\cal S}_{bulk}$ and $\tilde{\cal S}_{bulk}$
to yield exactly the GHY term of the formulation in $D=6$ dimensions.
Hence, we have now shown that we can match the two formulations of the theory:
\beqs
\label{eq:S6andS5actions}
{\cal S}_{bulk}+\sum_{i=1,2}{\cal S}_{GHY,i}+{\cal S}_{pot,i}
&=&
\int \di \eta \left(\frac{}{}\tilde{{\cal S}}_{bulk}+\sum_{i=1,2}\left(\tilde{{\cal S}}_{GHY,i}+\tilde{{\cal S}}_{pot,i}
\right)\right)\,.
\eeqs
Note that matching the formulations in $D=6$ and $D=5$ dimensions as in Eq.~\eqref{eq:S6andS5actions} does not require making use of the equations of motion.

\end{document}